\documentclass[a4paper, onecolumn, 10pt, accepted=2025-04-08]{quantumarticle}
 \pdfoutput=1
\usepackage{geometry}
\usepackage{amsfonts}
\usepackage{amsmath}
\usepackage{amssymb}
\usepackage{mathrsfs}
\usepackage{graphicx}
\usepackage{hyperref}
\usepackage{cancel}
\usepackage{cleveref}
\usepackage{bbm}
\usepackage{bm}
\usepackage{mathtools}
\usepackage{physics}
\usepackage{tensor}
\usepackage{array}
\usepackage{float}
\usepackage{wrapfig}
\usepackage{soul}
\usepackage{xcolor}
\usepackage{cite}
\usepackage{url}
\usepackage{tabu}
\usepackage{tikz}
\usepackage{anyfontsize}
\PassOptionsToPackage{compress}{natbib}
\usepackage[numbers]{natbib}

\newcommand{\ebd}{\vcentcolon =}

\newcommand\setItemnumber[1]{\setcounter{enumi}{\numexpr#1-1\relax}}

\newcommand{\upsigma}{\ket{\uparrow_\sigma}}

\newcommand{\downsigma}{\ket{\downarrow_\sigma}}

\newcommand{\up}{\ket{\uparrow}}

\newcommand{\down}{\ket{\downarrow}}

\newcommand{\one}{\mathbbm{1}}

\makeatletter
\newcommand{\fmarki}{*}
\newcommand{\fmarkii}{\ensuremath{\dagger}}
\newcommand{\fmarkiii}{\ensuremath{\ddagger}}
\newcommand{\fmarkiv}{\ensuremath{\mathsection}}

\def\@fnsymbol#1{{\ifcase#1\or \fmarki\or \fmarkii\or \fmarkiii \or \fmarkiv\else\@ctrerr\fi}}
\makeatother

\renewcommand{\fmarki}{$\natural$}
\renewcommand{\fmarkii}{$\star$}
\renewcommand{\fmarkiii}{$\diamond$}

\title{Doubly Quantum Mechanics}

\author[V. D'Esposito]{Vittorio D'Esposito}
\affiliation{Dipartimento di Fisica Ettore Pancini, Universit\`a di Napoli ``Federico II'', Complesso Univ. Monte S. Angelo, I-80126 Napoli, Italy}
\affiliation{INFN, Sezione di Napoli, I-80126 Napoli, Italy}
\email{vittorio.desposito@unina.it}

\author[G. Fabiano]{Giuseppe Fabiano}
\affiliation{Physics Division, Lawrence Berkeley National Laboratory, Berkeley, CA}
\affiliation{Department of Physics, University of California, Berkeley, CA 94720, USA}
\affiliation{Centro Ricerche Enrico Fermi—Museo Storico della Fisica e Centro Studi e Ricerche “Enrico Fermi”, Roma}
\email{gfabiano@lbl.gov}

\author[D. Frattulillo]{Domenico Frattulillo}
\affiliation{INFN, Sezione di Napoli, I-80126 Napoli, Italy}
\email{domenico.frattulillo@na.infn.it}

\author[F. Mercati]{Flavio Mercati}
\affiliation{Departamento de F\'isica, Universidad de Burgos, 09001 Burgos, Spain}
\email{flavio.mercati@gmail.com}

\begin{document}

\maketitle

\begin{abstract}
    Motivated by the expectation that relativistic symmetries might acquire quantum features in Quantum Gravity, we take the first steps towards a theory of ``Doubly'' Quantum Mechanics, a modification of Quantum Mechanics in which the geometrical configurations of physical systems, measurement apparata, and reference frame transformations are themselves quantized and described by ``geometry'' states in a Hilbert space. We develop the formalism for spin-$\frac{1}{2}$ measurements by promoting the group of spatial rotations $SU(2)$ to the quantum group $SU_q(2)$ and generalizing the axioms of Quantum Theory in a covariant way. As a consequence of our axioms, the notion of probability becomes a self-adjoint operator acting on the Hilbert space of geometry states, hence acquiring novel non-classical features. After introducing a suitable class of semi-classical geometry states, which describe near-to-classical geometrical configurations of physical systems, we find that probability measurements are affected, in these configurations, by intrinsic uncertainties stemming from the quantum properties of $SU_q(2)$. This feature translates into an unavoidable fuzziness for observers attempting to align their reference frames by exchanging qubits, even when the number of exchanged qubits approaches infinity, contrary to the standard $SU(2)$ case.
\end{abstract}

\section{Introduction}

When Quantum Theory and General Relativity are combined in any of the many approaches to Quantum Gravity \cite{Oriti:2009zz}, the common lore is that spacetime should acquire quantum properties, in one form or another. Most approaches to the Quantum Gravity problem suggest that some fundamental geometric notions that are pervasively used in physics might be ``quantized'', so that there are fundamental limitations to the measurability of the observables that depend on them.
The most studied possibility is that Quantum Gravity effects, which are supposedly characterized by a length scale of the order of the Planck length $\ell_P \sim 1.6 \times 10^{-35} m$, determine a small-scale discreteness or fuzzyness in dimensionful quantities like lengths (or distances), areas, and (possibly hyper) volumes \cite{Gross:1987ar,Amati:1987uf,Adler:1999bu,Rovelli:1994ge, Ashtekar:1996eg, Ashtekar:1997fb, Freidel:2002hx, Amelino-Camelia:2008fcv, Percacci:2010af}. One of the most natural settings to realize such a notion of ``quantum spacetime'' is provided by non-commutative geometry, where coordinates are promoted to operators satisfying non-trivial commutation relations \cite{Snyder:1947, Majid:2006xn, Szabo:2009tn, Pachoł_2013}, and the group structure of empty spacetime symmetries is replaced by the notion of quantum groups, where group parameters are also promoted to non-commutative operators \cite{majid_1995}.

Despite the fact that the physical regimes we have access to are far away from the Planck scale, decades of studies have now consolidated several phenomenological proposals of effects that are within our current experimental sensitivity \cite{Amelino-Camelia:1997ieq, Jacob:2006gn, Amelino-Camelia:2008aez, Addazi:2021xuf, AlvesBatista:2023wqm}, with preliminary but promising results \cite{Amelino-Camelia:2008aez, Amelino-Camelia:2009wvc, Stecker:2014oxa, Amelino-Camelia:2016fuh, Amelino-Camelia:2016ohi, Huang:2018ham, Amelino-Camelia:2022pja, DEsposito:2023psn}. Moreover, in recent years there has also been a growing interest in investigating putative Quantum Gravity effects in the regime of non-relativistic Quantum Mechanics. In fact, in this context it is possible to question the logical consistency of the very fundamental assumptions of General Relativity and Quantum Theory \cite{Hardy:2005fq, Oreshkov:2011er, Zych:2015fka, Belenchia:2018szb, Castro-Ruiz:2019nnl, Giacomini:2021gei, Foo:2021exb, Oppenheim:2022xjr, Foo:2022dnz, Galley:2023byb, Chen:2024xvm, delaHamette:2024xax, Goel:2024vtr}, as well as to concretely realize table-top experiments to test these assumptions at the interface between the two theories \cite{Kovachy:2015xcp, Rosi:2017ieh, Marletto:2017kzi, Bose:2017nin, Zych:2017tau, Goswami:2018rda, Christodoulou:2018cmk, Westphal:2020okx, Cepollaro:2021ccc, Christodoulou:2022mkf, Overstreet:2022zgq}.  In the following, we focus on this physical regime by investigating the effects of quantum group deformations on quantum mechanical systems, a largely unexplored area \cite{Arzano:2022nlo, Amelino-Camelia:2022dsj, Amelino-Camelia:2023rkg} since these deformations are typically considered in the contexts of classical relativistic mechanics \cite{Amelino-Camelia:2000stu, Bruno:2001mw, Kowalski-Glikman:2002iba, Amelino-Camelia:2002cqb, Kowalski-Glikman:2002oyi, Kowalski-Glikman:2004fsz, Amelino-Camelia:2007yca, Amelino-Camelia:2011lvm, Amelino-Camelia:2011gae, Amelino-Camelia:2023srg} and quantum field theory \cite{Fiore:2007vg,Szabo:2001kg,Lizzi:2021rlb,DiLuca:2022idu,Fabiano:2023xke}. 

The aim of this study is to lay the foundations of a ``Doubly Quantum Mechanics" (DQM) theory, where  the geometrical configuration of physical systems and measurement apparata, as well as the relation between reference frames, are described by elements of the Hilbert space relative to the quantum group describing the (deformed) symmetries of the quantum spacetime under study. Such a framework is doubly quantum since not only the phase space of physical systems is quantized, but also their geometrical configurations. A convenient setting for moving the first steps towards this formulation is the spin sector of standard Quantum Mechanics, where physical results concerning spin systems solely depend on the geometrical degrees of freedom of the spin system itself and the Stern-Gerlach apparata used to prepare and/or measure it. The doubly quantization procedure in this context is realized by replacing the standard $SU(2)$ rotation symmetry with its unique quantum deformation $SU_q(2)$ \cite{Woronowicz:1987vs},  where the dimensionless deformation parameter $q \in \mathbbm{C}$ is such that $q=1$ reproduces the standard $SU(2)$ group. In the context of Quantum Gravity, the study of quantum groups that involve a dimensionless deformation scale, such as $SU_q(2)$, is motivated by the observation that General Relativity does not only describe the geometrical properties of spacetime, which have to do with distances and volumes, but also its \emph{conformal} geometry, \emph{i.e.} angles. One piece of theoretical evidence in favour of quantum/non-classical angles \cite{Mikusch:2021kro, Amelino-Camelia:2022dsj} in Quantum Gravity is that the introduction of a cosmological constant in different approaches, such as Loop Quantum Gravity/Spin Foam models \cite{Major:1995yz, Freidel:1998pt, Smolin:2002sz} and Group Field Theory \cite{Girelli:2022foc}, requires the deformation of the local gauge group from $SU(2)$ to $SU_q(2)$. In these models, the deformation parameter $q$ is a function of the dimensionless ratio between the Planck length and the Hubble length scale associated with the cosmological constant such that $q \sim 1$. In fact, it has been argued that this reflects a minimal possible resolution in angular measurements, in a universe that is characterized by a fundamental discreteness (a short-distance cut-off) and a cosmological horizon (a large-distance cut-off) \cite{Bianchi:2011uq, Calmet:2021sws}.

We analyze the conceptual and phenomenological novelties introduced by promoting the standard rotation symmetry group $SU(2)$ to its quantum version $SU_q(2)$. This is implemented as a quantization of the complex coefficients characterizing a generic spin state and a generic Pauli matrix, which in standard Quantum Mechanics are completely specified by $SU(2)$ elements, thus requiring a doubly quantum mechanical description of the system. With the same line of reasoning, the complex coefficients of the $SU(2)$ elements describing the relation between different observers will also be affected by the same type of quantization. We will show that the formalism naturally yields a quantization of the Born rule, which requires measurement probabilities to be described by self-adjoint and positive semi-definite operators acting on the Hilbert space associated to the $SU_q(2)$ quantum group, introducing a novel operational meaning for probability measurements.

The paper is structured as follows. In \Cref{sec:spin measurement classic} we recall the formalism required to describe spin measurements in standard Quantum Mechanics, emphasizing the relation between the relevant physical observables and $SU(2)$. We review the necessary mathematical tools of $SU_q(2)$ in \Cref{sec:mathprel}, then present the framework of Doubly Quantum Mechanics for spin measurements in \Cref{sec:spin measurement doubly quantum} in an axiomatic way, and introduce the concept of quantum probability. In \Cref{sec:effective theory alice} we propose a class of semi-classical geometry states describing small deviations from classical geometrical configurations of physical systems. We employ these states in \Cref{sec:deformed protocol} to describe an alignment protocol between two observers whose reference frames are generally misaligned. Contrary to the classical case, where two observers can sharply align their reference frames by exchanging an infinite number of spins, we find that the quantum properties of $SU_q(2)$ yield a fundamental limitation to their alignment.

\section{Spin measurements in Quantum Mechanics}\label{sec:spin measurement classic}
In Quantum Mechanics, rotational symmetry is governed by the $SU(2)$ group. All the machinery required for quantum mechanical computations with spinors can be derived starting from three basic ingredients, the Pauli matrix $\sigma_z$ and its eigenstates $\ket{\uparrow},\ket{\downarrow}$, given by
\begin{equation}
    \sigma_z=\begin{pmatrix}
    1 & 0 \\
    0 & -1
\end{pmatrix}\, , \quad  \ket{\uparrow}=\begin{pmatrix}
    1 \\
    0
\end{pmatrix}\,,  \quad   \ket{\downarrow}=\begin{pmatrix}
    0 \\
    1 
\end{pmatrix}\;.
\end{equation}
Indeed, a generic spin up state $\ket{\uparrow_{\vec{n}}}$ oriented along a direction $\vec{n}=(\sin \theta_s \cos \omega_s, \sin \theta_s \sin \omega_s, \cos \theta_s)$ where $(\theta_s,\omega_s) \in [0,\pi]\times[0,2\pi]$, can be written as
\begin{equation}\label{eq:spin classic}
    \ket{\uparrow_{\vec{n}}} = U_s\ket{\uparrow} = \begin{pmatrix}
    x & -y^*\\
    y & x^*
\end{pmatrix}\begin{pmatrix}
    1 \\
    0
\end{pmatrix} = x \up + y \down\;\,,\,\;\;x,y \in \mathbb{C}\;,\;\;x^*x + y^*y =1 \;,
\end{equation}
namely by acting with an $SU(2)$ matrix on the spin up state.
These complex parameters can be represented in terms of angular variables as  
\begin{equation}
    x = e^{i \chi_s}\cos{\frac{\theta_s}{2}}\, , \qquad y = e^{i\phi_s} \sin{\frac{\theta_s}{2}}  \;, \qquad \chi_s,\phi_s\in [0,2\pi]\; ,
    \label{rappresentazione classica a e c methods}
\end{equation}
so that the resulting spin state can also be written as
\begin{equation}
    \label{eq:upn}\ket{\uparrow_{\vec{n}}}=e^{i \chi_s}\cos \frac{\theta_s}{2}\ket{\uparrow}+e^{i\phi_s}\sin\frac{\theta_s}{2}\ket{\downarrow} \cong \cos \frac{\theta_s}{2}\ket{\uparrow}+e^{i\omega_s}\sin\frac{\theta_s}{2}\ket{\downarrow}\;, 
\end{equation}
where $\cong$ indicates equivalence of rays in the Hilbert space, \emph{i.e.} up to global phases. Analogously, the generic spin down state is given by
\begin{equation} \label{eq:upd}
    \ket{\downarrow_{\vec{n}}} = -y^*\up + x^*\down \cong -e^{-i\omega_s}\sin\frac{\theta_s}{2}\ket{\uparrow}+\cos \frac{\theta_s}{2}\ket{\downarrow}\;.
\end{equation}
Since states are defined up to a global phase, we have defined $\omega_s=\phi_s-\chi_s$ to match the angle appearing in the generic direction $\vec{n}$ by omitting the irrelevant global phase $e^{i\chi_s}$.

It is possible to measure the spin along a generic direction $\vec{m}=(\sin \theta_a \cos \omega_a, \sin \theta_a \sin \omega_a, \cos \theta_a)$ on a generic spin state, using a Stern-Gerlach apparatus oriented in direction $\vec{m}$. The quantum mechanical observable associated with this measurement is described by the Pauli matrix $\sigma_{\vec{m}}$, obtained by acting on $\sigma_z$ with a $SU(2)$ matrix
\begin{equation}
\label{eq:sigmam}
    \sigma_{\Vec{m}}=U_a \sigma_z U_a^\dagger = \begin{pmatrix}
        a a^*-c^*c& 2ac^*\\ 2ca^*&cc^*-a^*a
    \end{pmatrix}=\begin{pmatrix}
       \cos\theta_a& e^{-i\omega_a}\sin\theta_a\\ e^{i\omega_a}\sin\theta_a&-\cos\theta_a
    \end{pmatrix} \;,
\end{equation}
where
\begin{equation}\label{eq:su2 stern gerlach classic}
   U_a = \begin{pmatrix}
    a & -c^*\\
    c & a^*
\end{pmatrix}\;\,,\;\;a = e^{i\chi_a}\cos{\frac{\theta_a}{2}} \;,\;\; c = e^{i\phi_a} \sin{\frac{\theta_a}{2}}\;.
\end{equation}
Notice that \eqref{eq:sigmam} can be rewritten in terms of projector operators $\Pi_{\uparrow_{\vec{m}}}$, $\Pi_{\downarrow_{\vec{m}}}$ as
\begin{equation}\label{Pauligenericprojectorsclassic}
     \sigma_{\vec{m}} =  \Pi_{\uparrow_{\vec{m}}}- \Pi_{\downarrow_{\vec{m}}} \ebd   \ketbra{\uparrow_{\vec{m}}} -\ketbra{\downarrow_{\vec{m}}}\;.
\end{equation}
These projectors define the probabilities of finding spin up and down in the direction $\vec{m}$ by performing a measurement on a spin up state along a direction $\Vec{n}$. These are given by

\begin{equation}\label{eq:probability classical} 
    \begin{aligned}
        &\begin{aligned}
            P_{\uparrow_{\Vec{m}}}\qty(\uparrow_{\vec{n}}) &=\mel{\uparrow_{\Vec{n}}}{\Pi_{\uparrow_{\vec{m}}}}{\uparrow_{\Vec{n}}} = x^*xaa^* + y^*ycc^*+ x^*yac^* + y^*xca^*=\\
            &=\frac{1}{2}\Big[1+\cos(\theta_a)\cos(\theta_s)+\cos(\omega_a-\omega_s)\sin(\theta_a)\sin(\theta_s)\Big]\;,
            \end{aligned}\\
            &\begin{aligned}
                P_{\downarrow_{\Vec{m}}}\qty(\uparrow_{\vec{n}})  &= \mel{\uparrow_{\Vec{n}}}{\Pi_{\downarrow_{\vec{m}}}}{\uparrow_{\Vec{n}}} = x^* x  c^*c + yy^* a^*a -  x^* y c^*a -  y^*x  a^*c = \\
                &=\frac{1}{2}\Big[1-\cos(\theta_a)\cos(\theta_s)-\cos(\omega_a-\omega_s)\sin(\theta_a)\sin(\theta_s)\Big] \,,
            \end{aligned}
    \end{aligned}
\end{equation}
so that the expectation value of $\sigma_{\vec{m}}$ in $\ket{\uparrow_{\vec{n}}}$ is

\begin{equation}\label{Pauliclassicamedia}
    \expval{\sigma_{\vec{m}}} = P_{\uparrow_{\Vec{m}}}\qty(\uparrow_{\vec{n}})-P_{\downarrow_{\Vec{m}}}\qty(\uparrow_{\vec{n}})\;.
\end{equation}
The quantities of interest for this result are the probabilities, which are expressed in terms of both couples of parameters $\{x,y\}$ and $\{a,c\}$, defining the directions of the spin state and the measurement device, respectively.  
With the same techniques, it is possible to calculate similar expressions when the spin is in a generic down state. 

The physical setup described above is relative to a single observer, Alice, in her reference frame. If we consider a second observer, Bob, whose axes are misaligned with Alice's, we can obtain Bob's description of Alice's physical results by applying a $SU(2)$ symmetry transformation, denoted by $U_g$, on both states and observables as

\begin{equation}
    \label{eq:covcomm}\ket{\uparrow_{\vec{n}}^B}=U_g\ket{\uparrow_{\vec{n}}^A}\;\,,\;\;\,\sigma_{\vec{m}}^{B}=U_g\sigma_{\vec{m}}^{A} {U_g}^{\dagger} \; ,
\end{equation}
where $\ket{\uparrow_{\Vec{m}}^A}$ and $\sigma_{\vec{m}}^{A}$ are the generic spin up state and Pauli matrix in Alice's description, respectively given by \eqref{eq:upn} and \eqref{eq:sigmam}. The matrix $U_g$ is given by
\begin{equation}
     U_g = \begin{pmatrix}\label{eq:classical relative orientation}
    u  & -v^*\\
    v & u^*
    \end{pmatrix} \;\,,\;\;u = e^{i\chi_g}\cos{\frac{\theta_g}{2}} \;,\;\; v = e^{i\phi_g} \sin{\frac{\theta_g}{2}}\;.
\end{equation}
Contrary to the definition of spins and Pauli matrices along generic directions for the single observer, both phases $\chi_g$ and $\phi_g$ are relevant since the relative orientation between two observers is defined by three angles, whereas a generic direction in space is specified only by two angles.
Of course, given that $U_g$ is unitary, the probabilities of spin measurements and the expectation value of the spin are invariant under this transformation. The covariance of the framework under $SU(2)$ transformations is explicitly verified by observing that the spin up state and the generic Pauli matrix in Bob's reference frame can be rewritten in the same form of Alice's spin up state and generic Pauli matrix. Indeed, by expanding \eqref{eq:covcomm}, it is possible to show that by defining
\begin{equation}
\begin{aligned}
 & x'= u x -  \,v^* y\qquad y'= v x + u^* y\\
&   a'= u a -  \,v^* c\qquad c'= v a + u^* c  
  \end{aligned}
\end{equation}
we can write
\begin{equation}
\begin{gathered}
    \ket{\uparrow_{\vec{n}}^B} =  x' \up   + y' \down  \;\\ \\
    \sigma_{\vec{m}}^{B}=\begin{pmatrix}   a' a'^*-c'^*c'& 2a'c'^*\\ 2c'a'^*&c'c'^*-a'^*a' \, ,
\end{pmatrix}\;,
    \end{gathered}
\end{equation}
which have the same form of \eqref{eq:upn} and \eqref{eq:sigmam}, respectively. This means that the eigenvalues of the transformed Pauli matrix are still given by $\{-1, 1\}$ and that the expressions to compute spin observables are the same in form for each observer.

\section{Mathematical preliminaries}\label{sec:mathprel}

In this paper, we are interested in developing a framework in which the $SU(2)$ rotational symmetry of \Cref{sec:spin measurement classic} is promoted to a quantum group symmetry implications
described by $SU_q(2)$ \cite{Xing-ChangSong_1992, Schmidt:2007yp, franz2013idempotent, Amelino-Camelia:2002cqb, Kowalski-Glikman:2004fsz}. Before diving into the physical construction, we present some basic features of this quantum group. First of all, a remark is in order, to clarify what we mean with ``quantum group'' in general, for the benefit of the readers that might be unfamiliar with the concept. A quantum group is a Lie group $G$ with an additional non-commutative product, besides the (non-commutative as well, in the case of non-abelian groups) group product. This non-commutative structure deforms the \textit{algebra of functions} on $G$, denoted by $C\qty(G)$, with its commutative pointwise product $(f_1 \cdot f_2) (g) = f_1(g) \,  f_2(g)$, by making said product non-commutative. This new non-commutative product needs to satisfy certain compatibility conditions with the Lie group structures (group product, inverse and identity element), which make $C\qty(G)$ into a \textit{Hopf algebra} \cite{Chari:1994pz, majid_1995}.
In the case of $SU(2)$, we can specify the new non-commutative structure that promotes it to a quantum group by working with the algebra $C\qty(SU(2))$, which is generated by two \textit{coordinate functions}, $a$ and $c$ in \eqref{eq:su2 stern gerlach classic} (and their complex conjugates). These coordinates will be promoted to the generators $\alpha$ and $\gamma$ of the following non-commutative algebra
\begin{equation}
\label{eq:algebrasuq2}
\begin{aligned}
& \alpha \gamma=q \, \gamma \alpha \qquad \alpha \gamma^*= q \,  \gamma^*\alpha \qquad \gamma\gamma^*=\gamma^*\gamma \\
& \gamma^*\gamma+\alpha^*\alpha = \mathbbm{1} \qquad \alpha \alpha^*-\alpha^*\alpha=(1-q^2)\gamma^*\gamma\;,
\end{aligned}
\end{equation}
where, in general, $q \in \mathbbm{C}$, although in the present work we will focus only on the real case $q\in(0,1)$,  and $*$ denotes the Hermitian conjugation. The deformed algebra is denoted by $C\qty(SU_q(2))$. In terms of these operators, the non-commutative counterpart of the 2-dimensional matrix representation of the group \eqref{eq:su2 stern gerlach classic} is written as \cite{Woronowicz:1987wr}
\begin{equation}
U^q=\begin{pmatrix}\label{su2quantum}
    \alpha & -q\gamma^*\\
    \gamma & \alpha^* 
\end{pmatrix} \; .
\end{equation}
In order to complete the Hopf algebra structure of $C\qty(SU_q(2))$, one has to specify how the group product, inverse and identity element are deformed. These can be found, for example, in \cite{franz2013idempotent}, but will not be used in our construction, for the moment. Instead, we will focus on a representation of the  algebra \eqref{eq:algebrasuq2}, which in turn enables the analysis of the states on the algebra. These represent the non-commutative generalization of the notion of points, and encode all the informations regarding the limitations to the localization of regions in the group manifold that are implied by the non-commutativity \cite{Murphy:1990, Lizzi:2018qaf, Lizzi:2019wto}. These structures will play a prominent role in the present paper, which is ultimately focused on the physical consequences of the relaxation of locality implied by quantum groups: if a certain point on the classical $SU(2)$ group cannot be localized perfectly in the $SU_q(2)$ quantum group, it means that the associated rotation between reference frames is not realizable with arbitrary precision. Notice that, when we talk about representations of $SU_q(2)$, we mean a realization of the algebra \eqref{eq:algebrasuq2} as linear operators acting on a Hilbert space (which is, indeed, the space of ``fuzzy points'' on the group). This notion of representation has nothing to do with the representations of the $SU(2)$ group (\textit{i.e.} spin 1/2, spin 1 \textit{etc.}), and the two should not be confused.

The representations of $\alpha,\gamma$ have been thoroughly studied in the literature, e.g. \cite{Xing-ChangSong_1992, franz2013idempotent}.
The Hilbert space containing the two unique irreducible representations of the $SU_q(2)$ algebra, when $q\in(0,1)$, is $\mathcal{H}_{SU_q(2)} = \mathcal{H}_{\pi} \oplus \mathcal{H}_\rho$ where $\mathcal{H}_{\pi} = \ell^2\otimes L^2(S^1)\otimes L^2(S^1)$ and $\mathcal{H}_{\rho} = L^2(S^1)$. If $\phi,\chi \in [0,2\pi[$ are coordinates on $S^1$ and $\ket{n}$ is the canonical basis of $\ell^2$, the algebra of functions on $SU_q(2)$ is represented as
\begin{equation}
    \rho(\alpha) \ket{\chi}=e^{i\chi}\ket{\chi}\qquad \rho(\alpha^*)\ket{\chi}=e^{-i\chi}\ket{\chi} \qquad \rho(\gamma)\ket{\chi}=\rho(\gamma^*)\ket{\chi}=0
    \label{rappresentazione rho e rhostar}
\end{equation}
\begin{gather}
    \pi(\alpha)\ket{n, \phi,\chi}=e^{i\chi}\sqrt{1-q^{2n}}\ket{n-1, \phi,\chi} \qquad \pi(\gamma)\ket{n, \phi,\chi}=e^{i\phi}q^{n}\ket{n, \phi,\chi}\nonumber
    \\ \label{rappresentazione pi} \\
    \pi(\alpha^*)\ket{n, \phi,\chi}=e^{-i\chi}\sqrt{1-q^{2n+2}}\ket{n+1, \phi,\chi} \qquad \pi(\gamma^*)\ket{n, \phi,\chi}=e^{-i\phi}q^{n}\ket{n, \phi,\chi}\;.\nonumber
\end{gather}
In the following sections, we will consider multiple copies of $C(SU_q(2))$ in which we will label the quantum numbers appearing in their representations by subscripts $\{s,a,g\}$ to distinguish the role of the various $SU_q(2)$ transformations, as done in \Cref{sec:spin measurement classic}. Moreover, from now on in all of our expressions we will implicitly refer to the representation of the $SU_q(2)$ operators on $\mathcal{H}_{SU_q(2)}$, thereby omitting the symbols $\rho$ and $\pi$ to simplify the notation.

\section{Spin measurements in Doubly Quantum Mechanics}\label{sec:spin measurement doubly quantum}

In classical mechanics, the state of a physical system at a given time is completely characterized by a point in phase space. In non-relativistic Quantum Mechanics phase-space points are replaced by rays in a Hilbert space. Both theories share the same classical Euclidean background: the quantum properties of a spin system are encoded in the fact that its state can be described in terms of a superposition of spin states, which, however, are always relative to some classical direction in space. 
When taking the quantum spacetime hypothesis at face value, it is plausible to imagine that a necessary step toward the understanding of the quantum gravity problem should require a transition from Quantum Mechanics to a novel theory, in which also the geometrical properties of physical systems and the transformation laws between observers are described by states in a Hilbert space. We present an example of such a \emph{Doubly Quantum Mechanical} model, limiting our attention to the physics of spin measurements. Specifically, the standard rotation symmetry under the $SU(2)$ group is promoted to a deformed symmetry under the quantum group $SU_q(2)$,  the only possible  quantum deformation of its classical counterpart, which has been employed in some models of quantum gravity \cite{Major:1995yz, Freidel:1998pt, Smolin:2002sz, Girelli:2022foc}.
The classical $SU(2)$ group parameters become operators satisfying the commutation relations \eqref{eq:algebrasuq2} and the states on which they act encode the relevant information concerning the orientation in space of a physical system.

From an operational point of view, in the commutative case, an observer can extract relevant physical information from a spin state only by performing some measurements on the latter with a Stern-Gerlach apparatus. A spin measurement is thus characterized by two independent three-vectors in space ($\vec m$ and $\vec n$ in \Cref{sec:spin measurement classic}): one defining the spin orientation and the other the Stern-Gerlach orientation. If we want to carry on this operational framework to the non-commutative case, we are led to characterize the physical information relevant to an observer by means of two independent geometry states. One of them encodes information about the spin orientation and the other contains information about the Stern-Gerlach orientation.

In this regard, a spin system becomes a doubly quantum mechanical system, since its full description now requires two different states living in different Hilbert spaces. The Stern-Gerlach apparatus, albeit being a classical measurement device from the point of view of standard Quantum Mechanics, acquires quantum properties in Doubly Quantum Mechanics, since its orientation in space becomes characterized by a geometry state.

In the following, we adopt an axiomatic approach to extend the basic principles of Quantum Mechanics when the quantization of rotation symmetry by means of $SU_q(2)$ is taken into account. By adapting the logical steps outlined in \Cref{sec:spin measurement classic} to our Doubly Quantum Mechanics formalism, we derive the fundamental quantities describing a spin measurement in this context.

\subsection{The axioms of Doubly Quantum Mechanics}\label{sec: axioms Alice}

\vspace{7pt}\noindent\textbf{Axiom 0} (Geometry)\quad
The information on the directions in space of physical systems and the relative orientation between reference frames is encoded in \emph{geometry states}, elements of the Hilbert space $\mathcal{H}_{SU_q(2)}$.

\vspace{7pt}\noindent\textbf{Axiom 1} (Pre-measurement states)\quad The states prepared by an observer, referred to as \emph{pre-measurement states}, are spin $\frac{1}{2}$ states with coefficients that are operators acting on the Hilbert space $\mathcal{H}_{SU_{q}(2)}$. These states are elements of $\widehat{\mathcal{H}}_{\frac{1}{2}} \ebd \mathcal{H}_{\frac{1}{2}}\otimes C(SU_q(2))$, where $\mathcal{H}_{\frac{1}{2}}$ is the Hilbert space of a single spin ($\mathbbm{C}^2$ with the standard inner product), and $ C(SU_q(2))$ is the algebra of functions on $SU_q(2)$. In the following, to simplify the notation, we omit the tensor product between the $\mathcal{H}_{\frac{1}{2}}$ component and the operator coefficients. 

The generic spin up pre-measurement state is obtained, analogously to the generic spin up state in \eqref{eq:spin classic}, by acting on $\up\mathbbm{1}$ with an $SU_q(2)$ matrix

\begin{equation}
\label{eq:genericstate}
    \ket{\psi} = U^q_{s}\Big(\up\mathbbm{1}\Big) = \up x  + \down  y\;,
\end{equation}
with
\begin{equation}
    U^q_{s} \ebd \begin{pmatrix}\label{su2quantumspin}
    x & -qy^*\\
    y & x^*
\end{pmatrix} = \ketbra{\uparrow}x  -q \ketbra{\uparrow}{\downarrow} y^* + \ketbra{\downarrow}{\uparrow}y + \ketbra{\downarrow}x^*\;,
\end{equation}
where $\mathbbm{1}$ is the identity operator in the $SU_q(2)$ algebra and the operators $x,y\in C(SU_q(2))$ and their Hermitian conjugates satisfy the $SU_q(2)$ algebra relations \eqref{eq:algebrasuq2}, with $x = \alpha$ and $y = \gamma$. 
The bra corresponding to \eqref{eq:genericstate} is given by
\begin{equation}
    \bra{\psi}=\bra{\uparrow}x^* + \bra{\downarrow} y^* \; ,
\end{equation}
and is such that
\begin{equation}
    \bra{\psi}\ket{\psi}= x^* x  +  y^*y = \one \; ,
\end{equation}
where the $\braket{\cdot}{\cdot}$ notation denotes the scalar product on the $\mathcal{H}_{\frac{1}{2}}$ Hilbert space and the pointwise product between the operator coefficients. The last equality follows from the commutation relations of the $SU_q(2)$ algebra \eqref{eq:algebrasuq2}. This is just the non-commutative analogue of the fact that a generic state is normalized in the commutative setting.

Analogously to \eqref{eq:upd}, the generic spin down state can be obtained by acting with $U^q_s$ on $\down\one$ as done for \eqref{eq:genericstate}
\begin{equation}\label{generic spin down state}
    \ket{\psi^\prime}=-q \up \, y^* +\down x^* \; ,
\end{equation}
which is also normalized since
\begin{equation}
    \braket{\psi^\prime}{\psi^\prime} = q^2 xx^* + yy^* = \one \;.
\end{equation}
Moreover, $\ket{\psi}$ and $\ket{\psi^\prime}$ satisfy $\braket{\psi^\prime}{\psi} = \braket{\psi}{\psi^\prime} = 0$.

\vspace{7pt}\noindent\textbf{Axiom 2} (Observables)\quad A generic Pauli matrix, representing the observable of a spin measurement, is an element of $\mathcal{D}(\mathcal{H}_{\frac{1}{2}})\otimes C(SU_q(2))$, where $\mathcal{D}(\mathcal{H}_{\frac{1}{2}})$ is the Hilbert space of linear operators on $\mathcal{H}_{\frac{1}{2}}$. Again we omit the tensor product between the $\mathcal{D}(\mathcal{H}_{\frac{1}{2}})$ component and the operator coefficients. The possible outcomes of a spin measurement are given by the eigenvalues of the corresponding self-adjoint Pauli matrix.

The generic Pauli matrix is obtained, analogously to the one in \eqref{eq:sigmam}, by conjugating the $q$-deformed Pauli $z$ matrix with an $SU_q(2)$ matrix 

\begin{equation}\label{eq:pauligeneric}
    \sigma^q=U^q_{a} \,\sigma_z^q\, {U^q_{a}}^\dagger =\begin{pmatrix}  q\qty(\mathbbm{1}-\qty(1+q^2)c^*c)
         & (q+q^{-1}) ac^* \\
        (q+q^{-1}) ca^* & -q^{-1}\qty(\mathbbm{1}-\qty(1+q^2)c^*c)
    \end{pmatrix}\;,\;\;
\end{equation}
where
\begin{eqnarray}
    U^q_{a} \ebd \begin{pmatrix}\label{su2quantumpauli}
     a & - q c^*\\
   c &   a^*
\end{pmatrix} = \ketbra{\uparrow}  a -q \ketbra{\uparrow}{\downarrow}c^* + \ketbra{\downarrow}{\uparrow} c + \ketbra{\downarrow}  a^*\;,
\end{eqnarray}
with the operators $a, c \in C\qty(SU_q(2))$ and their Hermitian conjugates defining a second copy of the $SU_q(2)$ algebra, where the generators $a$ and $c$ (and their Hermitian conjugates $a^*$, $c^*$) play the same role of $\alpha$ and $\gamma$ (resp. $\alpha^*$, $\gamma^*$) in \eqref{eq:algebrasuq2}. Notice that \eqref{eq:pauligeneric} is the non-commutative generalization of the generic Pauli matrix \eqref{eq:sigmam} that represents a Stern-Gerlach apparatus oriented along a generic direction and is self-adjoint in the sense that $\qty(\sigma^q)^*_{kh}=\qty(\sigma^q)_{hk}$.

The matrix $\sigma_z^q$, is the $q$-deformation of $\sigma_z$ and reads \cite{Xing-ChangSong_1992}
\begin{equation}
\label{eq:sigmaqz}
    \sigma^q_z=\begin{pmatrix}
        q & 0 \\
        0 & -q^{-1}
    \end{pmatrix}\mathbbm{1} = q \ketbra{\uparrow}\mathbbm{1}-q^{-1}\ketbra{\downarrow}\mathbbm{1}\;.
\end{equation}
We interpret this matrix as the one characterizing a Stern-Gerlach apparatus oriented along the positive $z$ direction. In the commutative case, $\sigma_z$ is traceless and the generic Pauli matrix obtained by conjugation with $U\in SU(2)$ is still traceless. In the present case, $\sigma_z^q$ is $q$-traceless \cite{Schmidt:2007yp} and the generic Pauli matrix in \eqref{eq:pauligeneric}, obtained by conjugation wtih $U^q_{a}\in SU_q(2)$, is still $q$-traceless, where the $q$-trace is defined as
\begin{equation}
    \Tr_q\big\{A\big\} \ebd \sum_iq^{2i}A_{ii}\;.
\end{equation}

Analogously to the classical case,  \eqref{eq:pauligeneric} can be rewritten in terms of the non-commutative generalization of projectors along directions specified by the generic Pauli matrix, which we denote by
$\Pi_{\uparrow_\sigma}$, $\Pi_{\downarrow_\sigma}$, as

\begin{equation}\label{Pauligenericprojectors}
     \sigma^q = q\, \Pi_{\uparrow_\sigma} -q^{-1}\, \Pi_{\downarrow_\sigma}\ebd q  \ketbra{\uparrow_\sigma} -q^{-1}\ketbra{\downarrow_\sigma}\;,
\end{equation}
where
\begin{equation}
\label{eq:upsigma}
    \upsigma=\up a + \down c,\quad \downsigma=-q\up c^* + \down  a^*\;, 
\end{equation}
with

\begin{equation}
    \braket{\uparrow_\sigma}{\uparrow_\sigma} = \braket{\downarrow_\sigma}{\downarrow_\sigma} = \mathbbm{1}\;,\;\; \braket{\uparrow_\sigma}{\downarrow_\sigma} = \braket{\downarrow_\sigma}{\uparrow_\sigma} = 0\;.
\end{equation}
From \eqref{Pauligenericprojectors}, it immediately follows that $\upsigma$ ($\downsigma$) is an eigenstate of $\sigma^q$ with eigenvalue $q$ ($-q^{-1}$). The explicit expressions for $\ketbra{\uparrow_\sigma}$ and $\ketbra{\downarrow_\sigma}$ are given by
\begin{equation}\label{projectors}
    \begin{aligned}
        &\Pi_{\uparrow_\sigma} = \ketbra{\uparrow_\sigma}=\ketbra{\uparrow}aa^* + \ketbra{\downarrow}   c c^* + \ketbra{\uparrow}{\downarrow} ac^*+\ketbra{\downarrow}{\uparrow} ca^* \\
        &\Pi_{\downarrow_\sigma} = \ketbra{\downarrow_\sigma}=\ketbra{\uparrow} q^2 cc^* + \ketbra{\downarrow}   a^*a - \ketbra{\uparrow}{\downarrow}  qc^*a-\ketbra{\downarrow}{\uparrow}  qa^*c
    \end{aligned}\;.
\end{equation}
Therefore the projectors $\Pi_{\uparrow_\sigma}$ and $\Pi_{\downarrow_\sigma}$ satisfy the following properties

\begin{equation}\label{properties projectors}
    \Pi_{\uparrow_\sigma}^2 = \Pi_{\uparrow_\sigma}\;,\;\; \Pi_{\downarrow_\sigma}^2 = \Pi_{\downarrow_\sigma}\;,\;\; \Pi_{\uparrow_\sigma}\Pi_{\downarrow_\sigma} = \Pi_{\downarrow_\sigma}\Pi_{\uparrow_\sigma} = 0\;,\;\; \Pi_{\uparrow_\sigma} + \Pi_{\downarrow_\sigma} = \mathbb{I} \otimes \one\;,
\end{equation}
where $\mathbb{I} \ebd \ketbra{\uparrow}+\ketbra{\downarrow}$.

This axiom introduces an important conceptual departure from standard Quantum Mechanics for what concerns the nature of measurement apparata as their geometrical configurations are now specified by quantum geometry states in the Hilbert space $\mathcal{H}_{SU_q(2)}$. This opens up the possibility of describing ``non-classical measurement devices", characterized by geometrical configurations that do not have a classical counterpart. We will not delve into this feature of our framework in this paper, postponing this task to future work.

\vspace{7pt}\noindent\textbf{Axiom 3} (Probabilities and expectation values)\quad The expectation value of an observable $\mathcal{O} \in \mathcal{D}\qty(\mathcal{H}_{\frac{1}{2}})\otimes C(SU_q(2))$ on a generic state $\ket{\psi} \in \mathcal{H}_{\frac{1}{2}}\otimes C(SU_q(2))$ is defined as a map
\begin{equation}\label{eq:map definition scalar product}
    \expval{\mathcal{O}} \vcentcolon \qty[\mathcal{H}_{\frac{1}{2}}\otimes C(SU_q(2))] \times \qty[\mathcal{D}\qty(\mathcal{H}_{\frac{1}{2}})\otimes C(SU_q(2))] \ni \ket{\psi}\times \mathcal{O} \mapsto \mel{\psi}{\mathcal{O}}{\psi} \in C(SU_q(2))\otimes C(SU_q(2))
\end{equation}
At the practical level, this map performs the standard expectation value in the $\mathcal{H}_{1/2}$ component, which is then multiplied by a coefficient that is given by the tensor product between the operator coefficients of $\mathcal{O}$ (on the right side of the tensor product) and the product between the operator coefficients of $\bra{\psi}$ and the operator coefficients of $\ket{\psi}$ (on the left side of the tensor product). Of course, the ordering of the tensor product is just a choice that does not affect any result. Analogously, this map can be defined on a generic spin state $\ket{\psi^\prime}$.

With this definition, the non-commutative generalizations of the probabilities of finding spin $\uparrow_\sigma$ or spin $\downarrow_\sigma$ on a generic spin up state $\ket{\psi}$ are elements of $C\qty(SU_q(2))\otimes C\qty(SU_q(2))$ and can be defined as
\begin{equation}
\label{eq:probabilityquantumdefinition}
    \begin{aligned}
        &P\qty(\uparrow_\sigma) \ebd \expval{\Pi_{\uparrow}} = \mel{\psi}{\Pi_{\uparrow_\sigma}}{\psi} = \braket{\psi}{\uparrow_\sigma}\braket{\uparrow_\sigma}{\psi}\;, \\ &P\qty(\downarrow_\sigma)\ebd \expval{\Pi_{\downarrow}} = \mel{\psi}{\Pi_{\downarrow_\sigma}}{\psi} = \braket{\psi}{\downarrow_\sigma}\braket{\downarrow_\sigma}{\psi}\;.
    \end{aligned}
\end{equation}
The non-commutative generalizations of the probabilities of finding spin $\uparrow_\sigma$ or spin $\downarrow_\sigma$ on a generic spin down state $\ket{\psi^\prime}$ are given by the same formula with $\ket{\psi^\prime}$ replacing $\ket{\psi}$. By definition, the probabilities defined in \eqref{eq:probabilityquantumdefinition} are self-adjoint and positive semi-definite operators, and from \eqref{properties projectors} it immediately follows that 
\begin{equation}
    P\qty(\uparrow_\sigma) + P\qty(\downarrow_\sigma) = \one \otimes \one\;.
\end{equation}
Therefore, they satisfy the desirable properties that probabilities must have. The explicit expressions for $P\qty(\uparrow_\sigma)$ and $P\qty(\downarrow_\sigma)$ are given by
\begin{equation}\label{probabilities Alice}
\begin{aligned}
    &P\qty(\uparrow_\sigma)=x^*x\otimes a a^* + y^*y\otimes cc^* + x^*y\otimes ac^*+y^*x\otimes ca^* \\
    &P\qty(\downarrow_\sigma)=q^2 x^* x \otimes c^*c + y^*y\otimes a^*a - q x^* y\otimes c^*a - q y^*x \otimes a^*c 
\end{aligned}\;.
\end{equation}
The properties listed above for these operators can also be explicitly verified using the defining rules of the $SU_q(2)$ algebra in \eqref{eq:algebrasuq2}. Notice that these expressions are the non-commutative generalizations of \eqref{eq:probability classical}, with the tensor product separating the spin and the Stern-Gerlach components.

From the definitions of probabilities, we see that the expectation value of the generic Pauli matrix \eqref{eq:pauligeneric} in a generic spin up pre-measurement state \eqref{eq:genericstate} is given by

\begin{equation}\label{expectation value sigma Alice}
    \expval{\sigma^q} = \mel{\psi}{\sigma^q}{\psi} = q\, P\qty(\uparrow_\sigma) -q^{-1}P\qty(\downarrow_\sigma)\;.
\end{equation}
It is possible to repeat the same steps and calculate the analogous of \eqref{probabilities Alice} and \eqref{expectation value sigma Alice} for $\ket{\psi^\prime}$.

\vspace{7pt}\noindent\textbf{Axiom 4} (Measurements)\quad Performing a measurement with a macroscopic Stern-Gerlach apparatus associated to \eqref{eq:pauligeneric} projects a pre-measurement state $\ket{\psi}$ or $\ket{\psi'}$ given by \eqref{eq:genericstate} and \eqref{generic spin down state} respectively, onto $\upsigma$ or $\downsigma$. As a consequence, the geometry state of the spin system is updated to the geometry state of the Stern-Gerlach apparatus, given that $\upsigma$ and $\downsigma$ are eigenstates of the generic Pauli matrix \eqref{eq:pauligeneric}. States $\upsigma$ and $\downsigma$ are \emph{post-measurement} states that can be also interpreted as the \emph{pre-measurement} states of a subsequent measurement, analogously to standard Quantum Mechanics, since the operators $a,c$ appearing in \eqref{eq:upsigma} satisfy the commutation relations of the $SU_q(2)$ algebra as also operators $x,y$ in \eqref{eq:genericstate} do. This means that we can make the identifications

\begin{equation}
    \begin{gathered}
         a = x^\prime \;,\;\;  c = y^\prime \;,\\
        - qc^* = -q{y^\prime}^* \;,\;\;  a^* = {x^\prime}^* \;,
    \end{gathered}
\end{equation}
when the measurement outcome is $\upsigma$ or $\downsigma$, respectively. The operators $x^\prime,y^\prime$ satisfy the commutation relations in \eqref{eq:algebrasuq2}, with $x^\prime = \alpha$ and $y^\prime = \gamma$. This ensures that the \emph{post-measurement} state has the same structure as a \emph{pre-measurement} state, meaning that the relevant physical quantities can be calculated as outlined in axiom 3. Of course, the observable used for a subsequent measurement is obtained as described in axiom 2 with a different copy of $SU_q(2)$ defined by the operators $a^\prime,c^\prime$.

\subsection{What about Bob?}\label{sec:Bob axiom}

The axioms stated above govern the measurement procedures performed by a single observer. We now wish to extend this formalism to the case in which two observers connected by a $SU_q(2)$ transformation are considered, to investigate the covariance properties of the framework we are proposing.

A change of reference frame between two observers Alice and Bob is described by an $SU_q(2)$ transformation that maps $\widehat{\mathcal{H}}^{(A)}_{\frac{1}{2}}$ in $\widehat{\mathcal{H}}^{(B)}_{\frac{1}{2}} \ebd \widehat{\mathcal{H}}_{\frac{1}{2}}^{(A)}\otimes  C\qty(SU_q(2)) = \mathcal{H}_{\frac{1}{2}}\otimes C\qty(SU_q(2))\otimes C\qty(SU_q(2))$. Similarly, Alice's observables are mapped into Bob's, which are elements of $\mathcal{D}\qty(\mathcal{H}_{\frac{1}{2}})\otimes C\qty(SU_q(2))\otimes C\qty(SU_q(2))$. The last copy of $ C(SU_q(2))$ encodes the information on the relative orientation between the two observers. The definition of change of reference frame presented here aligns with the one adopted in other studies investigating symmetry transformations when quantum aspects of geometry are taken into account \cite{Lizzi:2018qaf, Lizzi:2022hcq}. The transformed state of the system explicitly contains information about both the initial state of the system relative to Alice's reference frame and the relation between Alice and Bob. The type of change of reference frame here proposed thus realizes a relational description between observers, a feature also common in quantum reference frame studies \cite{Giacomini:2017zju, Vanrietvelde:2018pgb}.
In the following, the quantities relative to Alice refer to those introduced in axioms 1, 2, 3, formally extended with an additional $\one$ on the last copy of $C\qty(SU_q(2))$, and will be denoted by a superscript or subscript $A$, while the quantities relative to Bob are denoted by a superscript or subscript $B$.

The generic spin up state in Bob's reference frame can be obtained as

\begin{equation}
\label{eq:psibob}
    \ket{\psi^B}=U^q_{g} \ket{\psi^A}\;,
\end{equation}
where

\begin{equation}
    U^q_{g} \ebd \begin{pmatrix}\label{su2quantumrf}
    \mathbbm{1}\otimes u  & -q\mathbbm{1}\otimes v^*\\
    \mathbbm{1}\otimes v & \mathbbm{1}\otimes u^*
    \end{pmatrix} = \ketbra{\uparrow}\qty(\mathbbm{1}\otimes u) -q \ketbra{\uparrow}{\downarrow}\qty(\mathbbm{1}\otimes  v^*) + \ketbra{\downarrow}{\uparrow} \qty(\mathbbm{1}\otimes v) + \ketbra{\downarrow}\qty(\mathbbm{1}\otimes  u^*)\;,
\end{equation}
is the $SU_q(2)$ matrix connecting the two observers written in terms of a further copy of the $SU_q(2)$ algebra generators $u,v$, which still satisfy the commutation relations \eqref{eq:algebrasuq2}, with $u$ and $v$ taking the role of $\alpha$ and $\gamma$, respectively. The generic Pauli matrix can be written in Bob's reference fame as

\begin{equation}\label{eq:sigma Bob axiom}
    \sigma_B^q=U^q_{g}\sigma^q_A{U^q_{g}}^{\dagger}
\end{equation}

With these definitions, the probabilities and the expectation value in \eqref{probabilities Alice} and \eqref{expectation value sigma Alice} are invariant, namely

\begin{equation}
    \expval{\sigma_B^q}_B  = \expval{\sigma_A^q}_A \;,\;\;   P_B\qty(\uparrow_{\sigma}^B) = P_A\qty(\uparrow_\sigma^A) \;,\;\;   P_B\qty(\downarrow_{\sigma}^B) = P_A\qty(\downarrow_\sigma^A) \;,
\end{equation}
where
\begin{equation}
    \sigma_B^q = q \ketbra{\uparrow_{\sigma}^B} -q^{-1}\ketbra{\downarrow_{\sigma}^B} \ebd q\, U^q_{g}\ketbra{\uparrow_{\sigma}^A}{\uparrow_{\sigma}^A} {U^q_{g}}^\dagger -q^{-1}\, U^q_{g}\ketbra{\downarrow_{\sigma}^A}{\downarrow_{\sigma}^A} {U^q_{g}}^\dagger\;,
\end{equation}
\begin{equation}\label{eq:sigma expectation value and probability bob}
    \expval{\sigma_I^q}_J = \mel{\psi^{J}}{\sigma_I^q}{\psi^{J}}\;,\;\; P_I\qty(\uparrow_{\sigma}^J) = \braket{\psi^{I}}{\uparrow_\sigma^J}\braket{\uparrow_\sigma^J}{\psi^{I}} \;,\;\; I,J \in \{A,B\}\;,
\end{equation}
and the expectation values are elements of $C\qty(SU_q(2)) \otimes C\qty(SU_q(2)) \otimes C\qty(SU_q(2))$ defined as maps analogously to \eqref{eq:map definition scalar product}, where the third component of the tensor product contains products of elements of the copy of $C(SU_q(2))$ pertaining to the relative orientation, and the terms appearing in the first two copies of the tensor product are defined as in \eqref{eq:map definition scalar product}. This implies that different observers agree on the observed values of probabilities and expectation values of spin, which ensures the physical consistency of the theory as in the standard $SU(2)$ case.

Finally, upon explicitly expanding \eqref{eq:psibob} and \eqref{eq:sigma Bob axiom}, it is possible to show that by defining
\begin{equation}
\begin{aligned}
 & x'= u \otimes x - q \,v^* \otimes y\qquad y'= v\otimes x + u^*\otimes y\\
&   a'= u \otimes a - q \,v^* \otimes c\qquad c'= v\otimes a + u^*\otimes c  
  \end{aligned}
\end{equation}
the pair of operators $x',y'$ and $a',c'$ satisfy the commutations relations \eqref{eq:algebrasuq2}, thus realizing an isomorphism $C(SU_q(2))\otimes C(SU_q(2))\simeq C(SU_q(2))$~\cite{Demichev:1997sg,Chari:1994pz,majid_1995}. Moreover, we can write
\begin{equation}
\begin{gathered}
    \ket{\psi^B} =  \up x'  + \down  y'\;\\ \\
    \sigma^q_B=\begin{pmatrix}  q\qty(\mathbbm{1}-\qty(1+q^2)c'^*c')
         & (q+q^{-1}) a'c'^* \\
        (q+q^{-1}) c'a'^* & -q^{-1}\qty(\mathbbm{1}-\qty(1+q^2)c'^*c')
\end{pmatrix}\;,
    \end{gathered}
\end{equation}
Namely, the spin up state and the generic Pauli matrix in Bob's reference frame can be rewritten in the same form of Alice's spin up state and generic Pauli matrix, respectively, meaning that the eigenvalues of the transformed generic Pauli matrix are still given by $\big\{q, -q^{-1}\big\}$ and that the expressions to compute spin observables are the same in form for each observer, similar to what was observed also for the commutative case in \Cref{sec:spin measurement classic}. This property guarantees that the framework is covariant under $SU_q(2)$ transformations. Of course, with a similar argument, the same discussion also applies to the generic spin down state, $|{\psi^\prime}^B\rangle = U^q_{g} |{\psi^\prime}^A\rangle$.

\subsection{Quantum probabilities}\label{sec: quantum prob}
In standard Quantum Mechanics, observables are given by self-adjoint operators and measurement outcomes are described by probability distributions that depend on the form of the operator and of the state of the physical system on which measurements are performed. The probabilistic nature of measurement outcomes arises from the superposition principle. In the axiomatization of our framework, the implementation of quantum rotational symmetry has the natural consequence that probabilities themselves are not described by non-negative real numbers, rather they are described by positive semi-definite self-adjoint operators, so that the probabilities of observing given outcomes are characterized by probability distributions as well. The latter depend on the form of the probability operators and of the geometry states $\ket{\Phi} \in \mathcal{H}_{SU_q(2)}\otimes \mathcal{H}_{SU_q(2)} $, which codify information on the geometrical configuration of the spin state and of the Stern-Gerlach device. Measurement outcomes of the probability thus depend on these geometrical configurations, just as in standard Quantum Mechanics, with the additional feature that our novel non-commutative framework allows for the geometry states to be written in terms of superpositions of probability eigenstates, defined as
\begin{equation}\label{eq:probability eigenstates Alice}
P(\uparrow_\sigma)\ket{p,r}=p\ket{p,r} \;,
\end{equation}
where $r$ denotes the possible degeneracy of the eigenvalue $p$. In \Cref{app:probabilityeigenstates} we present a detailed analysis of such eigenstates. As we shall discuss in the next section, generic geometry states describing semi-classical scenarios are factorizable in $\mathcal{H}_{SU_q(2)}\otimes\mathcal{H}_{SU_q(2)}$. Therefore, these states will be written as $\ket{\Phi} = \ket{\Phi^S}\otimes\ket{\Phi^{SG}}$, where $\ket{\Phi^S}$ ($\ket{\Phi^{SG}}$) is the geometry state describing the spatial orientation of the spin state (Stern-Gerlach apparatus), and will be generally given by superpositions of basis states in its copy of $\mathcal{H}_{SU_q(2)}$. In \Cref{app:probabilityeigenstates} we show that, in general, these semi-classical states of the geometry are not eigenstates of the probability operator, so they must be written as a superposition thereof. The overlap between the geometry states $\ket{\Phi}$ and the probability eigenstates $\ket{p,r}$ defines the distribution of outcomes of a probability measurement, according to 
 
\begin{equation}\label{eq:probability distribution Alice}
    f(p)=\sum_r \braket{\Phi}{p,r}\braket{p,r}{\Phi} \; ,
\end{equation}
which is automatically normalized. This distribution is characterized by a mean value $p_0$ and a variance $\Delta^2_f$, which is in general nonzero. Of course, a classical probability is described by a distribution with vanishing variance, namely a delta function of the form $f(p) = \delta(p-p_0)$, obtained when the geometry state of the experimental setup is a probability eigenstate with eigenvalue $p_0$.

What is the meaning of a probability distribution for probabilities? To understand this, let us first discuss what we mean by a probability measurement in the context of Doubly Quantum Mechanics. In standard Quantum Mechanics, probability is not an observable of the theory, but rather something inferred from the data associated to an experimental procedure involving quantum mechanical measurements. For example, the probability of finding a particle with a given energy is something we infer from data obtained by repeating an experiment apt to measure the energy of many particles, all prepared in the same initial state. In Doubly Quantum Mechanics, probability is promoted to an observable, whose quantum mechanical features are encoded in geometry states. By definition, any observable must be characterized by a measurement procedure in order to compare its theoretical predictions with real-world data. In the context of our doubly quantum mechanical measurements of spin, the measurement apparatus consists of a Stern-Gerlach device and a beam of electrons. The output of this measurement device will be a number between $0$ and $1$ characterizing the fraction of spin up (down) electrons in the beam. 
In the ideal case, the probability measurement should be performed by measuring this frequency with an infinite number of electrons. The doubly quantum mechanical interpretation is that when the measurement is performed, the geometry state, which can be written as a superposition of probability eigenstates, collapses into the eigenstate associated to the probability value that has been measured. Of course, in a real-world experiment, the device outlined above is equipped with a finite number $N$ of electrons, and the resolution of the probability measurement will increase as the number $N$ of electrons increases. According to our formalism, for a high enough value of $N$, two probability measurements performed in identical geometrical configurations may yield incompatible outcomes within their experimental uncertainty, featuring a fundamental discrepancy due to the quantum deformation of $SU(2)$. By repeating the probability measurement in the same geometrical configuration several times (i.e. with the same geometry states for spin systems and Stern-Gerlach apparata) and with enough precision to be sensible to $q$-deformation effects, an observer will be able to reconstruct the ``probability distribution of probability'' $f(p)$ introduced in \eqref{eq:probability distribution Alice}. In general, $f(p)$ will be characterized by a non-vanishing variance, arising from the fact that the geometry states are superpositions of probability eigenstates. We emphasize the fact that a single determination of the probability requires a Stern-Gerlach device equipped with $N\gg 1$ electrons. Once the geometry state has collapsed, the measurement procedure is completed. Attempts to combine such a measurement with another measurement given by another batch of electrons to infer a single determination of the probability would be meaningless, as in standard Quantum Mechanics it would be meaningless to employ two different quantum mechanical position measurements on two different particles prepared in the same initial state to infer a single determination of the position.

\section{Semi-classical states}\label{sec:effective theory alice}

In the previous section, we have defined the non-commutative generalization of physical quantities related to the outcome of spin measurements and expressed them in terms of operators that act on geometry states. The effective outcome of actual measurements is described by a distribution with mean value and variance that are obtained by computing the average values and variances of these operators in the spinor and Stern-Gerlach geometry states. We will denote the average value in the geometry states with a bar. For instance, the average value of $\sigma^q$ in the geometry state $\ket{\Phi}$ is denoted by $\overline{\sigma^q} \ebd \expval{\sigma^q}{\Phi}$ and the average value in the geometry of the expectation value of $\sigma^q$ in a spin state is $\overline{\expval{\sigma^q}}$. The uncertainty on the measurement of a given observable, defined as the square root of the variance of the observable in the geometry states, is denoted by $\overline{\Delta}$. For instance, the uncertainty in the geometry states of the expectation value of the spin will be indicated by $\overline{\Delta}\big[\expval{\sigma^q}\big] = \sqrt{\,\overline{\Delta}^{\,2}\big[\expval{\sigma^q}\big]}  \ebd \sqrt{\,\overline{\expval{\sigma^q}^2} - \overline{\expval{\sigma^q}}^{\,2}}$.

\subsection{Semi-classical conditions}
In this section, we focus on \emph{semi-classical} geometry states that yield small deviations, which vanish in the limit $q\rightarrow 1$, with respect to the results of standard Quantum Mechanics. This is done by requiring the average values of the relevant physical quantities in the geometry states to differ by $\mathcal{O}\qty(1-q)$ with respect to the standard quantum mechanical counterparts and their variances to be $\mathcal{O}\qty(1-q)$. These requirements are consistent with the fact that the parameter $q$ is expected to be very close to $1$, $q\sim 1$, if the $SU_q(2)$ deformation is taken to be quantum-gravitational in origin, as in our case.

According to axiom 0, the states encoding information on the spin, Stern-Gerlach apparata, and relative orientation are elements of the Hilbert spaces of the representations of the $SU_q(2)$ algebras defined by $\{x,y\}$, $\{a,c\}$, and $\{u,v\}$ respectively. The Hilbert space is always given by $\mathcal{H}_{SU_q(2)} = \mathcal{H}_{\pi} \oplus \mathcal{H}_\rho$ defined in \Cref{sec:mathprel}. The representations are given by \eqref{rappresentazione rho e rhostar} and \eqref{rappresentazione pi} where the quantum numbers $\{n,\phi, \chi\}$ are replaced by $\{n_s,\phi_s,\chi_s\}$, $\{n_a,\phi_a,\chi_a\}$, and $\{n_g,\phi_g,\chi_g\}$ for the representations of $\{x,y\}$, $\{a,c\}$, and $\{u,v\}$, respectively.

As pointed out in \cite{Amelino-Camelia:2022dsj}, these representations can be interpreted as giving a quantum description of the angles that define directions and relative orientations in space. Specifically, the angles $\phi,\chi$ in the classical representations \eqref{rappresentazione classica a e c methods} and \eqref{eq:su2 stern gerlach classic} retain their classical nature, while the angle $\theta$ becomes discretized in the range $[0,\pi]$, according to
\begin{equation}\label{eq:discreteangle}
    \theta(n) = \left\{
	\begin{array}{ll}
  2 \arcsin q^n\;,\;\; n \in \mathbbm{N}_0 \\ \\ 0\;,\;\; n= \infty
	\end{array}
	\right. \; ,
\end{equation}
where we formally set $\theta\qty(\infty) = 0$. Specifically, the values in $]0,\pi]$ corresponding to $n \in \mathbbm{N}_0$ are derived from the $\mathcal{H}_\pi$ component of $\mathcal{H}_{SU_q(2)}$, while $\theta \qty(\infty) = 0$ arises from the $\mathcal{H}_{\rho}$ component.

We start by identifying a class of semi-classical states describing the directions of spin systems and Stern-Gerlach apparata in terms of these quantum numbers. Since we want to describe a physical setup in which spin states and Stern-Gerlach apparata can be prepared independently, we require that the semi-classical geometry state describing the experimental setup is separable, so that it can be written as the tensor product of a geometry state characterizing the spin direction and the geometry state describing the Stern-Gerlach direction. Each of these states is an element of $\mathcal{H}_{SU_q(2)}$ and will be specified by two angles $\theta = \theta(n)$ and $\omega=\phi-\chi$, indicating the classical counterpart of the direction along which spins and Stern-Gerlach apparata are aligned, where $\theta(n)$ is one of the allowed values in \eqref{eq:discreteangle}. These states will be denoted by $\ket{\Phi^S\qty(\theta_s, \omega_s)}$ and $\ket{\Phi^{SG}\qty(\theta_a, \omega_a)}$ and the full geometry state of the experimental setup will then be given by the tensor product $\ket{\Phi\qty(\theta_s, \omega_s,\theta_a, \omega_a)} = \ket{\Phi^S\qty(\theta_s, \omega_s)}\otimes\ket{\Phi^{SG}\qty(\theta_a, \omega_a)} \in \mathcal{H}_{SU_q(2)}\otimes\mathcal{H}_{SU_q(2)}$. Of course, since the angle $\theta(n)$ assumes discrete values, for a given classical direction specified by angles $(\theta,\omega)$ it is only possible to find, in general, a semi-classical state that describes a spin or a Stern-Gerlach aligned along a direction $(\theta\qty(n_\theta),\omega)$ that is as close as possible to $(\theta,\omega)$, where $n_\theta$ is the value of $n$ such that $\abs{\theta\qty(n)-\theta}$ is minimal. For values of $q$ closer to $1$, the gap between two consecutive angles becomes smaller, so that the values $\theta(n)$ are more dense in any given angular range.

The semi-classicality conditions for the geometry states $\ket{\Phi^S\qty(\theta_s, \omega_s)}$ and $\ket{\Phi^{SG}\qty(\theta_a, \omega_a)}$ read

\begin{equation}\label{eq:semi-classical factors}
\left\{\begin{array}{ll}
 \overline{\ket{\psi}}_{\theta_s,\omega_s}  = \qty(\cos \frac{\theta_s}{2}+\mathcal{O}(1-q))\ket{\uparrow}+\qty(e^{i\omega_s}\sin\frac{\theta_s}{2}+\mathcal{O}(1-q))\ket{\downarrow}\\ \\ 
\overline{\Delta}^{\,2}\big[\ket{\psi}\big]_{\theta_s,\omega_s}= \mathcal{O}(1-q)\ket{\uparrow}+\mathcal{O}(1-q)\ket{\downarrow}\\ \\
 \overline{\sigma^q}_{\theta_a,\omega_a} =        \begin{pmatrix}
       \cos\theta_a + \mathcal{O}(1-q) & e^{-i\omega_a}\sin\theta_a + \mathcal{O}(1-q)\\ e^{i\omega_a}\sin\theta_a + \mathcal{O}(1-q) &-\cos\theta_a + \mathcal{O}(1-q)
    \end{pmatrix} \\ \\
    
   \overline{\Delta}^{\,2}\big[\sigma^q\big]_{\theta_a,\omega_a}=  \begin{pmatrix}
       \mathcal{O}(1-q) & \mathcal{O}(1-q)\\ \mathcal{O}(1-q) & \mathcal{O}(1-q)
    \end{pmatrix}
\end{array} \right. \;,
\end{equation}
where the average values and variances are taken in the full geometry state $\ket{\Phi\qty(\theta_s, \omega_s,\theta_a, \omega_a)}$ and the subscripts indicate which of the two components of the full geometry state enters non-trivially in the computation. The variance of a non-Hermitian operator $O$ is defined as $ \overline{\Delta}^{\,2} \big[O\big] = \expval{O^\dagger O}{\Phi} - \expval{O^\dagger}{\Phi}\expval{O}{\Phi}$ following \cite{Pati_2015}. Additionally, the full state of the geometry $\ket{\Phi\qty(\theta_s, \omega_s,\theta_a, \omega_a)}$ has to satisfy

\begin{equation}\label{eq:semi-classical probabilities}
    \left\{\begin{array}{ll}
    \overline{P(\uparrow_\sigma)}_{\theta_s,\omega_s,\theta_a,\omega_a} = \frac{1}{2}\Big[1+\cos(\theta_a)\cos(\theta_s)+\cos(\omega_a-\omega_s)\sin(\theta_a)\sin(\theta_s)\Big]+ \mathcal{O}(1-q)\\ \\
   \overline{\Delta}^{\,2}\big[P(\uparrow_\sigma)\big]_{\theta_s,\omega_s,\theta_a,\omega_a} =
   \mathcal{O}(1-q)
    \end{array} \right. \;.
\end{equation}
We emphasize that the quantities in \eqref{eq:semi-classical factors} only serve as guidelines for identifying the states that describe semi-classical experimental setups, but do not enter directly in the physical predictions of the theory. As discussed in \Cref{sec: quantum prob}, the average value and variance of the probability are the observable quantities and are linked to the expectation value of the spin. The condition \eqref{eq:semi-classical probabilities} guarantees that they are affected by small corrections of order $(1-q)$ when the geometry states describe a semi-classical scenario.

When two observers are involved, we consider an additional copy of $C(SU_q(2))$ and the corresponding additional copy of $\mathcal{H}_{SU_q(2)}$ contains states describing the relative orientation between the observers (see \Cref{sec:Bob axiom}). The discussion of semi-classical states in this context is a simple extension of the one presented above. Specifically, the semi-classical conditions can be generalized by replacing the operators appearing in \eqref{eq:semi-classical factors}, \eqref{eq:semi-classical probabilities} by the corresponding ones for Bob, defined in \Cref{sec:Bob axiom}. The full state of the geometry now involves three states and is of the form 
$\ket{\Phi\qty(\theta_s, \omega_s,\theta_a, \omega_a, \theta_g,\chi_g,\phi_g)} = \ket{\Phi^S\qty(\theta_s, \omega_s)}\otimes\ket{\Phi^{SG}\qty(\theta_a, \omega_a)}\otimes\ket{\Phi^{RO}\qty(\theta_g, \chi_g, \phi_g)}$, \emph{i.e.} the full geometry state is factorizable. In \Cref{sec:numerical example protocol} and \Cref{appendix: semi-classical states} we show that states of the form $\ket{\Phi^{RO}\qty(\theta_g, \chi_g, \phi_g)}$ are the semi-classical counterpart of generic rotations $R_z(\alpha) R_x(\theta_g) R_z(\gamma)$, where $\chi_g = \frac{\alpha+\gamma}{2} $, $\phi_g = \frac{3}{2}\pi - \frac{\alpha-\gamma}{2}$, and $\theta_g = \theta\qty(n)$ is again one of the allowed values in \Cref{eq:discreteangle}. The discretized nature of $\theta(n)$ in the context of two observers implies that for a given classical rotation specified by angles $(\theta,\phi,\chi)$ it is only possible to find, in general, a semi-classical state that describes a rotation defined by $(\theta\qty(n_\theta),\phi,\chi)$, where $n_\theta$ has the same meaning as before. Notice that the states $\ket{\Phi^{RO}\qty(\theta_g, \chi_g, \phi_g)}$ depend on both angles $\chi_g$ and $\phi_g$ and not only on their difference, as is the case for semi-classical states describing the direction of physical systems, since rotating a reference frame (or, \emph{e.g.}, a solid object) requires specifying three angles, while rotating a vector (or specifying a direction) requires only two.

\subsection{Semi-classical states and probability eigenstates}\label{sec:semiclassical states and probability eigenstates}

We now comment on the connection between the semi-classical geometry states and probability eigenstates. As shown in \Cref{app:probabilityeigenstates}, the only factorizable eigenstates of the probability operator are of the form $\ket{\chi_s}\otimes\ket{\chi_a}$, $\ket{\chi_s}\otimes\ket{n_a,\phi_a,\chi_a}$, $\ket{n_s,\phi_s,\chi_s}\otimes\ket{\chi_a}$, $\ket{\mu_s,\omega_s}_s\otimes\ket{\mu_a,\omega_a}_a$, with
\begin{equation}
\label{eq:semiclassicalplanexy}
    \begin{aligned}
        &\ket{\mu_s,\omega_s}_s=N_s\sum_{n_s=0}^\infty f(\mu_s,n_s)\ket{n_s,\phi_s,\chi_s} =N_s\sum_{n_s=0}^\infty \frac{q^{\frac{n_s}{2}(n_s-1)}\sqrt{(1-q^2)^{n_s-1}}}{\sqrt{(q^4;q^2)_{n_s-1}}}\,\mu_s^{n_s}\ket{n_s,\phi_s,\chi_s}\;\\
        &\ket{\mu_a,\omega_a}_a=N_a\sum_{n_a=0}^\infty q^{n_a}f(\mu_a,n_a)\ket{n_a,\phi_a,\chi_a}=N_a \sum_{n_a=0}^\infty \frac{q^{\frac{n_a}{2}(n_a+1)}\sqrt{(1-q^2)^{n_a-1}}}{\sqrt{(q^4;q^2)_{n_a-1}}}\,\mu_a^{n_a}\ket{n_a,\phi_a,\chi_a}
    \end{aligned} \;,
\end{equation} 
where $\omega_{s,a} = \phi_{s,a}-\chi_{s,a}$, $N_{s,a}$ are normalization constants, and the q-Pochhammer symbol $(a;q)_{n}$ is defined in \Cref{app:probabilityeigenstates}. As discussed in \Cref{appendix: semi-classical states}, the only separable probability eigenstates that do not satisfy the requirements \eqref{eq:semi-classical factors}, \eqref{eq:semi-classical probabilities} are of the form $\ket{\chi_s}\otimes\ket{n_a,\phi_a,\chi_a}$ and $\ket{n_s,\phi_s,\chi_s}\otimes\ket{\chi_a}$ for $n_s,n_a\geq 1$. We also show that states of the form $\ket{\chi_{s,a}}$ and $\ket{\mu_{s,a},\omega_{s,a}}$ describe spin systems and Stern-Gerlach apparata aligned along directions $\qty(\theta_{s,a}(n),\omega_{s,a})$. In particular, $\ket{\Phi^{S,SG}\qty(0,\omega_{s,a})} = \ket{\chi_{s,a} = \omega_{s,a}}$ describe spin and Stern-Gerlach apparata aligned along the positive $z$ direction, while for a classical direction $(\theta,\omega)$ with $\theta \neq 0$, the states that semi-classically describe spin and Stern-Gerlach apparata aligned along the direction $(\theta\qty(n_\theta),\omega)$ which is the closest to $(\theta,\omega)$ are given by $\ket{\Phi^{S,SG}\qty(\theta_{s,a}\qty(n_\theta),\omega_{s,a})} = \ket{\mu_{s,a}\qty(\theta),\omega_{s,a}}_{s,a}$. In the previous relations, $n_\theta$ is the value of $n$ such that $\abs{\theta(n)-\theta}$ is minimal, where $\theta(n)$ is given by \eqref{eq:discreteangle}, and $\mu_{s,a}\qty(\theta)$ is such that the distributions $\abs{f(\mu_{s,a}\qty(\theta),n_{s,a})}^2$ have maximum in $n_\theta$. In general, a semi-classical full geometry state $\ket{\Phi^S\qty(\theta_s,\omega_s)} \otimes \ket{\Phi^{SG}\qty(\theta_a,\omega_a)}$ will not be a probability eigenstate, of course. Probability measurements involving these geometry states will exhibit doubly quantum mechanical behaviour, as discussed in \Cref{sec: quantum prob}. For what concerns the relative orientation, in \Cref{appendix: semi-classical states} we show that the states of the same form as those considered for the single observer satisfy the generalized semi-classical conditions, where states of the form \eqref{eq:semiclassicalplanexy} are denoted as $\ket{\mu_g,\phi_g,\chi_g}_g$ to make the dependence on both angles $\phi_g,\chi_g$ explicit. The states $\ket{\Phi^{RO}\qty(\theta_g\qty(\theta),\phi_g,\chi_g)}$ semi-classically describe the relative orientation between two observers. Classically, the latter is specified by the rotation matrix that relates the two observers, parameterized as $R_z\qty(\alpha)R_x\qty(\theta)R_z\qty(\gamma)$. In our quantum setting, for a given classical rotation specified by three angles $(\theta,\alpha,\gamma)$, the state that describes the rotation that is closest to $R_z\qty(\alpha)R_x\qty(\theta)R_z\qty(\gamma)$ is $\ket{\Phi^{RO}\qty(\theta_g\qty(n_\theta),\phi_g,\chi_g)} =\ket{\mu_g\qty(\theta),\phi_g,\chi_g}_g$, where $\chi_g = \frac{\alpha+\gamma}{2} $, $\phi_g = \frac{3}{2}\pi - \frac{\alpha-\gamma}{2}$, and $\theta_g\qty(n_\theta) \neq 0$ is such that $\abs{f(\mu_{g},n_{g})}^2$, which replaces $\abs{f(\mu_{s,a},n_{s,a})}^2$ in \eqref{eq:semiclassicalplanexy}, is peaked around $n_\theta$. States of the form $\ket{\Phi^{RO}\qty(0,\phi_g,\chi_g)} =\ket{\chi_g}_g$ semi-classically describe rotations $R_z(\alpha) R_x(0) R_z(\gamma) = R_z(\alpha+\gamma) = R(2\chi_g)$. As for the case of the single observer, a semi-classical full geometry state $\ket{\Phi^S\qty(\theta_s,\omega_s)} \otimes \ket{\Phi^{SG}\qty(\theta_a,\omega_a)}\otimes\ket{\Phi^{RO}\qty(\theta_g,\phi_g,\chi_g)}$ will not be a probability eigenstate, in general. In \Cref{table semi-classical states} we summarize all the semi-classical states that we are going to use in the following. Further details, as well as some numerical examples involving these states can be found in \Cref{appendix: semi-classical states}.

\begin{table}\centering
{\tabulinesep=1.2mm\begin{tabu}{c|c|c|}
\cline{2-3}
                            & $\pmb{(0,\omega)}$ \textbf{direction}          & $\pmb{(\theta,\omega)}$ \textbf{direction, $\pmb{\theta\neq 0}$}                 \\ \hline
\multicolumn{1}{|l|}{$\ket{\Phi^S\qty(\theta_s\qty(n_\theta),\omega_s)}$}  & $\ket{\chi_s = \omega_s}$        & $\ket{\mu_s\qty(\theta),\omega_s}_s$                \\ \hline
\multicolumn{1}{|l|}{$\ket{\Phi^{SG}\qty(\theta_a\qty(n_\theta),\omega_a)}$} & $\ket{\chi_a = \omega_a}$        &  $\ket{\mu_a\qty(\theta),\omega_a}_a$                \\ \hline
                            & $\pmb{R_z\qty(2\chi_g)}$ & $\pmb{R_z\qty(\alpha)R_x\qty(\theta)R_z\qty(\gamma)}$\textbf{, } $\pmb{\theta\neq 0}$\\ \hline
\multicolumn{1}{|l|}{$\ket{\Phi^{RO}\qty(\theta_g\qty(n_\theta),\phi_g,\chi_g)}$} &  $\ket{\chi_g}$     &       $\ket{\mu_g\qty(\theta),\phi_g,\chi_g}_g$       \\ \hline
\end{tabu}}\caption{Semi-classical states used in the rest of the paper. Further details about the notation adopted for these states can be found in \Cref{sec:semiclassical states and probability eigenstates}.}\label{table semi-classical states}
\end{table}

\section{(Non)-Alignment protocol between two observers}
\label{sec:deformed protocol}

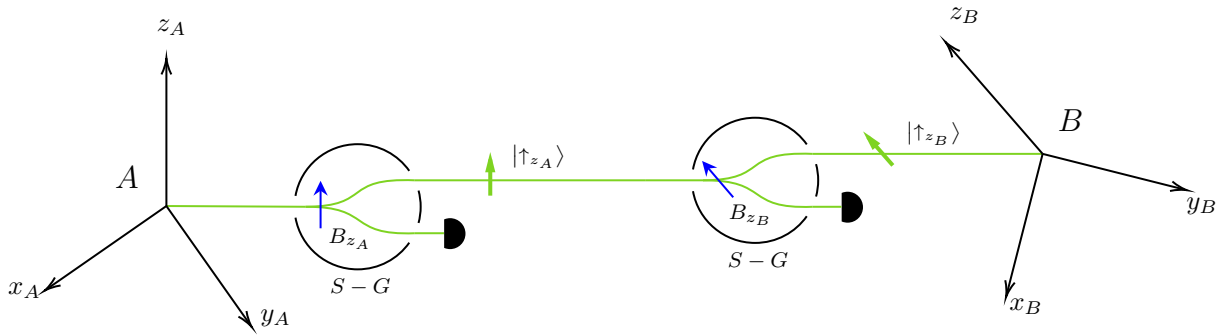
\begin{figure}
    \centering

\tikzset{every picture/.style={line width=0.75pt}} 

\begin{tikzpicture}[x=0.65pt,y=0.65pt,yscale=-1,xscale=1]

\draw [color={rgb, 255:red, 126; green, 211; blue, 33 }  ,draw opacity=1 ]   (437.51,125.29) -- (560.71,125.29) ;
\draw [color={rgb, 255:red, 126; green, 211; blue, 33 }  ,draw opacity=1 ]   (92.07,153.4) -- (166.53,153.8) ;
\draw    (92.07,153.4) -- (28.45,197.95) ;
\draw [shift={(26.81,199.09)}, rotate = 325] [color={rgb, 255:red, 0; green, 0; blue, 0 }  ][line width=0.75]    (8.74,-2.63) .. controls (5.56,-1.12) and (2.65,-0.24) .. (0,0) .. controls (2.65,0.24) and (5.56,1.12) .. (8.74,2.63)   ;
\draw    (92.07,153.4) -- (136.61,217.02) ;
\draw [shift={(137.76,218.66)}, rotate = 235] [color={rgb, 255:red, 0; green, 0; blue, 0 }  ][line width=0.75]    (8.74,-2.63) .. controls (5.56,-1.12) and (2.65,-0.24) .. (0,0) .. controls (2.65,0.24) and (5.56,1.12) .. (8.74,2.63)   ;
\draw [color={rgb, 255:red, 126; green, 211; blue, 33 }  ,draw opacity=1 ]   (166.53,153.8) .. controls (202.53,153.6) and (188.25,169.42) .. (224.91,168.06) ;
\draw  [fill={rgb, 255:red, 0; green, 0; blue, 0 }  ,fill opacity=1 ] (241.15,160.97) .. controls (241.98,160.65) and (242.87,160.48) .. (243.81,160.48) .. controls (248.05,160.48) and (251.49,163.98) .. (251.49,168.28) .. controls (251.49,172.59) and (248.05,176.08) .. (243.81,176.08) .. controls (242.87,176.08) and (241.98,175.91) .. (241.15,175.6) -- cycle ;
\draw [color={rgb, 255:red, 126; green, 211; blue, 33 }  ,draw opacity=1 ]   (166.53,153.8) .. controls (202.53,153.6) and (185.43,137.99) .. (224.91,139.54) ;
\draw [color={rgb, 255:red, 126; green, 211; blue, 33 }  ,draw opacity=1 ]   (224.91,168.06) -- (240.51,168.06) ;
\draw [color={rgb, 255:red, 126; green, 211; blue, 33 }  ,draw opacity=1 ]   (224.91,139.54) -- (348.11,139.54) ;
\draw    (92.07,153.4) -- (92.07,75.73) ;
\draw [shift={(92.07,73.73)}, rotate = 90] [color={rgb, 255:red, 0; green, 0; blue, 0 }  ][line width=0.75]    (8.74,-2.63) .. controls (5.56,-1.12) and (2.65,-0.24) .. (0,0) .. controls (2.65,0.24) and (5.56,1.12) .. (8.74,2.63)   ;
\draw    (560.71,125.29) -- (541.6,200.57) ;
\draw [shift={(541.11,202.51)}, rotate = 284.24] [color={rgb, 255:red, 0; green, 0; blue, 0 }  ][line width=0.75]    (8.74,-2.63) .. controls (5.56,-1.12) and (2.65,-0.24) .. (0,0) .. controls (2.65,0.24) and (5.56,1.12) .. (8.74,2.63)   ;
\draw    (560.71,125.29) -- (635.99,144.4) ;
\draw [shift={(637.93,144.89)}, rotate = 194.24] [color={rgb, 255:red, 0; green, 0; blue, 0 }  ][line width=0.75]    (8.74,-2.63) .. controls (5.56,-1.12) and (2.65,-0.24) .. (0,0) .. controls (2.65,0.24) and (5.56,1.12) .. (8.74,2.63)   ;
\draw    (560.71,125.29) -- (510,66.46) ;
\draw [shift={(508.69,64.95)}, rotate = 49.24] [color={rgb, 255:red, 0; green, 0; blue, 0 }  ][line width=0.75]    (8.74,-2.63) .. controls (5.56,-1.12) and (2.65,-0.24) .. (0,0) .. controls (2.65,0.24) and (5.56,1.12) .. (8.74,2.63)   ;
\draw [color={rgb, 255:red, 126; green, 211; blue, 33 }  ,draw opacity=1 ]   (348.11,139.54) -- (379.31,139.54) ;
\draw  [draw opacity=0] (161.11,147.94) .. controls (163.89,132.11) and (177.7,120.07) .. (194.33,120.07) .. controls (205.76,120.07) and (215.86,125.76) .. (221.96,134.45) -- (194.33,153.8) -- cycle ; \draw   (161.11,147.94) .. controls (163.89,132.11) and (177.7,120.07) .. (194.33,120.07) .. controls (205.76,120.07) and (215.86,125.76) .. (221.96,134.45) ;  
\draw  [draw opacity=0] (226.92,145.07) .. controls (227.66,147.85) and (228.06,150.78) .. (228.06,153.8) .. controls (228.06,156.82) and (227.66,159.75) .. (226.92,162.53) -- (194.33,153.8) -- cycle ; \draw   (226.92,145.07) .. controls (227.66,147.85) and (228.06,150.78) .. (228.06,153.8) .. controls (228.06,156.82) and (227.66,159.75) .. (226.92,162.53) ;  
\draw  [draw opacity=0] (221.96,173.15) .. controls (215.86,181.84) and (205.76,187.53) .. (194.33,187.53) .. controls (177.7,187.53) and (163.89,175.49) .. (161.11,159.66) -- (194.33,153.8) -- cycle ; \draw   (221.96,173.15) .. controls (215.86,181.84) and (205.76,187.53) .. (194.33,187.53) .. controls (177.7,187.53) and (163.89,175.49) .. (161.11,159.66) ;  
\draw [color={rgb, 255:red, 126; green, 211; blue, 33 }  ,draw opacity=1 ]   (379.13,139.55) .. controls (415.13,139.35) and (400.85,155.17) .. (437.51,153.81) ;
\draw  [fill={rgb, 255:red, 0; green, 0; blue, 0 }  ,fill opacity=1 ] (453.75,146.72) .. controls (454.58,146.4) and (455.47,146.23) .. (456.41,146.23) .. controls (460.65,146.23) and (464.09,149.73) .. (464.09,154.03) .. controls (464.09,158.34) and (460.65,161.83) .. (456.41,161.83) .. controls (455.47,161.83) and (454.58,161.66) .. (453.75,161.35) -- cycle ;
\draw [color={rgb, 255:red, 126; green, 211; blue, 33 }  ,draw opacity=1 ]   (379.13,139.55) .. controls (415.13,139.35) and (398.03,123.74) .. (437.51,125.29) ;
\draw [color={rgb, 255:red, 126; green, 211; blue, 33 }  ,draw opacity=1 ]   (437.51,153.81) -- (453.11,153.81) ;
\draw  [draw opacity=0] (373.71,133.69) .. controls (376.49,117.86) and (390.3,105.82) .. (406.93,105.82) .. controls (418.36,105.82) and (428.46,111.51) .. (434.56,120.2) -- (406.93,139.55) -- cycle ; \draw   (373.71,133.69) .. controls (376.49,117.86) and (390.3,105.82) .. (406.93,105.82) .. controls (418.36,105.82) and (428.46,111.51) .. (434.56,120.2) ;  
\draw  [draw opacity=0] (439.52,130.82) .. controls (440.26,133.6) and (440.66,136.53) .. (440.66,139.55) .. controls (440.66,142.57) and (440.26,145.5) .. (439.52,148.28) -- (406.93,139.55) -- cycle ; \draw   (439.52,130.82) .. controls (440.26,133.6) and (440.66,136.53) .. (440.66,139.55) .. controls (440.66,142.57) and (440.26,145.5) .. (439.52,148.28) ;  
\draw  [draw opacity=0] (434.56,158.9) .. controls (428.46,167.59) and (418.36,173.28) .. (406.93,173.28) .. controls (390.3,173.28) and (376.49,161.24) .. (373.71,145.41) -- (406.93,139.55) -- cycle ; \draw   (434.56,158.9) .. controls (428.46,167.59) and (418.36,173.28) .. (406.93,173.28) .. controls (390.3,173.28) and (376.49,161.24) .. (373.71,145.41) ;  
\draw [color={rgb, 255:red, 0; green, 0; blue, 255 }  ,draw opacity=1 ][fill={rgb, 255:red, 0; green, 255; blue, 255 }  ,fill opacity=1 ]   (395.23,148.7) -- (380.51,131.38) ;
\draw [shift={(378.56,129.09)}, rotate = 49.63] [fill={rgb, 255:red, 0; green, 0; blue, 255 }  ,fill opacity=1 ][line width=0.08]  [draw opacity=0] (8.04,-3.86) -- (0,0) -- (8.04,3.86) -- (5.34,0) -- cycle;
\draw [color={rgb, 255:red, 0; green, 0; blue, 255 }  ,draw opacity=1 ][fill={rgb, 255:red, 0; green, 255; blue, 255 }  ,fill opacity=1 ]   (174.53,165.4) -- (174.53,142.8) ;
\draw [shift={(174.53,139.8)}, rotate = 90] [fill={rgb, 255:red, 0; green, 0; blue, 255 }  ,fill opacity=1 ][line width=0.08]  [draw opacity=0] (8.04,-3.86) -- (0,0) -- (8.04,3.86) -- (5.34,0) -- cycle    ;
\draw [color={rgb, 255:red, 126; green, 211; blue, 33 }  ,draw opacity=1 ][line width=1.5]    (265.28,147.8) -- (265.28,128.3) ;
\draw [shift={(265.28,124.3)}, rotate = 90] [fill={rgb, 255:red, 126; green, 211; blue, 33 }  ,fill opacity=1 ][line width=0.08]  [draw opacity=0] (10.92,-2.73) -- (0,0) -- (10.92,2.73) -- cycle    ;
\draw [color={rgb, 255:red, 126; green, 211; blue, 33 }  ,draw opacity=1 ][line width=1.5]    (480.58,131.5) -- (467.29,116.19) ;
\draw [shift={(464.67,113.17)}, rotate = 49.04] [fill={rgb, 255:red, 126; green, 211; blue, 33 }  ,fill opacity=1 ][line width=0.08]  [draw opacity=0] (10.92,-2.73) -- (0,0) -- (10.92,2.73) -- cycle    ;

\draw (4,202) node [anchor=north west][inner sep=0.75pt]    {$x_{A}$};
\draw (140.8,207.4) node [anchor=north west][inner sep=0.75pt]    {$y_{A}$};
\draw (85.3,51.4) node [anchor=north west][inner sep=0.75pt]    {$z_{A}$};
\draw (178.07,190.93) node [anchor=north west][inner sep=0.75pt]  [font=\footnotesize]  {$S-G$};
\draw (177.22,163.68) node [anchor=north west][inner sep=0.75pt]  [font=\footnotesize]  {$B_{z_{A}}$};
\draw (272,110) node [anchor=north west][inner sep=0.75pt]  [font=\scriptsize]  {$\ket{\uparrow _{z_{A}}}$};
\draw (541.32,201.92) node [anchor=north west][inner sep=0.75pt]    {$x_{B}$};
\draw (636,147) node [anchor=north west][inner sep=0.75pt]    {$y_{B}$};
\draw (509.32,45) node [anchor=north west][inner sep=0.75pt]    {$z_{B}$};
\draw (391.27,150.93) node [anchor=north west][inner sep=0.75pt]  [font=\footnotesize]  {$B_{z_{B}}$};
\draw (481,98) node [anchor=north west][inner sep=0.75pt]  [font=\scriptsize]  {$\ket{\uparrow _{z_{B}}}$};
\draw (63,130.4) node [anchor=north west][inner sep=0.75pt]  [font=\large]  {$A$};
\draw (567.15,99.5) node [anchor=north west][inner sep=0.75pt]  [font=\large]  {$B$};
\draw (390.67,176.68) node [anchor=north west][inner sep=0.75pt]  [font=\footnotesize]  {$S-G$};

\end{tikzpicture}

    \caption{Schematic description of the alignment protocol. Alice sends spin systems aligned along her axes and Bob performs spin measurements on these spins with Stern-Gerlach apparata aligned along his axes. The spin expectation values computed with these measurements give the elements $R_{ij}$ of the rotation matrix relating the two observers. Notice that Alice prepares her spin states in the $\ket{\uparrow_{z_A}}$ state by blocking the lower path of her Stern-Gerlach apparatus, and Bob counts the number of spin up results by doing the same on his Stern-Gerlach apparatus. In the picture, the Stern-Gerlach apparata (\emph{i.e.} the magnetic fields $B_i$) used by Alice and Bob to prepare the spins and perform the measurements are aligned along their $z$ axes, $z_A$ and $z_B$ respectively, so that the picture describes the measurement of the $R_{33}$ component of the rotation matrix. By preparing spin systems aligned along the other directions $i$ and performing spin measurement along other directions $j$, all the $R_{ij}$ matrix elements can be measured.}
    \label{fig:protocol}
\end{figure}

In this section, we analyze a protocol in which two observers attempt to align themselves by exchanging spin systems, emphasizing the novelties introduced by the Doubly Quantum Mechanics framework and providing also some numerical examples using the semi-classical states introduced in the previous section.

In the classical setting, two observers can reconstruct the rotation matrix that relates their (in general misaligned) reference frames according to the following protocol. Suppose that Alice and Bob are two observers, each equipped with their own set of Stern-Gerlach apparata defining their Cartesian axes.
Alice prepares $3N$ spins oriented along the positive direction of each one of her axes, $x_A,y_A,z_A$ and sends them to Bob. Then Bob divides each group into three sets of $N$ spins, and with each set he performs spin measurements along each of his axes, $x_B,y_B,z_B$, by using the corresponding Stern-Gerlach apparata. This physical setup is depicted in \Cref{fig:protocol}. The spin expectation values along Bob's axes computed with the spin states prepared by Alice are the elements of the rotation matrix $R$ that relates the two reference frames \cite{Amelino-Camelia:2022dsj} according to
\begin{equation}\label{eq:measurements protocolo classic}
    R_{ij} = \mel{\uparrow_{\Vec{i}}^A}{\sigma_{\Vec{j}}^{B}}{\uparrow_{\Vec{i}}^A}\;,\;\; \;\;\; i,j=x,y,z\;.
\end{equation}
These expectation values are given by the difference between the probability of obtaining spin up and the probability of obtaining spin down, according to \eqref{Pauliclassicamedia}. To compute these probabilities, Bob will measure the frequencies of spin up (down) outcomes for every $i,j$ pair. For any of these pairs, the number of spin up results, $k$, on a total of $N$ measurements, will be distributed according to a binomial distribution of the form
\begin{equation}
    P(k,N) = \binom{N}{k}p_0^k(1-p_0)^{N-k}\;.
\end{equation}
where $p_0$ is the probability of obtaining spin up, and depends on the spin state used to perform the measurement, the direction along which the Stern -Gerlach is oriented and the relative orientation between the two observers. The expectation value of $k$ is given by $\mathbb{E}\qty[k] = Np_0$ and its uncertainty is given by the square root of the variance $\Delta\qty[k]=\sqrt{\mathbb{E}\qty[k^2]-\mathbb{E}\qty[k]^2} = \sqrt{Np_0(1-p_0)}$. Therefore, the expectation value for the frequency is given by $p_0$ with uncertainty $N^{-\frac{1}{2}}\sqrt{p_0(1-p_0)}$. For any given $i,j$ pair, denoting the probability of obtaining spin up by $p_{ij}$, Bob will then measure the expectation value of the spin to be distributed according to

\begin{equation}\label{eq:classical expval sigma bob measurement}
    \mel{\uparrow_{\Vec{i}}^A}{\sigma_{\Vec{j}}^{B}}{\uparrow_{\Vec{i}}^A}^{\text{meas}} \ebd \frac{\big(2\,\mathbb{E}\qty[k]-N\big) \pm 2\,\Delta[k]}{N} = \qty(2p_{ij}-1) \pm 2\sqrt{\frac{p_{ij}(1-p_{ij})}{N}}\xrightarrow[N \to \infty]{} 2p_{ij}-1\;.
\end{equation}
By means of this protocol, Bob can measure the elements of the rotation matrix with arbitrary precision, since the uncertainty scales as $N^{-\frac{1}{2}}$, hence the two observers can sharply align their reference frames when exchanging an infinite number of spin systems.

\subsection{The doubly quantum protocol}\label{sec:doubly quantum protocol theory}

\begin{figure}
    \centering

\tikzset{every picture/.style={line width=0.75pt}} 

\begin{tikzpicture}[x=0.69pt,y=0.69pt,yscale=-1,xscale=1]

\draw [color={rgb, 255:red, 126; green, 211; blue, 33 }  ,draw opacity=1 ]   (436.19,108.29) -- (559.39,108.29) ;
\draw [color={rgb, 255:red, 126; green, 211; blue, 33 }  ,draw opacity=1 ]   (90.75,136.4) -- (165.22,136.8) ;
\draw    (90.75,136.4) .. controls (90.34,138.72) and (88.97,139.68) .. (86.65,139.27) .. controls (84.33,138.86) and (82.97,139.82) .. (82.56,142.14) .. controls (82.15,144.46) and (80.78,145.41) .. (78.46,145) .. controls (76.14,144.59) and (74.78,145.55) .. (74.37,147.87) .. controls (73.96,150.19) and (72.59,151.15) .. (70.27,150.74) .. controls (67.95,150.33) and (66.59,151.29) .. (66.18,153.61) .. controls (65.77,155.93) and (64.4,156.89) .. (62.08,156.48) .. controls (59.76,156.07) and (58.39,157.02) .. (57.98,159.34) .. controls (57.57,161.66) and (56.21,162.62) .. (53.89,162.21) .. controls (51.57,161.8) and (50.2,162.76) .. (49.79,165.08) .. controls (49.38,167.4) and (48.02,168.36) .. (45.7,167.95) .. controls (43.38,167.54) and (42.01,168.49) .. (41.6,170.81) .. controls (41.19,173.13) and (39.83,174.09) .. (37.51,173.68) -- (33.68,176.36) -- (27.13,180.95) ;
\draw [shift={(25.49,182.09)}, rotate = 325] [color={rgb, 255:red, 0; green, 0; blue, 0 }  ][line width=0.75]    (8.74,-2.63) .. controls (5.56,-1.12) and (2.65,-0.24) .. (0,0) .. controls (2.65,0.24) and (5.56,1.12) .. (8.74,2.63)   ;
\draw    (90.75,136.4) .. controls (93.07,136.81) and (94.03,138.18) .. (93.62,140.5) .. controls (93.21,142.82) and (94.17,144.18) .. (96.49,144.59) .. controls (98.81,145) and (99.76,146.37) .. (99.35,148.69) .. controls (98.94,151.01) and (99.9,152.37) .. (102.22,152.78) .. controls (104.54,153.19) and (105.5,154.56) .. (105.09,156.88) .. controls (104.68,159.2) and (105.64,160.56) .. (107.96,160.97) .. controls (110.28,161.38) and (111.24,162.75) .. (110.83,165.07) .. controls (110.42,167.39) and (111.37,168.76) .. (113.69,169.17) .. controls (116.01,169.58) and (116.97,170.94) .. (116.56,173.26) .. controls (116.15,175.58) and (117.11,176.95) .. (119.43,177.36) .. controls (121.75,177.77) and (122.71,179.13) .. (122.3,181.45) .. controls (121.89,183.77) and (122.84,185.14) .. (125.16,185.55) .. controls (127.48,185.96) and (128.44,187.32) .. (128.03,189.64) -- (130.71,193.47) -- (135.3,200.02) ;
\draw [shift={(136.44,201.66)}, rotate = 235] [color={rgb, 255:red, 0; green, 0; blue, 0 }  ][line width=0.75]    (8.74,-2.63) .. controls (5.56,-1.12) and (2.65,-0.24) .. (0,0) .. controls (2.65,0.24) and (5.56,1.12) .. (8.74,2.63)   ;
\draw [color={rgb, 255:red, 126; green, 211; blue, 33 }  ,draw opacity=1 ]   (165.22,136.8) .. controls (201.22,136.6) and (186.93,152.42) .. (223.59,151.06) ;
\draw  [fill={rgb, 255:red, 0; green, 0; blue, 0 }  ,fill opacity=1 ] (239.83,143.97) .. controls (240.66,143.65) and (241.56,143.48) .. (242.49,143.48) .. controls (246.73,143.48) and (250.17,146.98) .. (250.17,151.28) .. controls (250.17,155.59) and (246.73,159.08) .. (242.49,159.08) .. controls (241.56,159.08) and (240.66,158.91) .. (239.83,158.6) -- cycle ;
\draw [color={rgb, 255:red, 126; green, 211; blue, 33 }  ,draw opacity=1 ]   (165.22,136.8) .. controls (201.22,136.6) and (184.11,120.99) .. (223.59,122.54) ;
\draw [color={rgb, 255:red, 126; green, 211; blue, 33 }  ,draw opacity=1 ]   (223.59,151.06) -- (239.19,151.06) ;
\draw [color={rgb, 255:red, 126; green, 211; blue, 33 }  ,draw opacity=1 ]   (223.59,122.54) -- (346.79,122.54) ;
\draw    (90.75,136.4) .. controls (89.08,134.73) and (89.08,133.07) .. (90.75,131.4) .. controls (92.42,129.73) and (92.42,128.07) .. (90.75,126.4) .. controls (89.08,124.73) and (89.08,123.07) .. (90.75,121.4) .. controls (92.42,119.73) and (92.42,118.07) .. (90.75,116.4) .. controls (89.08,114.73) and (89.08,113.07) .. (90.75,111.4) .. controls (92.42,109.73) and (92.42,108.07) .. (90.75,106.4) .. controls (89.08,104.73) and (89.08,103.07) .. (90.75,101.4) .. controls (92.42,99.73) and (92.42,98.07) .. (90.75,96.4) .. controls (89.08,94.73) and (89.08,93.07) .. (90.75,91.4) .. controls (92.42,89.73) and (92.42,88.07) .. (90.75,86.4) .. controls (89.08,84.73) and (89.08,83.07) .. (90.75,81.4) .. controls (92.42,79.73) and (92.42,78.07) .. (90.75,76.4) .. controls (89.08,74.73) and (89.08,73.07) .. (90.75,71.4) -- (90.75,66.73) -- (90.75,58.73) ;
\draw [shift={(90.75,56.73)}, rotate = 90] [color={rgb, 255:red, 0; green, 0; blue, 0 }  ][line width=0.75]    (8.74,-2.63) .. controls (5.56,-1.12) and (2.65,-0.24) .. (0,0) .. controls (2.65,0.24) and (5.56,1.12) .. (8.74,2.63)   ;
\draw    (559.39,108.29) .. controls (560.6,110.32) and (560.19,111.93) .. (558.16,113.14) .. controls (556.13,114.35) and (555.72,115.96) .. (556.93,117.99) .. controls (558.14,120.02) and (557.73,121.63) .. (555.7,122.83) .. controls (553.67,124.04) and (553.26,125.65) .. (554.47,127.68) .. controls (555.68,129.71) and (555.27,131.32) .. (553.24,132.53) .. controls (551.21,133.73) and (550.8,135.34) .. (552.01,137.37) .. controls (553.22,139.4) and (552.81,141.01) .. (550.78,142.22) .. controls (548.75,143.42) and (548.34,145.03) .. (549.55,147.06) .. controls (550.76,149.09) and (550.35,150.7) .. (548.32,151.91) .. controls (546.29,153.12) and (545.88,154.73) .. (547.09,156.76) .. controls (548.3,158.79) and (547.89,160.4) .. (545.86,161.6) .. controls (543.83,162.81) and (543.42,164.42) .. (544.63,166.45) .. controls (545.84,168.48) and (545.43,170.09) .. (543.4,171.3) -- (542.25,175.82) -- (540.29,183.57) ;
\draw [shift={(539.79,185.51)}, rotate = 284.24] [color={rgb, 255:red, 0; green, 0; blue, 0 }  ][line width=0.75]    (8.74,-2.63) .. controls (5.56,-1.12) and (2.65,-0.24) .. (0,0) .. controls (2.65,0.24) and (5.56,1.12) .. (8.74,2.63)   ;
\draw    (559.39,108.29) .. controls (561.42,107.08) and (563.03,107.49) .. (564.24,109.52) .. controls (565.44,111.55) and (567.05,111.96) .. (569.08,110.75) .. controls (571.11,109.54) and (572.72,109.95) .. (573.93,111.98) .. controls (575.14,114.01) and (576.75,114.42) .. (578.78,113.21) .. controls (580.81,112) and (582.42,112.41) .. (583.62,114.44) .. controls (584.83,116.47) and (586.44,116.88) .. (588.47,115.67) .. controls (590.5,114.46) and (592.11,114.87) .. (593.31,116.9) .. controls (594.52,118.93) and (596.13,119.34) .. (598.16,118.13) .. controls (600.19,116.92) and (601.8,117.33) .. (603.01,119.36) .. controls (604.21,121.39) and (605.82,121.8) .. (607.85,120.59) .. controls (609.88,119.38) and (611.49,119.79) .. (612.7,121.82) .. controls (613.91,123.85) and (615.52,124.26) .. (617.55,123.05) .. controls (619.58,121.84) and (621.19,122.25) .. (622.39,124.28) -- (626.92,125.43) -- (634.67,127.4) ;
\draw [shift={(636.61,127.89)}, rotate = 194.24] [color={rgb, 255:red, 0; green, 0; blue, 0 }  ][line width=0.75]    (8.74,-2.63) .. controls (5.56,-1.12) and (2.65,-0.24) .. (0,0) .. controls (2.65,0.24) and (5.56,1.12) .. (8.74,2.63)   ;
\draw    (559.39,108.29) .. controls (557.04,108.12) and (555.96,106.86) .. (556.13,104.51) .. controls (556.3,102.16) and (555.21,100.89) .. (552.86,100.72) .. controls (550.51,100.55) and (549.42,99.28) .. (549.6,96.93) .. controls (549.77,94.58) and (548.68,93.31) .. (546.33,93.14) .. controls (543.98,92.97) and (542.9,91.71) .. (543.07,89.36) .. controls (543.24,87.01) and (542.15,85.74) .. (539.8,85.57) .. controls (537.45,85.4) and (536.36,84.13) .. (536.54,81.78) .. controls (536.71,79.43) and (535.62,78.17) .. (533.27,78) .. controls (530.92,77.83) and (529.83,76.56) .. (530.01,74.21) .. controls (530.19,71.86) and (529.1,70.59) .. (526.75,70.42) .. controls (524.4,70.25) and (523.31,68.98) .. (523.48,66.63) .. controls (523.65,64.28) and (522.57,63.02) .. (520.22,62.85) .. controls (517.87,62.68) and (516.78,61.41) .. (516.95,59.06) -- (513.91,55.52) -- (508.68,49.46) ;
\draw [shift={(507.38,47.95)}, rotate = 49.24] [color={rgb, 255:red, 0; green, 0; blue, 0 }  ][line width=0.75]    (8.74,-2.63) .. controls (5.56,-1.12) and (2.65,-0.24) .. (0,0) .. controls (2.65,0.24) and (5.56,1.12) .. (8.74,2.63)   ;
\draw [color={rgb, 255:red, 126; green, 211; blue, 33 }  ,draw opacity=1 ]   (346.79,122.54) -- (377.99,122.54) ;
\draw  [draw opacity=0] (159.8,130.94) .. controls (162.57,115.11) and (176.39,103.07) .. (193.02,103.07) .. controls (204.44,103.07) and (214.55,108.76) .. (220.65,117.45) -- (193.02,136.8) -- cycle ; \draw   (159.8,130.94) .. controls (162.57,115.11) and (176.39,103.07) .. (193.02,103.07) .. controls (204.44,103.07) and (214.55,108.76) .. (220.65,117.45) ;  
\draw  [draw opacity=0] (225.6,128.07) .. controls (226.35,130.85) and (226.74,133.78) .. (226.74,136.8) .. controls (226.74,139.82) and (226.35,142.75) .. (225.6,145.53) -- (193.02,136.8) -- cycle ; \draw   (225.6,128.07) .. controls (226.35,130.85) and (226.74,133.78) .. (226.74,136.8) .. controls (226.74,139.82) and (226.35,142.75) .. (225.6,145.53) ;  
\draw  [draw opacity=0] (220.65,156.15) .. controls (214.55,164.84) and (204.44,170.53) .. (193.02,170.53) .. controls (176.39,170.53) and (162.57,158.49) .. (159.8,142.66) -- (193.02,136.8) -- cycle ; \draw   (220.65,156.15) .. controls (214.55,164.84) and (204.44,170.53) .. (193.02,170.53) .. controls (176.39,170.53) and (162.57,158.49) .. (159.8,142.66) ;  
\draw [color={rgb, 255:red, 126; green, 211; blue, 33 }  ,draw opacity=1 ]   (377.82,122.55) .. controls (413.82,122.35) and (399.53,138.17) .. (436.19,136.81) ;
\draw  [fill={rgb, 255:red, 0; green, 0; blue, 0 }  ,fill opacity=1 ] (452.43,129.72) .. controls (453.26,129.4) and (454.16,129.23) .. (455.09,129.23) .. controls (459.33,129.23) and (462.77,132.73) .. (462.77,137.03) .. controls (462.77,141.34) and (459.33,144.83) .. (455.09,144.83) .. controls (454.16,144.83) and (453.26,144.66) .. (452.43,144.35) -- cycle ;
\draw [color={rgb, 255:red, 126; green, 211; blue, 33 }  ,draw opacity=1 ]   (377.82,122.55) .. controls (413.82,122.35) and (396.71,106.74) .. (436.19,108.29) ;
\draw [color={rgb, 255:red, 126; green, 211; blue, 33 }  ,draw opacity=1 ]   (436.19,136.81) -- (451.79,136.81) ;
\draw  [draw opacity=0] (372.4,116.69) .. controls (375.17,100.86) and (388.99,88.82) .. (405.62,88.82) .. controls (417.04,88.82) and (427.15,94.51) .. (433.25,103.2) -- (405.62,122.55) -- cycle ; \draw   (372.4,116.69) .. controls (375.17,100.86) and (388.99,88.82) .. (405.62,88.82) .. controls (417.04,88.82) and (427.15,94.51) .. (433.25,103.2) ;  
\draw  [draw opacity=0] (438.2,113.82) .. controls (438.95,116.6) and (439.34,119.53) .. (439.34,122.55) .. controls (439.34,125.57) and (438.95,128.5) .. (438.2,131.28) -- (405.62,122.55) -- cycle ; \draw   (438.2,113.82) .. controls (438.95,116.6) and (439.34,119.53) .. (439.34,122.55) .. controls (439.34,125.57) and (438.95,128.5) .. (438.2,131.28) ;  
\draw  [draw opacity=0] (433.25,141.9) .. controls (427.15,150.59) and (417.04,156.28) .. (405.62,156.28) .. controls (388.99,156.28) and (375.17,144.24) .. (372.4,128.41) -- (405.62,122.55) -- cycle ; \draw   (433.25,141.9) .. controls (427.15,150.59) and (417.04,156.28) .. (405.62,156.28) .. controls (388.99,156.28) and (375.17,144.24) .. (372.4,128.41) ;  
\draw [color={rgb, 255:red, 126; green, 211; blue, 33 }  ,draw opacity=1 ][line width=1.5]    (264.47,133.67) .. controls (262.8,132) and (262.8,130.34) .. (264.47,128.67) .. controls (266.14,127) and (266.14,125.34) .. (264.47,123.67) .. controls (262.8,122) and (262.8,120.34) .. (264.47,118.67) -- (264.47,115.3) -- (264.47,107.3) ;
\draw [shift={(264.47,103.3)}, rotate = 90] [fill={rgb, 255:red, 126; green, 211; blue, 33 }  ,fill opacity=1 ][line width=0.08]  [draw opacity=0] (10.92,-2.73) -- (0,0) -- (10.92,2.73) -- cycle    ;
\draw [color={rgb, 255:red, 0; green, 0; blue, 255 }  ,draw opacity=1 ]   (400.67,138.17) .. controls (398.32,137.92) and (397.28,136.62) .. (397.53,134.27) .. controls (397.78,131.92) and (396.74,130.63) .. (394.39,130.38) .. controls (392.05,130.13) and (391.01,128.83) .. (391.26,126.49) .. controls (391.51,124.14) and (390.47,122.84) .. (388.12,122.59) .. controls (385.77,122.34) and (384.73,121.05) .. (384.98,118.7) -- (384.15,117.66) -- (379.13,111.43) ;
\draw [shift={(377.25,109.09)}, rotate = 51.15] [fill={rgb, 255:red, 0; green, 0; blue, 255 }  ,fill opacity=1 ][line width=0.08]  [draw opacity=0] (8.04,-3.86) -- (0,0) -- (8.04,3.86) -- (5.34,0) -- cycle    ;
\draw [color={rgb, 255:red, 0; green, 0; blue, 255 }  ,draw opacity=1 ]   (171.72,155.17) .. controls (170.05,153.5) and (170.05,151.84) .. (171.72,150.17) .. controls (173.39,148.5) and (173.39,146.84) .. (171.72,145.17) .. controls (170.05,143.5) and (170.05,141.84) .. (171.72,140.17) .. controls (173.39,138.5) and (173.39,136.84) .. (171.72,135.17) .. controls (170.05,133.5) and (170.05,131.84) .. (171.72,130.17) -- (171.72,129.3) -- (171.72,121.3) ;
\draw [shift={(171.72,118.3)}, rotate = 90] [fill={rgb, 255:red, 0; green, 0; blue, 255 }  ,fill opacity=1 ][line width=0.08]  [draw opacity=0] (8.04,-3.86) -- (0,0) -- (8.04,3.86) -- (5.34,0) -- cycle    ;
\draw [color={rgb, 255:red, 126; green, 211; blue, 33 }  ,draw opacity=1 ][line width=1.5]    (488.19,116.93) .. controls (485.84,116.82) and (484.71,115.59) .. (484.81,113.24) .. controls (484.92,110.89) and (483.79,109.66) .. (481.44,109.55) .. controls (479.09,109.45) and (477.96,108.22) .. (478.06,105.87) -- (475.57,103.15) -- (470.17,97.25) ;
\draw [shift={(467.47,94.3)}, rotate = 47.51] [fill={rgb, 255:red, 126; green, 211; blue, 33 }  ,fill opacity=1 ][line width=0.08]  [draw opacity=0] (10.92,-2.73) -- (0,0) -- (10.92,2.73) -- cycle    ;

\draw (103.67,50.9) node [anchor=north west][inner sep=0.75pt]  [font=\scriptsize]  {$\ket{\Phi ^{RO}}$};
\draw (176.75,173.93) node [anchor=north west][inner sep=0.75pt]  [font=\footnotesize]  {$S-G$};
\draw (61.68,113.4) node [anchor=north west][inner sep=0.75pt]  [font=\large]  {$A$};
\draw (565.83,82.5) node [anchor=north west][inner sep=0.75pt]  [font=\large]  {$B$};
\draw (389.35,159.68) node [anchor=north west][inner sep=0.75pt]  [font=\footnotesize]  {$S-G$};
\draw (229.67,89.23) node [anchor=north west][inner sep=0.75pt]  [font=\scriptsize]  {$\ket{\Phi ^{S}}$};
\draw (430,70) node [anchor=north west][inner sep=0.75pt]  [font=\scriptsize]  {$\ket{\Phi ^{SG}}$};
\draw (545.67,43.4) node [anchor=north west][inner sep=0.75pt]  [font=\scriptsize]  {$\ket{\Phi ^{RO}}$};

\end{tikzpicture}

    \caption{The doubly quantum version of the protocol in \Cref{fig:protocol}. The schematic setup is similar to the classic case, the difference being that the geometrical degrees of freedom are now described by quantum states in the Hilbert space $\mathcal{H}_{SU_q(2)}$. Specifically, the states $\ket{\Phi^S}$ and $\ket{\Phi^{SG}}$ replace the notion of alignment of spins and Stern-Gerlach apparata along the observers' axes, while the state $\ket{\Phi^{RO}}$ replaces the notion of coordinate axes and describes the relative orientation between Alice and Bob, whose reference frames will not be sharp anymore. The quantum nature of these geometrical degrees of freedom is represented by curly lines for the reference frames, the magnetic fields inside the Stern-Gerlach apparata, and the spin systems in place of the straight lines of \Cref{fig:protocol}.}
    \label{fig:protocol quantum}
\end{figure}
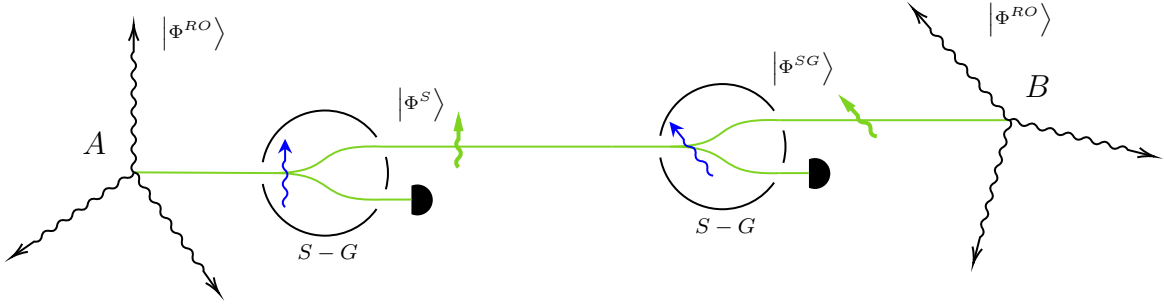

We now investigate how the protocol outlined above changes in our non-commutative framework, as a consequence of the quantization of the geometry of the spin states, Stern-Gerlach apparata, and relative orientation. In doing so, we have to describe how Bob performs measurements with his apparata on the states that Alice prepares, according to \eqref{eq:measurements protocolo classic}, which we want to generalize to the non-commutative setting. This means that measurements are performed on the states $\ket{\psi^A}\otimes \one$, where $\ket{\psi^A}$ is given by \eqref{eq:genericstate}, and the observable is $\sigma^q_B$, given by \eqref{eq:sigma Bob axiom}, which describes a rotation of Alice's observable. 

In light of the properties of quantum probabilities discussed in \Cref{sec: quantum prob}, the probability that Bob measures spin up or down in a measurement is itself described by a probability distribution $f(p)$. This means that the number of spin up outcomes $k$ in $N$ spin measurements is distributed according to the joint probability density
\begin{equation}\label{eq:deformed binomial}
    \mathcal{P}(k,N,p) = \binom{N}{k}p^k(1-p)^{N-k}f(p)\;,
\end{equation}
obtained by multiplying the binomial distribution $P(k,N)$ with the probability density $f(p)$\footnote{This is nothing more than the probability density of measuring the value $p$ for the observable $P_A\qty(\uparrow^B_\sigma)$ and observing $k$ spin up outcomes within $N$ spin measurements, which is given by the product between the conditional probability for the latter event given that the value measured for the probability is $p$, $P(k,N)$, and the probability density of measuring the value $p$, $f(p)$. This is discussed in more detail in \Cref{appendix: interpretation of joint probability}.}, where $p$ is the probability of obtaining spin up in a measurement and $f(p)$ is given by
\begin{equation}\label{eq:probability distribution protocol}
    f(p) = \sum_{r} \braket{\Phi}{p,r}\braket{p,r}{\Phi}\;,\;\; \ket{\Phi} = \ket{\Phi^S}\otimes\ket{\Phi^{SG}}\otimes\ket{\Phi^{RO}} \; ,
\end{equation}
which generalizes \eqref{eq:probability distribution Alice} to the case in which also the relative orientation between Alice and Bob is taken into account, and $\ket{p,r}$ are now the eigenstates of the probability operator $P_A\qty(\uparrow_\sigma^B)$ defined in \eqref{eq:sigma expectation value and probability bob}. We will focus on the case in which these three states are semi-classical (in the sense of \Cref{sec:effective theory alice}) and can be assigned independently. 

With these ingredients, we can generalize the classical measurement procedure. The experimental setup is analogous to the commutative case (see \Cref{fig:protocol quantum}), but the results obtained by Bob will be different because the number of spin up outcomes is distributed according to \eqref{eq:deformed binomial}. For this reason, the expectation value of $k$ and its uncertainty, as shown in \Cref{appendix: joint probability distribution}, read

\begin{equation}\label{eq:expval outcomes e uncertainty deformed}
    \mathbb{E}\qty[k] = N p_0\;,\;\; \Delta[k] = \sqrt{N(N-1)\Delta_f^2+Np_0(1-p_0)}
\end{equation}
where $p_0$ and $\Delta_f^2$ are the mean value and variance of $f(p)$, respectively. Recalling that $\expval{\sigma_B^q}_A = q P_A\qty(\uparrow_\sigma^B) - q^{-1}P_A\qty(\downarrow_\sigma^B) = \qty(q+q^{-1})P_A\qty(\uparrow_\sigma^B) -q^{-1}\mathbbm{1}$, the measurement outcomes of the expectation value of the spin will be distributed, up to constant factors, according to $f(p)$, as shown in \Cref{appendix: variance protocol}. Therefore, analogously to \eqref{eq:classical expval sigma bob measurement}, Bob will measure the expectation value of the spin $\expval{\sigma^q_B}_A^\text{meas} \ebd \overline{\expval{\sigma^q_B}}_A \pm \overline{\Delta}\big[\expval{\sigma_B^q}_A\big]$ to be distributed as
\begin{equation}\label{eq:fundamental uncertainty theory}
    \begin{aligned}
    {\expval{\sigma^q_B}}_A^\text{ meas} &= \frac{\Big[\qty(q+q^{-1})\mathbb{E}\qty[k]-Nq^{-1}\Big]\,\pm \qty(q+q^{-1})\Delta[k]}{N}\\
    &\Longrightarrow{\expval{\sigma^q_B}}_A^\text{ meas} \xrightarrow[N \to \infty]{} \Big[\qty(q+q^{-1}) \,p_0-q^{-1}\Big]\pm \qty(q+q^{-1})\Delta_f\;.
    \end{aligned}
\end{equation}

We thus see that in the non-commutative case, Bob cannot measure the elements of the rotation matrix with arbitrary precision. Indeed, a fundamental uncertainty on these matrix elements, that depends on the states $\ket{\Phi^S}$, $\ket{\Phi^{SG}}$ and $\ket{\Phi^{RO}}$, arises in our framework. Of course, the more the distribution function $f(p)$ is peaked around the value $p_0$, the more precise the measurements performed by Bob will be. In standard Quantum Mechanics, this function becomes $f(p)=\delta\qty(p-p_0)$, and we recover the classical result \eqref{eq:classical expval sigma bob measurement}, since $\Delta_f =0$. However, in our setting, this uncertainty is typically non-vanishing, since geometry states describing semi-classical experimental configurations are not probability eigenstates, in general. This yields a fundamental uncertainty on the measured expectation value of the spin that does not vanish even when performing an infinite number of measurements, contrary to the standard result in \eqref{eq:classical expval sigma bob measurement}. In the following subsection, we will consider concrete examples of the theoretical result outlined above to numerically quantify the fundamental uncertainty appearing in \eqref{eq:fundamental uncertainty theory}.

\subsection{Numerical analysis of the doubly quantum protocol}\label{sec:numerical example protocol}

In the classical protocol, the directions along which spins and Stern-Gerlach apparata are aligned coincide with the Cartesian axes of the two observers involved, which define the rows and columns indexes of the rotation matrix that connects them, see \eqref{eq:measurements protocolo classic}. 
In the non-commutative setting, these directions are quantized, so that the rows and column indexes of the rotation matrix are replaced by geometry states that semi-classically describe the directions $x_q,y_q,z_q$ that are closest to three Cartesian axes $x,y,z$, in the spirit of \Cref{sec:effective theory alice}. These states are given by

\begin{equation}
    \begin{aligned}
    &x_q \;\vcentcolon\; \ket{x_q^S}\ebd \ket{\Phi^{S}\qty(\theta\qty(n_{\frac{\pi}{2}}),0)}=\ket{\mu_s\qty(\frac{\pi}{2}),0}_s\;,\; \ket{x_q^{SG}}\ebd\ket{\Phi^{SG}\qty(\theta\qty(n_{\frac{\pi}{2}}),0)}=\ket{\mu_a\qty(\frac{\pi}{2}),0}_a \; , \\
    &y_q \;\vcentcolon\;\ket{y_q^S}\ebd \ket{\Phi^{S}\qty(\theta\qty(n_{\frac{\pi}{2}}),\frac{\pi}{2})}=\ket{\mu_s\qty(\frac{\pi}{2}),\frac{\pi}{2}}_s\;,\; \ket{y_q^{SG}}\ebd\ket{\Phi^{SG}\qty(\theta\qty(n_{\frac{\pi}{2}}),\frac{\pi}{2})}=\ket{\mu_a\qty(\frac{\pi}{2}),\frac{\pi}{2}}_a \; ,\\
    &z_q \;\vcentcolon\;\ket{z_q^S}\ebd \ket{\Phi^{S}\qty(0,\omega_s)}=\ket{\chi_s=\omega_s}\;,\; \ket{z_q^{SG}}\ebd\ket{\Phi^{SG}\qty(0,\omega_a)}=\ket{\chi_a=\omega_a} \; ,\\
    \end{aligned}
\end{equation}
For the subsequent numerical examples, we take $q=0.99$ so that $\theta\qty(n_{\frac{\pi}{2}}) = \theta(34) = 1.006\,\frac{\pi}{2}$ is the $\theta(n)$ that is closest to $\frac{\pi}{2}$, corresponding to $\mu\qty(\frac{\pi}{2})=7$, with the prescriptions described in \Cref{appendix: semi-classical states} and \Cref{sec:semiclassical states and probability eigenstates}.

The building blocks needed to construct the non-commutative analogue of the matrix elements describing the relative orientation between Alice and Bob are
\begin{equation}
    \expval{\sigma_B^q}_A = \mel{\psi^{A}}{\sigma_B^q}{\psi^{A}}= \mel{\psi^{A}}{U^q_{g}\sigma^q_A {U^q_{g}}^{\dagger}}{\psi^{A}}\;,\;
\end{equation}
as is also the case in the classical protocol. Expectation values and variances of this operator in the states of the geometry yield the mean value and uncertainty of the distribution associated to the rotation matrix connecting the two observers. In particular, we define
\begin{equation}
\label{eq:matelements}
\begin{aligned}
    &\overline{R^q_{ij}}\qty(\theta_g,\phi_g,\chi_g)\ebd \expval{\expval{\sigma_B^q}_A}{\Phi_{ij}\qty(\theta_g,\phi_g,\chi_g)} \\
    &\overline{\Delta}^{\,2}\big[R^q_{ij}\big]\qty(\theta_g,\phi_g,\chi_g)\ebd \expval{\expval{\sigma_B^q}_A^2}{\Phi_{ij}\qty(\theta_g,\phi_g,\chi_g)}  -\Big[\overline{R^q_{ij}}\qty(\theta_g,\phi_g,\chi_g)\Big]^2\;, \;\; i,j=x,y,z \; ,
\end{aligned}
\end{equation}
where
\begin{equation}
\ket{\Phi_{ij}\qty(\theta_g,\phi_g,\chi_g)}=\ket{i_q^S}\otimes\ket{j_q^{SG}}\otimes\ket{\Phi^{RO}\qty(\theta_g,\phi_g,\chi_g)}\, , \qquad i,j=x,y,z \; .
\end{equation}

As we will concretely show in the following examples, the doubly quantum nature of the alignment protocol generally yields a deformed rotation matrix, with elements specified by \eqref{eq:matelements}, that replaces the standard rotation matrix. The elements in \eqref{eq:matelements} are characterized, in general, by a non-vanishing $\Delta_f$ that limits the possibility of the two observers to learn about their relative orientation with infinite precision.
We will consider three examples given by the states
\begin{equation}
    \begin{aligned}
        &\ket{\Phi^{RO}\qty(\theta_g,\phi_g,\chi_g)} = \ket{\chi_{g} = 0}\;,\\
        &\ket{\Phi^{RO}\qty(\theta_g,\phi_g,\chi_g)} = \ket{\mu_g\qty(\frac{\pi}{2}), \phi_g=\frac{3}{2}\pi, \chi_g = 0}_g\;,\\
        &\ket{\Phi^{RO}\qty(\theta_g,\phi_g,\chi_g)} = \ket{\mu_g(\pi), \phi_g = \frac{3}{2}\pi, \chi_g = 0}_g = \ket{n_g = 0, \phi_g = \frac{3}{2}\pi,\chi_g = 0}\;.
    \end{aligned}
\end{equation}
We denote the measurement outcomes of the alignment protocol as deformed rotation matrix elements $R^q_{ij}\qty(\theta_g,\phi_g,\chi_g)$, characterized by their mean values and uncertainties, collected in a $3\cross 3$ matrix denoted by $R^q\qty(\theta_g,\phi_g,\chi_g)$. Using the geometry states listed above, with the same numerical approximations outlined in \Cref{appendix: semi-classical states} and up to two decimal places, we obtain 

\begin{equation}
R^q\qty(0,0,0)=\left(
\begin{array}{ccc}
 0.99 & 0.00 & -0.03 \\
 0.00 & 0.99 & -0.03 \\
 0.00 & 0.00 & 0.99 \\
\end{array}
\right)\pm \left(
\begin{array}{ccc}
 0.00 & 0.14 & 0.10 \\
 0.14 & 0.00 & 0.10 \\
 0.10 & 0.10 & 0.00 \\
\end{array}
\right)
\end{equation}\\ \vspace{-15pt}

\begin{equation}
    R^q\qty(1.006\,\frac{\pi}{2},\frac{3}{2}\pi,0)=\left(
\begin{array}{ccc}
 0.98 & 0.02 & 0.00 \\
 0.00 & 0.01 & -0.99 \\
 0.00 & 0.97 & 0.00 \\
\end{array}
\right)\pm \left(
\begin{array}{ccc}
 0.03 & 0.17 & 0.10 \\
 0.17 & 0.17 & 0.02 \\
 0.14 & 0.02 & 0.10 \\
\end{array}
\right)
\end{equation}\\ \vspace{-15pt}

\begin{equation}
     R^q\qty(\pi,\frac{3}{2}\pi,0)=\left(
\begin{array}{ccc}
 0.98 & 0.00 & 0.02 \\
 0.00 & -0.98 & 0.02 \\
 0.00 & 0.00 & -0.97 \\
\end{array}
\right)\pm \left(
\begin{array}{ccc}
 0.03 & 0.14 & 0.17 \\
 0.14 & 0.03 & 0.17 \\
 0.17 & 0.17 & 0.00 \\
\end{array}
\right)
\end{equation}\\
The values in these matrices differ by at most $\mathcal{O}[(1-q)]$ from their classical counterparts, elements of the matrix $R_z\qty(\alpha)R_x(\theta_g)R_z(\gamma)$, with $\chi_g = \frac{\alpha+\gamma}{2}$ and $\phi_g = \frac{3}{2}\pi -\frac{\alpha-\gamma}{2}$. The uncertainties are the square roots of the elements $\overline{\Delta}^{\,2}[R_{ij}]$ appearing in \Cref{eq:matelements} and are at most $\mathcal{O}[(1-q)^{1/2}]$. Analogously to the examples for the single observer presented in \Cref{appendix: semi-classical states}, one can numerically check that as $q$ increases, the uncertainties become closer to $0$.
Notice that the result obtained for the first example, where the relative orientation state effectively describes the identity transformation, still predicts an intrinsic uncertainty in the alignment procedure. This is in agreement with the fact that, in our framework, the geometrical degrees of freedom describing the spin and Stern-Gerlach orientations and the relative orientation between reference frames acquire a quantum nature. Therefore, the directions associated with the very messengers of the protocol and the measurement apparata are quantum, so that an intrinsic uncertainty arises even when the state describing the relative orientation between the two observers corresponds to the identity transformation. 

The numerical examples analyzed in this subsection confirm what expected from the general theory presented in \Cref{sec:doubly quantum protocol theory}. We have chosen a particular class of semi-classical states and other choices are possible, which do not affect significantly the numerical results. Indeed, in \Cref{app:probabilityeigenstates} we have shown that there are no separable eigenstates of the probability operator that describe measurements of spin systems misaligned with the Stern-Gerlach apparata. For this reason, some of the elements of the quantum rotation matrix are inevitably affected by a non-vanishing uncertainty, for any choice of semi-classical states. This yields a fundamental limit on the observers' knowledge regarding other observers' orientation in space when executing the alignment protocol described above, preventing them from sharply aligning their reference frames.

\section{Discussion}

In this paper we moved the first steps toward the formulation of a DQM theory, namely a theory of Quantum Mechanics in which the geometrical degrees of freedom of physical systems are themselves quantized, formulating its axioms by specifying them for spin measurements. This is achieved by deforming the rotation symmetry group $SU(2)$ into the quantum group $SU_q(2)$. The striking prediction of this framework is the notion of quantum probabilities, namely probabilities that are operator-valued. The probability distribution of a given measurement depends on the geometrical configuration of the experimental setup, specified by its geometry states, and exhibits, in general, non-classical features, such as a non-vanishing variance. We have focused on semi-classical states, which present small deviations from classical geometrical configurations, to analyze an alignment protocol between two observers that are, in general, misaligned. Such a protocol allows them to sharply measure their relative orientation (the rotation matrix relating their reference frames) in the standard case. The quantumness of probabilities deriving from the DQM framework prevents them to do so: an intrinsic, in general non-vanishing, variance affects the matrix elements of the rotation matrix relating their reference frames. This feature could be a hint that the amount of information that two observers can exchange when they do not know the relation between their reference frames is limited, even if the number of exchanged messages is infinite. In the standard $SU(2)$ case, there is no such limit, as the number of logical bits (and qubits) encoded per physical qubit approaches unity for large numbers of exchanged physical qubits \cite{Kempe:2001, Bartlett_2003}, and the two observers can then align their reference frames with infinite precision \cite{Bartlett:2006tzx}. This is prevented when $SU(2)$ is replaced by $SU_q(2)$ in our DQM framework, and this could signal the emergence of an intrinsic limit on the amount of information that two observers can exchange.

At the practical level, the axiomatic approach outlined in our work promotes complex numbers, appearing in linear combinations of spin states, in the expression of the generic Pauli matrix, and in the matrix defining the relative orientation between two observers, to operators satisfying the commutation relations of $SU_q(2)$. This is the only technical difficulty introduced in our framework, as relevant physical quantities are then computed using formal prescriptions analogous to those adopted in Quantum Mechanics. Nevertheless, this is already enough to introduce, for the first time, the notion of quantum probability. Typically, the geometric configurations of experimental setups are given by superpositions of probability eigenstates, hence the probability distribution for a given measurement is non-classical, \emph{i.e.} it has a non-vanishing variance. This opens up the possibility of defining non-classical measurement apparata in a formal way, but also would require a rethinking of the operational meaning of probability itself.

Our work shares some features with \cite{Mikusch:2021kro}, in which quantum reference frames for spin systems are introduced. The main analogy lies in the fact that both frameworks allow for non-classical notions of directions in space and rotation angles, even though the origin and the nature of these two types of fuzziness are different in the two contexts. Moreover, in \cite{Mikusch:2021kro} the physical regime of ``unlimited resources" is considered for measuring orientations, thus allowing the definition of sharp classical directions. This is done by considering spin coherent states with large spin to define axes in space. In our framework, directions of physical systems (spins and Stern-Gerlach apparata), as well as the relative orientation between different observers, are specified by quantum states in Hilbert spaces that do not describe any physical system, rather they characterize the intrinsic (quantum) structure of space. These states are such that, in general, classical directions cannot be defined. It would be interesting to investigate the possible connections between our work and the ``limited resources" regime of the framework presented in \cite{Mikusch:2021kro}, as well as \cite{Poulin:2006ryq} in which the problem of limited resources for quantum reference frames for spin measurement is investigated. This would offer fertile ground for investigating the possible connections \cite{Girelli:2007xn} between the quantum reference frames \cite{Aharonov1984, Giacomini:2017zju, Vanrietvelde:2018pgb, Giacomini:2021gei} and the quantum groups approaches to quantum spacetime. This topic has attracted growing interest in recent years and preliminary studies are focusing on trying to find a (quantum) group structure for quantum reference frames transformations \cite{Toller:1996ki, Ballesteros:2020lgl, Ballesteros:2025ypr}.

The formalism developed in this paper can be regarded as a template for incorporating quantum spacetime effects in Quantum Mechanics in a consistent way, enforcing the relativistic invariance of the theory and starting from first principles. Interestingly, one of the axioms we have formulated attributes quantum properties to macroscopic measuring devices. While they are still ``classical" in the sense of standard Quantum Mechanics, they are described by geometry states in a Hilbert space in the sense of quantum spacetime. This suggests the possibility of realizing ``non-classical measurement devices", intended as measurement apparata that are characterized by fully quantum geometrical configurations, such as superpositions of semi-classical geometry states. We did not consider such possibilities in our work, postponing this task to future works to investigate the consequences of this property that, as of current knowledge, is exclusive to the DQM framework. The latter could also be generalized to accommodate other quantum symmetries and to characterize observables for quantum systems in a quantum spacetime invariant under such symmetries. Some preliminary works have studied geometry states of some non-commutative spacetime models and their relative quantum group transformations \cite{Lizzi:2018qaf,Lizzi:2019wto,Lizzi:2022hcq} and it would be interesting to incorporate those results in a DQM model, in order to also explore the boost and translation sector and define a DQM with general spacetime (quantum) symmetries. It could even be conceivable to implement a general formalism of DQM valid for any quantum group, analogously to \cite{delaHamette:2020dyi} in which a framework for general symmetry groups has been developed formally for quantum reference frames. A DQM framework for general quantum symmetries of spacetime would also allow to formalize dynamical models compatible with the quantum symmetries \cite{Amelino-Camelia:2023rkg}.

Finally, the DQM framework allows for a deeper discussion about the fundamental principles underlying the formulation of a Quantum Gravity theory, particularly concerning the nature of probabilities that acquire quantum features in DQM. Effectively, it is conceivable to try to encompass such a prediction in the broader area of generalized probabilistic theories \cite{Muller_2021}, possibly signaling that the DQM framework describes a model beyond Quantum Theory. To investigate this latter possibility, the DQM for spin measurements that we developed in our work could be already sufficient. Indeed, it might be possible to implement a doubly quantum version of the CHSH game \cite{CHSH} to test if the Tsirelson's bound \cite{Cirelson1980QuantumGO}, which sets the maximum violation of the CHSH inequality in Quantum Theory, is violated. This would lead to the violation of some fundamental principles of Quantum Theory, such as Information Causality \cite{Pawlowski_2009} (a generalization of the no-signaling principle), and provide an insight for going beyond them, in laying down the foundations of a Quantum Gravity theory.

All of the future perspectives presented above offer an opportunity for enriching both phenomenological and theoretical aspects of studies in Quantum Gravity, a much needed effort in light of the recent phenomenological opportunities apt to test \cite{Kovachy:2015xcp, Rosi:2017ieh, Marletto:2017kzi, Bose:2017nin, Zych:2017tau, Goswami:2018rda, Christodoulou:2018cmk, Westphal:2020okx, Cepollaro:2021ccc, Christodoulou:2022mkf, Overstreet:2022zgq} the quantum nature of spacetime and probe the limits of the assumptions underlying General Relativity and Quantum Mechanics at the interface between them.

\section*{Acknowledgements}
The authors would like to acknowledge the contribution of the COST Action CA23130 ``Bridging high and low energies in search of quantum gravity (BridgeQG)''.

F.M. acknowledges support by the Agencia Estatal de Investigación (Spain) under grants {\fontsize{9pt}{9pt}\selectfont CNS2023-143760} and {\fontsize{9pt}{9pt}\selectfont PID2023-148373NB-I00} funded by {\fontsize{9pt}{9pt}\selectfont MCIN/AEI/10.13039/501100011033/FEDER – UE}, and by the Q-CAYLE Project funded by the Regional Government of Castilla y León (Junta de Castilla y León) and by the Ministry of Science and Innovation MICIN through NextGenerationEU (PRTR C17.I1). V.D.E. acknowledges financial support from the Pauli Center for Theoretical Studies and hospitality from the Institute for Theoretical Physics of ETH Z{\"u}rich. G.F. and D.F. thank the Universidad de Burgos for their hospitality. G.F. acknowledges financial support from ``Fondazione Angelo Della Riccia'' and from the ``Foundation Blanceflor''. G.F. thanks UC Berkeley and Lawrence Berkeley National Lab for hospitality during the final stages of this project.

\newpage

\appendix

\section{Eigenstates of the probability operator}\renewcommand{\theequation}{\thesection.\arabic{equation}}\setcounter{equation}{0}
\label{app:probabilityeigenstates}
By explicit computations involving the probability operator \eqref{eq:probabilityquantumdefinition}, we see that the states $\big\{\ket{\chi_s}\otimes\ket{\chi_a},\ket{\chi_s}\otimes\ket{n_a,\phi_a,\chi_a},\ket{n_s,\phi_s,\chi_s}\otimes\ket{\chi_a}\big\}$ are eigenstates of $P(\uparrow_\sigma)$ with eigenvalues given by$\{1,1-q^{2n_a+2},1-q^{2n_s}\}$, respectively. Recalling that $n_a,n_s\in[0,\infty)$, these probability eigenvalues are discrete and take values in $[0,1]$. These states span the $\mathcal{H}_{\rho}\otimes\mathcal{H}_{\rho}$, $\mathcal{H}_{\rho}\otimes\mathcal{H}_{\pi}$, and $\mathcal{H}_{\pi}\otimes\mathcal{H}_{\rho}$ components of the Hilbert space $\mathcal{H}_{SU_q(2)}\otimes\mathcal{H}_{SU_q(2)}$, given that they are basis states. By means of \eqref{eq:probability eigenstates Alice}, we show that there is only one other class of non-entangled eigenstates in $\mathcal{H}_\pi\otimes\mathcal{H}_\pi$. To show this, consider a generic factorizable probability eigenstate independent from the ones written above, given by
\begin{equation}
    \label{eq:genericeigenstate}\ket{p,r}=\sum_{n_s,n_a=0}^\infty h_{n_s} k_{n_a} \ket{n_s,\phi_s,\chi_s}\otimes \ket{n_a,\phi_a,\chi_a}
\end{equation}
The eigenstate equation $P(\uparrow_\sigma)\ket{p,r}=p\ket{p,r}$ yields

\begin{equation}\label{eq:general equation eigenstates prob}
    \begin{aligned}
        &h_{n_s}k_{n_a}\qty[(1-q^{2n_s})(1-q^{2n_a+2})+q^{2(n_s+n_a)}-p]+\\ &+h_{n_s-1}k_{n_a+1}e^{i(\omega_s-\omega_a)}q^{n_s+n_a}\sqrt{(1-q^{2n_s})(1-q^{2n_a+2})}+\\
       &+h_{n_s+1}k_{n_a-1}e^{-i(\omega_s-\omega_a)}q^{n_s+n_a}\sqrt{(1-q^{2n_a})(1-q^{2n_s+2})}= 0  \;,
    \end{aligned}
\end{equation}
where $\omega_{s,a} = \phi_{s,a}-\chi_{s,a}$. 

Setting $n_s=0$ and $n_a = 0$ in \eqref{eq:general equation eigenstates prob}, we obtain

\begin{equation}\label{eq:general equation eigenstates prob ns=0}
    h_0k_{n_a}(q^{2n_a}-p)+h_1k_{n_a-1}e^{-i(\omega_s-\omega_a)}q^{n_a}\sqrt{(1-q^{2n_a})(1-q^{2})}=0\;,
\end{equation}
\begin{equation}\label{eq:general equation eigenstates prob na=0}
    h_{n_s}k_{0}\qty(1+q^{2n_s+2}-q^2-p)+h_{n_s-1}k_{0}e^{i(\omega_s-\omega_a)}q^{n_s}\sqrt{(1-q^{2n_s})(1-q^{2})}=0\;.
\end{equation}\\
Inspecting \eqref{eq:general equation eigenstates prob ns=0}, consider the case in which $h_0=0$, so that  either $h_1=0$ or $k_{n_a}=0\;\;\forall\,n_a$. In the latter case, the state \eqref{eq:genericeigenstate} is a null vector. In the former, we consider \eqref{eq:general equation eigenstates prob} for $n_s=1$, which yields $h_2k_{n_a-1}=0$. Therefore, either $h_2=0$ or $k_{n_a}=0 \;\;\forall\, n_a$. Again, in the latter case, \eqref{eq:genericeigenstate} is the null vector, while in the former we move on to  consider \eqref{eq:general equation eigenstates prob} for $n_s = 3$. This argument repeats in the same way for all the other values of $n_s$, ultimately yielding that either $h_{n_s} = 0\;\;\forall\, n_s$ or $k_{n_a} = 0\;\;\forall\, n_a$. Therefore if $h_0 = 0$ the vector in \eqref{eq:genericeigenstate} is the null vector and an analogous reasoning applies also for the case in which $k_0=0$. This means that if a separable state of the form \eqref{eq:genericeigenstate} exist, necessarily $h_0 \neq 0$ and $k_0 \neq 0$. 
\begin{figure}
    \centering
    \includegraphics[width=0.8\textwidth]{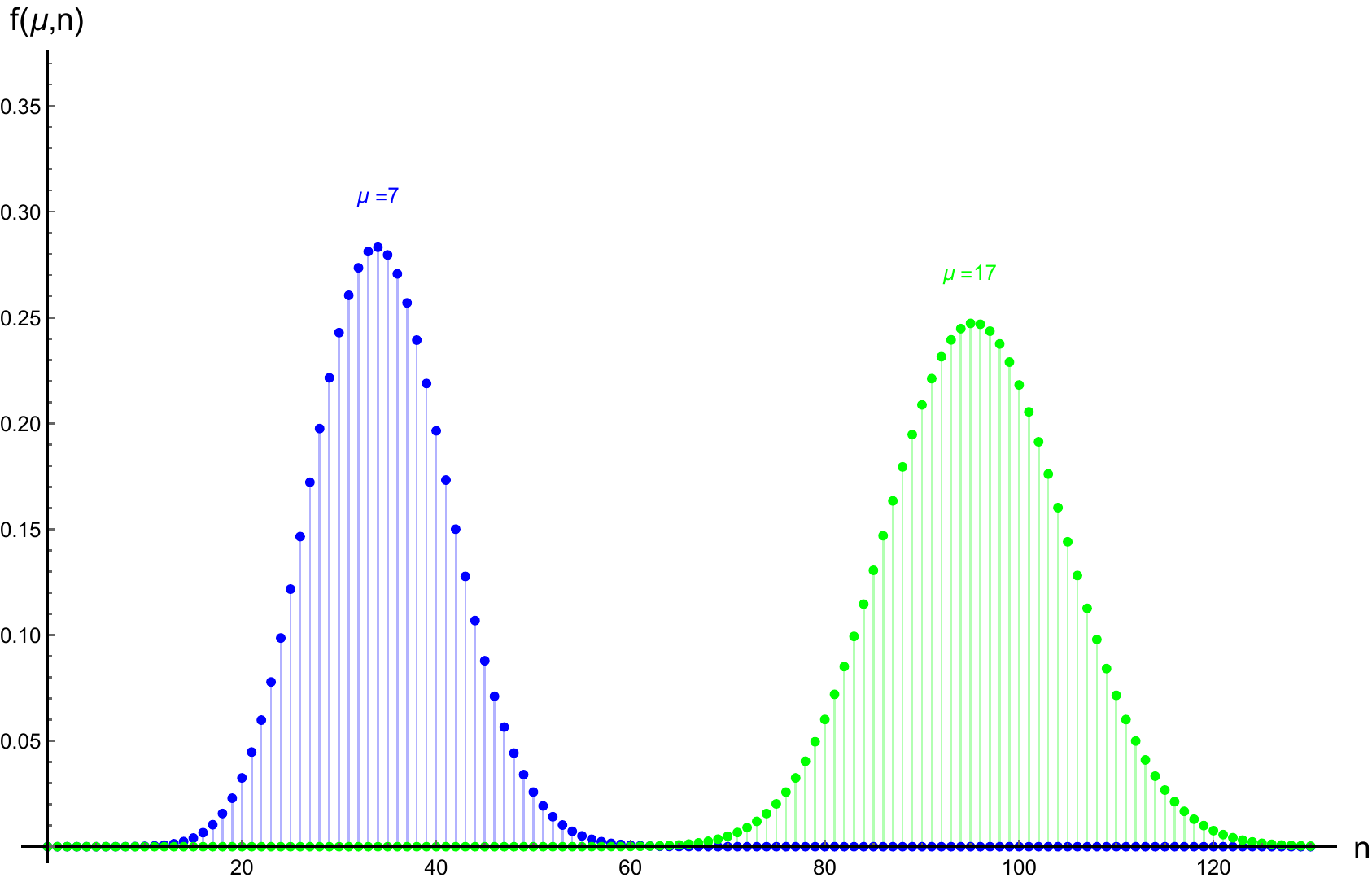}
    \caption{Distribution $f(\mu,n)$ with $n \in [0,130] \cap \mathbb{N}_0$ for two different values of $\mu$, $\mu=7$ (blue plot) and $\mu=17$ (green plot), corresponding to distributions centred in $n=34$ and $n=96$ (or $\theta(n) \approx \frac{\pi}{2}$ and $\theta(n) \approx \frac{\pi}{4}$), respectively, for $q=0.99$. These distributions are well approximated by the Gaussians that we used in \cite{Amelino-Camelia:2022dsj}. Notice also that the variance increases as the value of $n$ on which they are peaked increases, in the same way as the Gaussian states in \cite{Amelino-Camelia:2022dsj} do. Therefore, these states, derived by searching for the semi-classical eigenstates of the probability operator, formally justify the numerical construction of Gaussian states that we proposed in \cite{Amelino-Camelia:2022dsj}, since the characteristic behaviour of the physical quantities we consider there is not altered by the small differences between the states that we found here and the Gaussian states.}
    \label{fig: distributions coefficients appendix}
\end{figure}

Among the eigenstates with $h_0,k_0\neq 0$, let us first focus on the cases in which $p\neq 1$. If we set $n_a = 0$ in \eqref{eq:general equation eigenstates prob ns=0} we get $h_0 k_{n_a} = 0$, since $p\neq 1$. Therefore, since $h_0 \neq 0$, we have $k_{n_a} = 0\;\;\forall\, n_a$, in which case the state in \eqref{eq:genericeigenstate}  is the null vector. Let us then move on to the case in which $h_0,k_0\neq 0$ and $p=1$. Equations \eqref{eq:general equation eigenstates prob ns=0} and \eqref{eq:general equation eigenstates prob na=0} are solved by 
\begin{equation}
\label{eq:solutionsp=1}
    \begin{aligned}
        &h_{n_s}=N_sf(\mu_s,n_s)=N_s\frac{q^{\frac{n_s}{2}(n_s-1)}\sqrt{(1-q^2)^{n_s-1}}}{\sqrt{(q^4;q^2)_{n_s-1}}}\,\mu_s^{n_s} \; ,\\
        &k_{n_a}=N_aq^{n_a}f(\mu_a,n_a)=N_a\frac{q^{\frac{n_a}{2}(n_a+1)}\sqrt{(1-q^2)^{n_a-1}}}{\sqrt{(q^4;q^2)_{n_a-1}}}\,\mu_a^{n_a} \;,\\
        &\mu_s = e^{i \qty(\omega_s-\omega_a)}\,\mu_a\;,
    \end{aligned}
\end{equation}
where $(a;q)_n$ is the q-Pochhammer symbol defined as $(a;q)_n\ebd \prod_{k=0}^{n-1}(1-aq^k)\,,\;n>0$, $\qty(a;q)_0 = 1$, $\mu_s \ebd h_1h_0^{-1}$ ($\mu_a \ebd k_1k_0^{-1}$) determines the value of $n_s$ ($n_a$) on which the distribution of $\abs{f(\mu_s,n_s)}^2 (\abs{f(\mu_a,n_a)}^2)$ is centered as shown in \Cref{fig: distributions coefficients appendix}, and $N_s = h_0$ ($N_a=k_0$) is a normalization constant. We will denote with $\ket{\mu_s,\omega_s}$ and $\ket{\mu_a,\omega_a}$ states of the form 
\begin{equation}\label{eq:semi-classical states qpochhammer}
    \begin{aligned}
        &\ket{\mu_s,\omega_s}_s \ebd \sum_{n_s=0}^\infty h_{n_s}\ket{n_s,\phi_s,\chi_s}\;,\\
        &\ket{\mu_a,\omega_a}_a \ebd \sum_{n_a=0}^\infty k_{n_a}\ket{n_a,\phi_a,\chi_a}\;.    \end{aligned}
\end{equation}
In the case where $\mu_s$ and $\mu_a$ are complex, $\mu_s = \abs{\mu_s}e^{i\beta_s}$, $\mu_a = \abs{\mu_a}e^{i\beta_a}$, it is easy to show that the additional phase can be absorbed into the angles $\omega_s$ and $\omega_a$, meaning that the computation of all quantities of interest (expectation values in the geometry state of spin states, Pauli matrices and probabilities) gives the same result when performed with either $\ket{\abs{\mu_s}, \omega_s-\beta_s}\otimes \ket{\abs{\mu_a}, \omega_a-\beta_a}$ or $\ket{\abs{\mu_s}e^{i\beta_s}, \omega_s}\otimes \ket{\abs{\mu_a}e^{i\beta_a}, \omega_a}$. Given the generality of the angles $\omega_s$ and $\omega_a$, we consider $\mu_s,\mu_a\in \mathbbm{R}^+$. Therefore, the probability eigenstates in \eqref{eq:genericeigenstate} with eigenvalue $p=1$ can be written as $\ket{\mu,\omega}_s\otimes\ket{\mu,\omega}_a\,$, and describe spin systems and Stern-Gerlach apprata aligned along the same semi-classical directions.

\section{Numerical examples with semi-classical states}\label{appendix: semi-classical states}\renewcommand{\theequation}{\thesection.\arabic{equation}}\setcounter{equation}{0}
We consider some examples involving explicit computations of average values in the geometry states of the generic spin state \eqref{eq:genericstate} and of the generic Pauli matrix \eqref{eq:pauligeneric}, as well as average values and uncertainties in the geometry states of the probability operator $P(\uparrow_\sigma)$ in \eqref{probabilities Alice}.
These examples will give an indication on which geometry states may be regarded as those semi-classically describing the directions of spin systems and Stern-Gerlach apparata, in the sense of conditions \eqref{eq:semi-classical factors}, \eqref{eq:semi-classical probabilities}. 
The results concerning the average value in the geometry of the generic spin states and of the generic Pauli matrices should not be regarded as the starting points for the computation of quantum mechanical predictions for a spin system. We remind the reader that the physical predictions of our model are mainly encoded in the measurement outcomes of the probability $P(\uparrow_\sigma)$ (and consequently of $\expval{\sigma^q}$), according to the axioms outlined in \Cref{sec:spin measurement doubly quantum}. Let us begin by exploring geometry states inspired by the probability eigenstates derived in \Cref{app:probabilityeigenstates}. Once we have obtained a class of semi-classical geometry states, we will conclude with an example showing that these are also suitable for semi-classically describing the relative orientation between two observers, satisfying the requirements outlined at the end of \Cref{sec:effective theory alice}. 
Some of the following examples and computations in \Cref{sec:numerical example protocol} involve states of the form $\eqref{eq:semi-classical states qpochhammer}$ that require a numerical analysis, where we set $q=0.99$ for definiteness. For practical purposes, we truncate the infinite series originating from the computations involving these states so that we have control up to the second decimal place.

\begin{enumerate}
    \item Consider the state $\ket{\Phi^{S}}\otimes\ket{\Phi^{SG}} =\ket{\chi_s}\otimes\ket{\chi_a}\in\mathcal{H}_\rho\otimes\mathcal{H}_\rho$, which is an eigenstate of $P(\uparrow_\sigma)$ with eigenvalue $1$. The corresponding geometry states for the spin state and the Stern-Gerlach apparatus are $\ket{\Phi^{S}} = \ket{\chi_s}$ and $\ket{\Phi^{SG}} = \ket{\chi_a}$, respectively. The average values in these states of the generic spin state \eqref{eq:genericstate} and of the generic Pauli matrix \eqref{eq:pauligeneric} are given by
    \begin{equation}
    \label{eq:esempiozpositivo}
        \overline{\ket{\psi}} =\ket{\uparrow } \; , \;\;\overline{\Delta}\big[\ket{\psi}\big]=0\; ,  \;\; \overline{\sigma^q}=\begin{pmatrix}
        q & 0 \\
        0 & -q^{-1}
        \end{pmatrix}  \;,\;\; \overline{\Delta}\big[\sigma^q\big]=0_{2\cross 2}\;.
    \end{equation}
    This case semi-classically describes a spin and a Stern-Gerlach both oriented along the positive $z$ direction, in the sense of requirements \eqref{eq:semi-classical factors} and \eqref{eq:semi-classical probabilities}. We thus make the identification $\ket{\Phi^S(0,\omega_s)}\equiv \ket{\chi_s=\omega_s}$, $\ket{\Phi^{SG}(0,\omega_a)}\equiv \ket{\chi_a=\omega_a}$.

    \item Consider now $\ket{\Phi^{S}}\otimes\ket{\Phi^{SG}} = \ket{\chi_s}\otimes\ket{n_a,\phi_a,\chi_a}$ in $\mathcal{H}_\rho\otimes\mathcal{H}_\pi$, eigenstate of the probability with eigenvalue $1-q^{2n_a
+2}$. It may be explicitly verified that for $n_a\geq 1$, $\ket{n_a,\phi_a,\chi_a}$ does not satisfy requirements \eqref{eq:semi-classical factors} and \eqref{eq:semi-classical probabilities}. When $n_a=0$, we obtain
    \begin{equation}
         \overline{\ket{\psi}} =\ket{\uparrow } \; , \;\;\overline{\Delta}\big[\ket{\psi}\big]=0\; ,  \;\; \overline{\sigma^q}=\begin{pmatrix}
        -q^3 & 0 \\
        0 & q
        \end{pmatrix}  \; ,\;\; \overline{\Delta}\big[\sigma^q\big]=\begin{pmatrix}
        0 & 0  \\
        (1+q^2)^2(1-q^2) & 0
        \end{pmatrix}\;. 
    \end{equation}
     The state $\ket{0,\phi_a,\chi_a}$ semi-classically describes a Stern-Gerlach apparatus along the negative $z$ direction, so we make the identification $\ket{\Phi^{SG}(\pi,\omega_a)}\equiv \ket{0,\phi_a,\chi_a}$.

     Analogously, we may consider $\ket{\Phi^{S}}\otimes\ket{\Phi^{SG}} = \ket{n_s,\phi_s,\chi_s}\otimes\ket{\chi_s}$, eigenstate of the probability with eigenvalue $1-q^{2n_s}$. Similarly to the previous case, it may be explicitly verified that for $n_s\geq 1$, $\ket{n_s,\phi_s,\chi_s}$ does not satisfy requirements \eqref{eq:semi-classical factors} and \eqref{eq:semi-classical probabilities}. When $n_s=0$, we obtain
    \begin{equation}
         \overline{\ket{\psi}} =\ket{\downarrow } \; , \;\;\overline{\Delta}\big[\ket{\psi}\big]=0\; ,  \;\; \overline{\sigma^q}=\begin{pmatrix}
        q & 0 \\
        0 & -q^{-1}
        \end{pmatrix}  \; ,\;\; \overline{\Delta}\big[\sigma^q\big]=0_{2\times 2} \; .
    \end{equation}
    The state $\ket{0,\phi_s,\chi_s}$ thus semi-classically describes a spin along the negative $z$ direction, and we make the identification $\ket{\Phi^S(\pi,\omega_s)}\equiv \ket{0,\phi_s,\chi_s}$.
     \item We now consider states of the form $\ket{\Phi^{S}}\otimes\ket{\Phi^{SG}} =\ket{\mu_s,\omega_s}_s\otimes\ket{\mu_a,\omega_a}_a$, which are eigenstates of the probability in $\mathcal{H}_\pi\otimes\mathcal{H}_\pi$ with eigenvalue $1$. To exhibit a specific example involving these states, we resort to numerical computations by setting $q=0.99$, for definiteness. 
     We choose $\mu=7$, so that the $\abs{f(\mu,n)}^2$ function is peaked around $n_{\frac{\pi}{2}}=34$, corresponding to $\theta\qty(n_{\frac{\pi}{2}})\approx\frac{\pi}{2}$, in the spirit of \Cref{sec:effective theory alice}. The average values and uncertainties in the geometry states for the generic spin state and generic Pauli matrix read
     \begin{equation}
    \label{eq:esempio x}
        \begin{gathered}
         \overline{\ket{\psi}} =0.71\ket{\uparrow } + 0.71 e^{i\omega_s}\down\; , \;\;\overline{\Delta}\big[\ket{\psi}\big]=0.04\qty(\up+\down)\; ,  \;\; \\
        \overline{\sigma^q}=\begin{pmatrix}
        0.00 & 0.99 e^{-i\omega_a} \\
        0.99 e^{i\omega_a} & 0.00
        \end{pmatrix}  \;,\;\; \overline{\Delta}\big[\sigma^q\big]=\begin{pmatrix}
        0.10 & 0.10 \\
        0.10 & 0.10
        \end{pmatrix}\;.
        \end{gathered}
    \end{equation}
    corresponding to acceptable values for the states $\ket{7,\omega_s}_s$ and $\ket{7,\omega_s}_a$ to semi-classically describe directions that differ from those in the $x-y$ plane by quantities of order $\mathcal{O}(1-q)$, both for the spin system and the Stern-Gerlach apparatus, respectively. We therefore make the identification $\ket{\Phi^S\qty(\frac{1.006\pi}{2},\omega_s)}\equiv \ket{7,\omega_s}_s$, $\ket{\Phi^{SG}\qty(\frac{1.006\pi}{2},\omega_a)}\equiv \ket{7,\omega_a}_a\,$. In particular, by setting $\omega=0,\pi/2$, we obtain a semi-classical description of the axes closest to the $x,y$ axes, respectively, which will be employed in the deformed alignment protocol in \Cref{sec:numerical example protocol}.
    
    By varying the value of $\mu$, so that $\abs{f(\mu_{s,a},n_{s,a})}^2$  is peaked around a certain $n_\theta$, it can be shown that the state $\ket{\mu_{s,a},\omega_{s,a}}_{s,a}$ is the semi-classical counterpart of other directions corresponding to $\qty(\theta\qty(n_\theta),\omega)$.
\end{enumerate}
So far, we have focused on eigenstates of the probability. With similar computations, one can show that by combining geometry states representative of semi-classical directions of spin systems and Stern-Gerlach apparata defined in the three examples above, it is possible to construct states in $\mathcal{H}_{SU_q(2)}\otimes\mathcal{H}_{SU_q(2)}$ which are not necessarily probability eigenstates but still satisfy the semi-classical conditions \eqref{eq:semi-classical factors}, \eqref{eq:semi-classical probabilities}.
We close the discussion concerning a single observer with such an example.

\begin{enumerate}
    \setItemnumber{4}
    \item We consider $\ket{\Phi^S} = \ket{\chi_s}$ and $\ket{\Phi^{SG}} = \ket{7,0}_a$, 
    describing a semi-classical scenario in which the spin is oriented along the $z$ axis and the Stern-Gerlach apparatus is oriented along the $x$ axis, as indicated by the computations in \eqref{eq:esempiozpositivo} and \eqref{eq:esempio x}.
    The probability distribution function will therefore be characterized by 
    \begin{equation}
        \overline{P(\uparrow_\sigma)}=0.50 \; , \; \; \overline{\Delta}\big[P(\uparrow_\sigma)\big]= 0.05\;.
    \end{equation}
    As expected, the variance assumes a non-zero value, given that this particular geometry state is a superposition of probability eigenstates.
    
\end{enumerate}

We have provided some examples concerning the physical quantities of interest for a single observer and this allowed us to identify some states as suitable semi-classical descriptions of directions in space for both spin states and the Stern-Gerlach apparata. We now discuss how the same states can semi-classically describe relative orientations in space between two observers, approximating classical rotations  parameterized as 
\begin{equation}\label{eq:paramrotation}
    R_z(\alpha) R_x(\theta) R_z(\gamma)\;,\;\;\chi_g = \frac{\alpha+\gamma}{2}\;,\;\; \phi_g = \frac{3}{2}\pi - \frac{\alpha-\gamma}{2}\;.
\end{equation}
Taking into account the generalized requirements at the end of \Cref{sec:effective theory alice} and \eqref{eq:paramrotation}, one can show that states $\ket{\chi_g}$ semi-classically describe rotations around the $z$ axis of an angle $2\chi_g$, so we make the identification $\ket{\Phi^{RO}(0,0,\chi_g)}\equiv\ket{\chi_g}$. On the other hand, states of the form
\begin{equation}
    \ket{\mu_g,\phi_g,\chi_g}_g \ebd \sum_{n_g=0}^\infty h_{n_g}\ket{n_g,\phi_g,\chi_g}\;,
\end{equation}
semi-classically describe rotations \eqref{eq:paramrotation} with $\theta \,\in \;]0,\pi]$, where the value of $\mu_g$ is determined by $\theta$ according to \Cref{table semi-classical states}. We therefore make the identification $\ket{\Phi^{RO}(\theta_g=\theta\qty(n_\theta),\phi_g,\chi_g)} \equiv \ket{\mu_g,\phi_g,\chi_g}_g$, where $n_\theta$ has the same meaning as in the final part of example 3. In particular, the state $\ket{\mu_g=0,\phi_g,\chi_g}_g$ coincides with $\ket{n_g=0, \phi_g, \chi_g}$, and it can be shown that $\ket{0,\frac{3\pi}{2},0}$ semi-classically describes a rotation around the $x$ axis of an angle $\pi$. We conclude this appendix with an example involving a generic rotation around the $x$ axis.

\begin{enumerate}
    \setItemnumber{5}
    \item Consider $\ket{\Phi^S} = \ket{7,\omega_s}_{s}$, $\ket{\Phi^{SG}} = \ket{\chi_a}$ and $\ket{\Phi^{RO}} = \ket{7,\phi_g,\chi_g}_g$ (as in examples 3 and 4 we consider $q=0.99$ for definiteness). The average values and uncertainties in the geometry state for the generic spin state and the generic Pauli matrix read   
    \begin{equation}
  \begin{gathered}
  \overline{\ket{\psi}} =0.50\qty(1-e^{i \qty(\phi_g -\chi_g-\omega_s)})\ket{\uparrow } + e^{i\qty(\phi_g-\chi_g)}0.50\qty(1+ e^{-i \qty(\phi_g +\chi_g-\omega_s)})\down\; , \\\overline{\Delta}\big[\ket{\psi}\big]=0.1\abs{\sin\qty(\frac{\phi_g+\chi_g-\omega_s}{2})}\qty(\up+\down)\; , \\\overline{\sigma^q}=\begin{pmatrix}
        0.00 & 0.99 e^{-i\qty(\phi_g-\chi_g)} \\
        0.99 e^{i\qty(\phi_g-\chi_g)} & 0.00
        \end{pmatrix}  \;,\;\; \overline{\Delta}\big[\sigma^q\big]=\begin{pmatrix}
        0.10 & 0.10 \\
        0.10 & 0.10
        \end{pmatrix}
        \end{gathered}
\end{equation}
In the above, if we set $\phi_g=\frac{3}{2}\pi,\chi_g=0$, we conclude that $\ket{\Phi^{RO}}=\ket{7,\frac{3}{2}\pi,0}_g$ is a suitable state to semi-classically describe a counterclockwise rotation of $\theta(34)=1.006\frac{\pi}{2}$ around the $x$ axis. This is particularly evident when also setting $\omega_s=0$. Thus, we make the identification $\ket{7,\frac{3}{2}\pi,0}_g\equiv \ket{\Phi^{RO}(\frac{1.006\pi}{2},\frac{3}{2}\pi,0)}$. In this case, the probability is distributed according to
\begin{equation}
    \overline{P_A\qty(\uparrow_\sigma^B)}=0.51 \; , \; \; \overline{\Delta}\big[P_A\qty(\uparrow_\sigma^B)\big]=0.04 \; .
\end{equation}
The relative orientation state discussed in this example is also employed in the protocol described in \Cref{sec:numerical example protocol}, where in one of the examples Alice and Bob are connected by a rotation of $1.006\frac{\pi}{2}$ around the $x$ axis.
\end{enumerate}

\section{Further discussions on the joint probability distribution}\renewcommand{\theequation}{\thesection.\arabic{equation}}\setcounter{equation}{0}

In this appendix, we provide two (simple) proofs. First, we derive \eqref{eq:expval outcomes e uncertainty deformed}. Then, we show that the uncertainties written in \eqref{eq:matelements} are those appearing in \eqref{eq:fundamental uncertainty theory}. Finally, in the last part, we also discuss some interpretational aspects of the joint probability distribution \eqref{eq:deformed binomial}.

\subsection{Expectation value and variance of joint probability density}\label{appendix: joint probability distribution}

We want to compute the mean value and variance of a joint probability density given by
\begin{equation}
    \mathcal{P}(m,N,p) = \binom{N}{m} p^m (1-p)^{N-m}F(p)\;,
\end{equation}
where $F(p)$ is the probability distribution of $p$ with mean value and variance given by
\begin{equation}\label{eq:expval p and variance f}
    p_0 = \int_{0}^{1}\dd{p}\, p F(p)\;,\;\; \Delta^2_F = \int_{0}^{1}\dd{p}\, p^2 F(p)-p_0^2\;.
\end{equation}

Starting with $\mathbb{E}\qty[k]$, applying the law of total expectation, we have 
\begin{equation}
    \mathbb{E}\qty[k] = \int_{0}^1\dd{p} \sum_{m=0}^N m\,\mathcal{P}(m,N,p) = \int_{0}^1\dd{p}\sum_m \binom{N}{m} m\,p^m (1-p)^{N-m}F(p) = N\int_{0}^1\dd{p}\, p F(p) = Np_0\;,
\end{equation}
where we have used the definition of $p_0$ and

\begin{equation}
    \sum_m \binom{N}{m} m\,p^m (1-p)^{N-m} = Np\;.
\end{equation}
The proof for the variance takes similar steps. Again, by the law of total expectation we have
\begin{eqnarray}
    \Delta_k^2 &=& \mathbb{E}\qty[k^2]-\mathbb{E}\qty[k]^2 = \int_{0}^1\dd{p} \sum_{m=0}^N m^2\,\mathcal{P}(m,N,p) - N^2p_0^2 = \nonumber\\ &=& \int_{0}^1\dd{p}\sum_m \binom{N}{m} m^2\,p^m (1-p)^{N-m}F(p) -N^2p_0^2 \nonumber \\ 
    &=& N \int_{0}^1\dd{p}\, p(1-p) F(p) -N^2p_0^2 + N^2 \int_0^1\dd{p}\,p^2 F(p) =\nonumber \\
    &=& N^2 \Delta_F^2 + Np_0 - N \int_{0}^1\dd{p} p^2 F(p) = \nonumber\\
    &=& N(N-1)\Delta_F^2 + Np_0(1-p_0)\;,
\end{eqnarray}
where $p_0$ and $\Delta^2_F$ are the mean value and the variance of $F$ defined above and we have used

\begin{equation}
    \sum_m \binom{N}{m} m^2\,p^m (1-p)^{N-m} = Np(1-p)+n^2p^2\;.
\end{equation}

Notice that the probability operator $P_A\qty(\uparrow_\sigma^B)$ defined in \eqref{eq:sigma expectation value and probability bob} has a spectrum that has both a discrete and a continuous part, with degenerate eigenvalues. Therefore, the above integrals should actually be performed on the spectrum of $P_A\qty(\uparrow_\sigma^B)$, denoted as $\Lambda\qty(P)$. For this reason, we set the probability distribution $F(p)$ to be given by
\begin{equation}
    F(p) = \left\{
    \begin{array}{ll}
         f(p)\;,\;\; p \in \Lambda\qty(P) \\ \\ 0\;\;\;\;\;\;,\;\; p \not\in \Lambda\qty(P)
    \end{array}\right. \; ,
\end{equation}
where $f(p) = \sum_r \braket{\Phi}{p,r}\braket{p,r}{\Phi}$ is the probability distribution of the probability $P_A\qty(\uparrow_\sigma^B)$ in the geometry state $\ket{\Phi}$ and $r$ denotes the possible degeneracy of $p$. Of course, with this definition, the variance $\Delta_F^2$ is the same as the variance $\Delta_f^2$ used in the main text.

\subsection{Equivalence between the variances of the joint probability distributions and the variance of $\expval{\sigma_B^q}_A$}\label{appendix: variance protocol}

Starting from the variances in \eqref{eq:matelements}, we now provide the second proof. The variances in \eqref{eq:matelements} in a generic state of the geometry $\ket{\Phi}$ are explicitly given by

\begin{equation}
    \begin{aligned}
        \overline{\Delta}^{\,2}\big[\expval{\sigma^q_B}_A\big] &= (q+q^{-1})^2\mel{\Phi}{P_B^2\qty(\uparrow_\sigma^A)}{\Phi}+q^{-2}-2(1+q^{-2})\mel{\Phi}{P_B\qty(\uparrow_\sigma^A)}{\Phi} +\\ &- \Big[(q+q^{-1})\mel{\Phi}{P_B\qty(\uparrow_\sigma^A)}{\Phi}-q^{-1}\Big]^2\;.
    \end{aligned}
\end{equation}
From the definition of $f(p)$, $f(p) = \sum_{r} \braket{\Phi}{p,r}\braket{p,r}{\Phi}$, by decomposing the identity as

\begin{equation}
    \mathbbm{1} = \int_{\Lambda(P)}\dd{p}\sum_{r}\ketbra{p,r}\;,
\end{equation}
we have

\begin{equation}
    \mel{\Phi}{P_B^2\qty(\uparrow_\sigma^A)}{\Phi} = \int_{\Lambda(P)}\dd{p} p^2 f(p)\;,\;\;\mel{\Phi}{P_B\qty(\uparrow_\sigma^A)}{\Phi} = \int_{\Lambda(P)}\dd{p} p f(p)\;.
\end{equation}
Therefore, recalling the definitions of $\Delta_f$ and $p_0$ in \eqref{eq:expval p and variance f}, we get

\begin{equation}
     \overline{\Delta}^{\,2}\big[\expval{\sigma^q_B}_A\big] = \qty(q+q^{-1})^2\Delta_f^2 +q^{-2}-2\qty(1+q^{-2})p_0 -q^{-2} + 2\qty(1+q^{-2})p_0 = \qty(q+q^{-1})^2\Delta_f^2\;.
\end{equation}
Of course, the integral over $\Lambda(P)$ is intended to be an integral (discrete sum) over the continuous (discrete) part of the spectrum.

\subsection{The interpretation of the joint probability distribution}
\label{appendix: interpretation of joint probability}

Here, we want to delve into some interpretational points regarding the joint probability distribution in \eqref{eq:deformed binomial}. As extensively discussed in \Cref{sec:deformed protocol}, the quantity to be measured in the doubly quantum protocol is the expectation value in the spin state of the Pauli matrix operator, given by
\begin{equation}
    \expval{\sigma_B^q}_A = \qty(q+q^{-1})P_A\qty(\uparrow_\sigma^B) -q^{-1}\mathbbm{1}\;.
\end{equation}
Therefore, the measurement of this quantity effectively amounts to a measurement of the quantum probability observable $P_A\qty(\uparrow_\sigma^B)$. As discussed in \Cref{sec:effective theory alice}, in our framework, the effective result of actual measurements is described by the average value and the variance in the geometry states of the operators describing the observable that is being measured. As noted above, in the present case, the quantity measured in the alignment protocol is the probability of observing spin up in a spin measurement $P_A\qty(\uparrow_\sigma^B)$. Contrary to typical observables in standard QM, the measurement of the probability observable in \eqref{eq:sigma expectation value and probability bob} requires $N$ separate spin measurements, since it is itself defined by an expectation value in the spin states. This means that the experimental setup for the measurement of a single component of the rotation matrix in the alignment protocol is made of the Stern-Gerlach apparata used to prepare and measure the spins and the $N$ spins used to perform the measurements. At the end of the measurement procedure, a given value $p$, drawn from the spectrum of $P_A\qty(\uparrow_\sigma^B)$, will be observed, with a probability density $f(p)$ that depends on the geometry states. Therefore, the probability of observing $k$ spin up results out of $N$ measurements will be conditioned on the observation of the value $p$ as a result of the probability measurement. This conditional probability will be given by the binomial distribution $P(k,N) = \binom{N}{k}p^k(1-p)^{N-k}$, hence the probability density of observing $k$ spin up outcomes in $N$ measurements and observing $p$ as a result of the probability measurement is given, from the definition of conditional probability, by the product between this binomial and the distribution $f(p)$, which is exactly \eqref{eq:deformed binomial}. Notice also that the expectation value and the variance of this joint distribution coincide, in the large $N$ limit, with the expectation value and variance of $P_A\qty(\uparrow_\sigma^B)$ in the geometry states.


\newpage

\begin{thebibliography}{100}

\bibitem{Oriti:2009zz}
Daniele Oriti, editor.
\newblock ``Approaches to quantum gravity: Toward a new understanding of space,
  time and matter''.
\newblock \href{https://dx.doi.org/10.1017/CBO9780511575549}{Cambridge
  University Press}. ~(2009).

\bibitem{Gross:1987ar}
David~J. Gross and Paul~F. Mende.
\newblock ``{String Theory Beyond the Planck Scale}''.
\newblock \href{https://dx.doi.org/10.1016/0550-3213(88)90390-2}{Nucl. Phys. B
  {\bf 303}, 407--454}~(1988).

\bibitem{Amati:1987uf}
D.~Amati, M.~Ciafaloni, and G.~Veneziano.
\newblock ``{Classical and Quantum Gravity Effects from Planckian Energy
  Superstring Collisions}''.
\newblock \href{https://dx.doi.org/10.1142/S0217751X88000710}{Int. J. Mod.
  Phys. A {\bf 3}, 1615--1661}~(1988).

\bibitem{Adler:1999bu}
Ronald~J. Adler and David~I. Santiago.
\newblock ``{On gravity and the uncertainty principle}''.
\newblock \href{https://dx.doi.org/10.1142/S0217732399001462}{Mod. Phys. Lett.
  A {\bf 14}, 1371}~(1999).
\newblock  \href{http://arxiv.org/abs/gr-qc/9904026}{arXiv:gr-qc/9904026}.

\bibitem{Rovelli:1994ge}
Carlo Rovelli and Lee Smolin.
\newblock ``{Discreteness of area and volume in quantum gravity}''.
\newblock \href{https://dx.doi.org/10.1016/0550-3213(95)00150-Q}{Nucl. Phys. B
  {\bf 442}, 593--622}~(1995).
\newblock  \href{http://arxiv.org/abs/gr-qc/9411005}{arXiv:gr-qc/9411005}.

\bibitem{Ashtekar:1996eg}
Abhay Ashtekar and Jerzy Lewandowski.
\newblock ``{Quantum theory of geometry. 1: Area operators}''.
\newblock \href{https://dx.doi.org/10.1088/0264-9381/14/1A/006}{Class. Quant.
  Grav. {\bf 14}, A55--A82}~(1997).
\newblock  \href{http://arxiv.org/abs/gr-qc/9602046}{arXiv:gr-qc/9602046}.

\bibitem{Ashtekar:1997fb}
Abhay Ashtekar and Jerzy Lewandowski.
\newblock ``{Quantum theory of geometry. 2. Volume operators}''.
\newblock \href{https://dx.doi.org/10.4310/ATMP.1997.v1.n2.a8}{Adv. Theor.
  Math. Phys. {\bf 1}, 388--429}~(1998).
\newblock  \href{http://arxiv.org/abs/gr-qc/9711031}{arXiv:gr-qc/9711031}.

\bibitem{Freidel:2002hx}
Laurent Freidel, Etera~R. Livine, and Carlo Rovelli.
\newblock ``{Spectra of length and area in (2+1) Lorentzian loop quantum
  gravity}''.
\newblock \href{https://dx.doi.org/10.1088/0264-9381/20/8/304}{Class. Quant.
  Grav. {\bf 20}, 1463--1478}~(2003).
\newblock  \href{http://arxiv.org/abs/gr-qc/0212077}{arXiv:gr-qc/0212077}.

\bibitem{Amelino-Camelia:2008fcv}
Giovanni Amelino-Camelia, Giulia Gubitosi, and Flavio Mercati.
\newblock ``{Discreteness of area in noncommutative space}''.
\newblock \href{https://dx.doi.org/10.1016/j.physletb.2009.04.045}{Phys. Lett.
  B {\bf 676}, 180--183}~(2009).
\newblock  \href{http://arxiv.org/abs/0812.3663}{arXiv:0812.3663}.

\bibitem{Percacci:2010af}
Roberto Percacci and Gian~Paolo Vacca.
\newblock ``{Asymptotic Safety, Emergence and Minimal Length}''.
\newblock \href{https://dx.doi.org/10.1088/0264-9381/27/24/245026}{Class.
  Quant. Grav. {\bf 27}, 245026}~(2010).
\newblock  \href{http://arxiv.org/abs/1008.3621}{arXiv:1008.3621}.

\bibitem{Snyder:1947}
Hartland~S. Snyder.
\newblock ``Quantized space-time''.
\newblock \href{https://dx.doi.org/10.1103/PhysRev.71.38}{Phys. Rev. {\bf 71},
  38--41}~(1947).

\bibitem{Majid:2006xn}
Shahn Majid.
\newblock ``{Algebraic approach to quantum gravity. II: Noncommutative
  spacetime}''~(2006).
\newblock  \href{http://arxiv.org/abs/hep-th/0604130}{arXiv:hep-th/0604130}.

\bibitem{Szabo:2009tn}
Richard~J. Szabo.
\newblock ``{Quantum Gravity, Field Theory and Signatures of Noncommutative
  Spacetime}''.
\newblock \href{https://dx.doi.org/10.1007/s10714-009-0897-4}{Gen. Rel. Grav.
  {\bf 42}, 1--29}~(2010).
\newblock  \href{http://arxiv.org/abs/0906.2913}{arXiv:0906.2913}.

\bibitem{Pachoł_2013}
A~Pachoł.
\newblock ``Short review on noncommutative spacetimes''.
\newblock \href{https://dx.doi.org/10.1088/1742-6596/442/1/012039}{Journal of
  Physics: Conference Series {\bf 442}, 012039}~(2013).

\bibitem{majid_1995}
Shahn Majid.
\newblock ``Foundations of quantum group theory''.
\newblock \href{https://dx.doi.org/10.1017/CBO9780511613104}{Cambridge
  University Press}. ~(1995).

\bibitem{Amelino-Camelia:1997ieq}
G.~Amelino-Camelia, John~R. Ellis, N.~E. Mavromatos, Dimitri~V. Nanopoulos, and
  Subir Sarkar.
\newblock ``{Tests of quantum gravity from observations of gamma-ray bursts}''.
\newblock \href{https://dx.doi.org/10.1038/31647}{Nature {\bf 393},
  763--765}~(1998).
\newblock
  \href{http://arxiv.org/abs/astro-ph/9712103}{arXiv:astro-ph/9712103}.

\bibitem{Jacob:2006gn}
Uri Jacob and Tsvi Piran.
\newblock ``{Neutrinos from gamma-ray bursts as a tool to explore
  quantum-gravity-induced Lorentz violation}''.
\newblock \href{https://dx.doi.org/10.1038/nphys506}{Nature Phys. {\bf 3},
  87--90}~(2007).
\newblock  \href{http://arxiv.org/abs/hep-ph/0607145}{arXiv:hep-ph/0607145}.

\bibitem{Amelino-Camelia:2008aez}
Giovanni Amelino-Camelia.
\newblock ``{Quantum-Spacetime Phenomenology}''.
\newblock \href{https://dx.doi.org/10.12942/lrr-2013-5}{Living Rev. Rel. {\bf
  16}, 5}~(2013).
\newblock  \href{http://arxiv.org/abs/0806.0339}{arXiv:0806.0339}.

\bibitem{Addazi:2021xuf}
A.~Addazi et~al.
\newblock ``{Quantum gravity phenomenology at the dawn of the multi-messenger
  era\textemdash{}A review}''.
\newblock \href{https://dx.doi.org/10.1016/j.ppnp.2022.103948}{Prog. Part.
  Nucl. Phys. {\bf 125}, 103948}~(2022).
\newblock  \href{http://arxiv.org/abs/2111.05659}{arXiv:2111.05659}.

\bibitem{AlvesBatista:2023wqm}
R.~Alves~Batista et~al.
\newblock ``{White paper and roadmap for quantum gravity phenomenology in the
  multi-messenger era}''.
\newblock \href{https://dx.doi.org/10.1088/1361-6382/ad605a}{Class. Quant.
  Grav. {\bf 42}, 032001}~(2025).
\newblock  \href{http://arxiv.org/abs/2312.00409}{arXiv:2312.00409}.

\bibitem{Amelino-Camelia:2009wvc}
Giovanni Amelino-Camelia, Claus Lämmerzahl, Flavio Mercati, and Guglielmo~M.
  Tino.
\newblock ``{Constraining the Energy-Momentum Dispersion Relation with
  Planck-Scale Sensitivity Using Cold Atoms}''.
\newblock \href{https://dx.doi.org/10.1103/PhysRevLett.103.171302}{Phys. Rev.
  Lett. {\bf 103}, 171302}~(2009).
\newblock  \href{http://arxiv.org/abs/0911.1020}{arXiv:0911.1020}.

\bibitem{Stecker:2014oxa}
Floyd~W. Stecker, Sean~T. Scully, Stefano Liberati, and David Mattingly.
\newblock ``{Searching for Traces of Planck-Scale Physics with High Energy
  Neutrinos}''.
\newblock \href{https://dx.doi.org/10.1103/PhysRevD.91.045009}{Phys. Rev. D
  {\bf 91}, 045009}~(2015).
\newblock  \href{http://arxiv.org/abs/1411.5889}{arXiv:1411.5889}.

\bibitem{Amelino-Camelia:2016fuh}
Giovanni Amelino-Camelia, Leonardo Barcaroli, Giacomo D'Amico, Niccol\'o Loret,
  and Giacomo Rosati.
\newblock ``{IceCube and GRB neutrinos propagating in quantum spacetime}''.
\newblock \href{https://dx.doi.org/10.1016/j.physletb.2016.07.075}{Phys. Lett.
  B {\bf 761}, 318--325}~(2016).
\newblock  \href{http://arxiv.org/abs/1605.00496}{arXiv:1605.00496}.

\bibitem{Amelino-Camelia:2016ohi}
Giovanni Amelino-Camelia, Giacomo D'Amico, Giacomo Rosati, and Niccol\'o Loret.
\newblock ``{In-vacuo-dispersion features for GRB neutrinos and photons}''.
\newblock \href{https://dx.doi.org/10.1038/s41550-017-0139}{Nature Astron. {\bf
  1}, 0139}~(2017).
\newblock  \href{http://arxiv.org/abs/1612.02765}{arXiv:1612.02765}.

\bibitem{Huang:2018ham}
Yanqi Huang and Bo-Qiang Ma.
\newblock ``{Lorentz violation from gamma-ray burst neutrinos}''.
\newblock \href{https://dx.doi.org/10.1038/s42005-018-0061-0}{Communications
  Physics {\bf 1}, 62}~(2018).
\newblock  \href{http://arxiv.org/abs/1810.01652}{arXiv:1810.01652}.

\bibitem{Amelino-Camelia:2022pja}
Giovanni Amelino-Camelia, Maria~Grazia Di~Luca, Giulia Gubitosi, Giacomo
  Rosati, and Giacomo D'Amico.
\newblock ``{Could quantum gravity slow down neutrinos?}''.
\newblock \href{https://dx.doi.org/10.1038/s41550-023-01993-z}{Nature Astron.
  {\bf 7}, 996--1001}~(2023).
\newblock  \href{http://arxiv.org/abs/2209.13726}{arXiv:2209.13726}.

\bibitem{DEsposito:2023psn}
Vittorio D'Esposito and Giulia Gubitosi.
\newblock ``{Constraints on quantum spacetime-induced decoherence from neutrino
  oscillations}''.
\newblock \href{https://dx.doi.org/10.1103/PhysRevD.110.026004}{Phys. Rev. D
  {\bf 110}, 026004}~(2024).
\newblock  \href{http://arxiv.org/abs/2306.14778}{arXiv:2306.14778}.

\bibitem{Hardy:2005fq}
Lucien Hardy.
\newblock ``{Probability theories with dynamic causal structure: A New
  framework for quantum gravity}''~(2005).
\newblock  \href{http://arxiv.org/abs/gr-qc/0509120}{arXiv:gr-qc/0509120}.

\bibitem{Oreshkov:2011er}
Ognyan Oreshkov, Fabio Costa, and Caslav Brukner.
\newblock ``{Quantum correlations with no causal order}''.
\newblock \href{https://dx.doi.org/10.1038/ncomms2076}{Nature Commun. {\bf 3},
  1092}~(2012).
\newblock  \href{http://arxiv.org/abs/1105.4464}{arXiv:1105.4464}.

\bibitem{Zych:2015fka}
Magdalena Zych and \v{C}aslav Brukner.
\newblock ``{Quantum formulation of the Einstein Equivalence Principle}''.
\newblock \href{https://dx.doi.org/10.1038/s41567-018-0197-6}{Nature Phys. {\bf
  14}, 1027--1031}~(2018).
\newblock  \href{http://arxiv.org/abs/1502.00971}{arXiv:1502.00971}.

\bibitem{Belenchia:2018szb}
Alessio Belenchia, Robert~M. Wald, Flaminia Giacomini, Esteban Castro-Ruiz,
  \v{C}aslav Brukner, and Markus Aspelmeyer.
\newblock ``{Quantum Superposition of Massive Objects and the Quantization of
  Gravity}''.
\newblock \href{https://dx.doi.org/10.1103/PhysRevD.98.126009}{Phys. Rev. D
  {\bf 98}, 126009}~(2018).
\newblock  \href{http://arxiv.org/abs/1807.07015}{arXiv:1807.07015}.

\bibitem{Castro-Ruiz:2019nnl}
Esteban Castro-Ruiz, Flaminia Giacomini, Alessio Belenchia, and \v{C}aslav
  Brukner.
\newblock ``{Quantum clocks and the temporal localisability of events in the
  presence of gravitating quantum systems}''.
\newblock \href{https://dx.doi.org/10.1038/s41467-020-16013-1}{Nature Commun.
  {\bf 11}, 2672}~(2020).
\newblock  \href{http://arxiv.org/abs/1908.10165}{arXiv:1908.10165}.

\bibitem{Giacomini:2021gei}
Flaminia Giacomini.
\newblock ``{Spacetime Quantum Reference Frames and superpositions of proper
  times}''.
\newblock \href{https://dx.doi.org/10.22331/q-2021-07-22-508}{Quantum {\bf 5},
  508}~(2021).
\newblock  \href{http://arxiv.org/abs/2101.11628}{arXiv:2101.11628}.

\bibitem{Foo:2021exb}
Joshua Foo, Cemile~Senem Arabaci, Magdalena Zych, and Robert~B. Mann.
\newblock ``{Quantum Signatures of Black Hole Mass Superpositions}''.
\newblock \href{https://dx.doi.org/10.1103/PhysRevLett.129.181301}{Phys. Rev.
  Lett. {\bf 129}, 181301}~(2022).
\newblock  \href{http://arxiv.org/abs/2111.13315}{arXiv:2111.13315}.

\bibitem{Oppenheim:2022xjr}
Jonathan Oppenheim, Carlo Sparaciari, Barbara \v{S}oda, and Zachary
  Weller-Davies.
\newblock ``{Gravitationally induced decoherence vs space-time diffusion:
  testing the quantum nature of gravity}''.
\newblock \href{https://dx.doi.org/10.1038/s41467-023-43348-2}{Nature Commun.
  {\bf 14}, 7910}~(2023).
\newblock  \href{http://arxiv.org/abs/2203.01982}{arXiv:2203.01982}.

\bibitem{Foo:2022dnz}
Joshua Foo, Cemile~Senem Arabaci, Magdalena Zych, and Robert~B. Mann.
\newblock ``{Quantum superpositions of Minkowski spacetime}''.
\newblock \href{https://dx.doi.org/10.1103/PhysRevD.107.045014}{Phys. Rev. D
  {\bf 107}, 045014}~(2023).
\newblock  \href{http://arxiv.org/abs/2208.12083}{arXiv:2208.12083}.

\bibitem{Galley:2023byb}
Thomas~D. Galley, Flaminia Giacomini, and John~H. Selby.
\newblock ``{Any consistent coupling between classical gravity and quantum
  matter is fundamentally irreversible}''.
\newblock \href{https://dx.doi.org/10.22331/q-2023-10-16-1142}{Quantum {\bf 7},
  1142}~(2023).
\newblock  \href{http://arxiv.org/abs/2301.10261}{arXiv:2301.10261}.

\bibitem{Chen:2024xvm}
Lin-Qing Chen and Flaminia Giacomini.
\newblock ``{Quantum effects in gravity beyond the Newton potential from a
  delocalised quantum source}''~(2024).
\newblock  \href{http://arxiv.org/abs/2402.10288}{arXiv:2402.10288}.

\bibitem{delaHamette:2024xax}
Anne-Catherine de~la Hamette, Viktoria Kabel, and \v{C}aslav Brukner.
\newblock ``{What an event is not: unravelling the identity of events in
  quantum theory and gravity}''~(2024).
\newblock  \href{http://arxiv.org/abs/2404.00159}{arXiv:2404.00159}.

\bibitem{Goel:2024vtr}
Lakshay Goel, Everett~A. Patterson, Mar\'\i{}a~Rosa Preciado-Rivas, Mahdi
  Torabian, Robert~B. Mann, and Niayesh Afshordi.
\newblock ``{Accelerated detector in a superposed spacetime}''.
\newblock \href{https://dx.doi.org/10.1103/PhysRevD.111.025015}{Phys. Rev. D
  {\bf 111}, 025015}~(2025).
\newblock  \href{http://arxiv.org/abs/2409.06818}{arXiv:2409.06818}.

\bibitem{Kovachy:2015xcp}
T.~Kovachy, P.~Asenbaum, C.~Overstreet, C.~A. Donnelly, S.~M. Dickerson,
  A.~Sugarbaker, J.~M. Hogan, and M.~A. Kasevich.
\newblock ``{Quantum superposition at the half-metre scale}''.
\newblock \href{https://dx.doi.org/10.1038/nature16155}{Nature {\bf 528},
  530--533}~(2015).

\bibitem{Rosi:2017ieh}
G.~Rosi, G.~D'Amico, L.~Cacciapuoti, F.~Sorrentino, M.~Prevedelli, M.~Zych,
  C.~Brukner, and G.~M. Tino.
\newblock ``{Quantum test of the equivalence principle for atoms in
  superpositions of internal energy eigenstates}''.
\newblock \href{https://dx.doi.org/10.1038/ncomms15529}{Nature Commun. {\bf 8},
  5529}~(2017).
\newblock  \href{http://arxiv.org/abs/1704.02296}{arXiv:1704.02296}.

\bibitem{Marletto:2017kzi}
Chiara Marletto and Vlatko Vedral.
\newblock ``{Gravitationally-induced entanglement between two massive particles
  is sufficient evidence of quantum effects in gravity}''.
\newblock \href{https://dx.doi.org/10.1103/PhysRevLett.119.240402}{Phys. Rev.
  Lett. {\bf 119}, 240402}~(2017).
\newblock  \href{http://arxiv.org/abs/1707.06036}{arXiv:1707.06036}.

\bibitem{Bose:2017nin}
Sougato Bose, Anupam Mazumdar, Gavin~W. Morley, Hendrik Ulbricht, Marko
  Toro\v{s}, Mauro Paternostro, Andrew Geraci, Peter Barker, M.~S. Kim, and
  Gerard Milburn.
\newblock ``{Spin Entanglement Witness for Quantum Gravity}''.
\newblock \href{https://dx.doi.org/10.1103/PhysRevLett.119.240401}{Phys. Rev.
  Lett. {\bf 119}, 240401}~(2017).
\newblock  \href{http://arxiv.org/abs/1707.06050}{arXiv:1707.06050}.

\bibitem{Zych:2017tau}
Magdalena Zych, Fabio Costa, Igor Pikovski, and \v{C}aslav Brukner.
\newblock ``{Bell\textquoteright{}s theorem for temporal order}''.
\newblock \href{https://dx.doi.org/10.1038/s41467-019-11579-x}{Nature Commun.
  {\bf 10}, 3772}~(2019).
\newblock  \href{http://arxiv.org/abs/1708.00248}{arXiv:1708.00248}.

\bibitem{Goswami:2018rda}
K.~Goswami, C.~Giarmatzi, M.~Kewming, F.~Costa, C.~Branciard, J.~Romero, and
  A.~G. White.
\newblock ``{Indefinite Causal Order in a Quantum Switch}''.
\newblock \href{https://dx.doi.org/10.1103/PhysRevLett.121.090503}{Phys. Rev.
  Lett. {\bf 121}, 090503}~(2018).
\newblock  \href{http://arxiv.org/abs/1803.04302}{arXiv:1803.04302}.

\bibitem{Christodoulou:2018cmk}
Marios Christodoulou and Carlo Rovelli.
\newblock ``{On the possibility of laboratory evidence for quantum
  superposition of geometries}''.
\newblock \href{https://dx.doi.org/10.1016/j.physletb.2019.03.015}{Phys. Lett.
  B {\bf 792}, 64--68}~(2019).
\newblock  \href{http://arxiv.org/abs/1808.05842}{arXiv:1808.05842}.

\bibitem{Westphal:2020okx}
Tobias Westphal, Hans Hepach, Jeremias Pfaff, and Markus Aspelmeyer.
\newblock ``{Measurement of gravitational coupling between millimetre-sized
  masses}''.
\newblock \href{https://dx.doi.org/10.1038/s41586-021-03250-7}{Nature {\bf
  591}, 225--228}~(2021).
\newblock  \href{http://arxiv.org/abs/2009.09546}{arXiv:2009.09546}.

\bibitem{Cepollaro:2021ccc}
Carlo Cepollaro and Flaminia Giacomini.
\newblock ``{Quantum generalisation of Einstein\textquoteright{}s equivalence
  principle can be verified with entangled clocks as quantum reference
  frames}''.
\newblock \href{https://dx.doi.org/10.1088/1361-6382/ad6d26}{Class. Quant.
  Grav. {\bf 41}, 185009}~(2024).
\newblock  \href{http://arxiv.org/abs/2112.03303}{arXiv:2112.03303}.

\bibitem{Christodoulou:2022mkf}
Marios Christodoulou, Andrea Di~Biagio, Markus Aspelmeyer, \v{C}aslav Brukner,
  Carlo Rovelli, and Richard Howl.
\newblock ``{Locally Mediated Entanglement in Linearized Quantum Gravity}''.
\newblock \href{https://dx.doi.org/10.1103/PhysRevLett.130.100202}{Phys. Rev.
  Lett. {\bf 130}, 100202}~(2023).
\newblock  \href{http://arxiv.org/abs/2202.03368}{arXiv:2202.03368}.

\bibitem{Overstreet:2022zgq}
Chris Overstreet, Joseph Curti, Minjeong Kim, Peter Asenbaum, Mark~A. Kasevich,
  and Flaminia Giacomini.
\newblock ``{Inference of gravitational field superposition from quantum
  measurements}''.
\newblock \href{https://dx.doi.org/10.1103/PhysRevD.108.084038}{Phys. Rev. D
  {\bf 108}, 084038}~(2023).
\newblock  \href{http://arxiv.org/abs/2209.02214}{arXiv:2209.02214}.

\bibitem{Arzano:2022nlo}
Michele Arzano, Vittorio D'Esposito, and Giulia Gubitosi.
\newblock ``{Fundamental decoherence from quantum spacetime}''.
\newblock \href{https://dx.doi.org/10.1038/s42005-023-01159-3}{Commun. Phys.
  {\bf 6}, 242}~(2023).
\newblock  \href{http://arxiv.org/abs/2208.14119}{arXiv:2208.14119}.

\bibitem{Amelino-Camelia:2022dsj}
Giovanni Amelino-Camelia, Vittorio D'Esposito, Giuseppe Fabiano, Domenico
  Frattulillo, Philipp~A. Höhn, and Flavio Mercati.
\newblock ``{Quantum Euler angles and agency-dependent space-time}''.
\newblock \href{https://dx.doi.org/10.1093/ptep/ptae015}{PTEP {\bf 2024},
  033A01}~(2024).
\newblock  \href{http://arxiv.org/abs/2211.11347}{arXiv:2211.11347}.

\bibitem{Amelino-Camelia:2023rkg}
Giovanni Amelino-Camelia, Giuseppe Fabiano, and Domenico Frattulillo.
\newblock ``{Total momentum and other Noether charges for particles interacting
  in a quantum spacetime}''.
\newblock \href{https://dx.doi.org/10.3390/sym17020227}{Symmetry {\bf 17},
  227}~(2025).
\newblock  \href{http://arxiv.org/abs/2302.08569}{arXiv:2302.08569}.

\bibitem{Amelino-Camelia:2000stu}
Giovanni Amelino-Camelia.
\newblock ``{Relativity in space-times with short distance structure governed
  by an observer independent (Planckian) length scale}''.
\newblock \href{https://dx.doi.org/10.1142/S0218271802001330}{Int. J. Mod.
  Phys. D {\bf 11}, 35--60}~(2002).
\newblock  \href{http://arxiv.org/abs/gr-qc/0012051}{arXiv:gr-qc/0012051}.

\bibitem{Bruno:2001mw}
N.~R. Bruno, G.~Amelino-Camelia, and J.~Kowalski-Glikman.
\newblock ``{Deformed boost transformations that saturate at the Planck
  scale}''.
\newblock \href{https://dx.doi.org/10.1016/S0370-2693(01)01264-3}{Phys. Lett. B
  {\bf 522}, 133--138}~(2001).
\newblock  \href{http://arxiv.org/abs/hep-th/0107039}{arXiv:hep-th/0107039}.

\bibitem{Kowalski-Glikman:2002iba}
J.~Kowalski-Glikman and S.~Nowak.
\newblock ``{Doubly special relativity theories as different bases of
  $\kappa$-Poincaré algebra}''.
\newblock \href{https://dx.doi.org/10.1016/S0370-2693(02)02063-4}{Phys. Lett. B
  {\bf 539}, 126--132}~(2002).
\newblock  \href{http://arxiv.org/abs/hep-th/0203040}{arXiv:hep-th/0203040}.

\bibitem{Amelino-Camelia:2002cqb}
Giovanni Amelino-Camelia.
\newblock ``{Doubly special relativity}''.
\newblock \href{https://dx.doi.org/10.1038/418034a}{Nature {\bf 418},
  34--35}~(2002).
\newblock  \href{http://arxiv.org/abs/gr-qc/0207049}{arXiv:gr-qc/0207049}.

\bibitem{Kowalski-Glikman:2002oyi}
Jerzy Kowalski-Glikman.
\newblock ``{De sitter space as an arena for doubly special relativity}''.
\newblock \href{https://dx.doi.org/10.1016/S0370-2693(02)02762-4}{Phys. Lett. B
  {\bf 547}, 291--296}~(2002).
\newblock  \href{http://arxiv.org/abs/hep-th/0207279}{arXiv:hep-th/0207279}.

\bibitem{Kowalski-Glikman:2004fsz}
Jerzy Kowalski-Glikman.
\newblock ``{Introduction to doubly special relativity}''.
\newblock \href{https://dx.doi.org/10.1007/11377306_5}{Lect. Notes Phys. {\bf
  669}, 131--159}~(2005).
\newblock  \href{http://arxiv.org/abs/hep-th/0405273}{arXiv:hep-th/0405273}.

\bibitem{Amelino-Camelia:2007yca}
Giovanni Amelino-Camelia, Giulia Gubitosi, Antonino Marciano, Pierre
  Martinetti, and Flavio Mercati.
\newblock ``{A No-pure-boost uncertainty principle from spacetime
  noncommutativity}''.
\newblock \href{https://dx.doi.org/10.1016/j.physletb.2008.12.032}{Phys. Lett.
  B {\bf 671}, 298--302}~(2009).
\newblock  \href{http://arxiv.org/abs/0707.1863}{arXiv:0707.1863}.

\bibitem{Amelino-Camelia:2011lvm}
Giovanni Amelino-Camelia, Laurent Freidel, Jerzy Kowalski-Glikman, and Lee
  Smolin.
\newblock ``{The principle of relative locality}''.
\newblock \href{https://dx.doi.org/10.1103/PhysRevD.84.084010}{Phys. Rev. D
  {\bf 84}, 084010}~(2011).
\newblock  \href{http://arxiv.org/abs/1101.0931}{arXiv:1101.0931}.

\bibitem{Amelino-Camelia:2011gae}
Giovanni Amelino-Camelia.
\newblock ``{On the fate of Lorentz symmetry in relative-locality momentum
  spaces}''.
\newblock \href{https://dx.doi.org/10.1103/PhysRevD.85.084034}{Phys. Rev. D
  {\bf 85}, 084034}~(2012).
\newblock  \href{http://arxiv.org/abs/1110.5081}{arXiv:1110.5081}.

\bibitem{Amelino-Camelia:2023srg}
Giovanni Amelino-Camelia, Domenico Frattulillo, Giulia Gubitosi, Giacomo
  Rosati, and Suzana Bedi\'c.
\newblock ``{Phenomenology of DSR-relativistic in-vacuo dispersion in FLRW
  spacetime}''.
\newblock \href{https://dx.doi.org/10.1088/1475-7516/2024/01/070}{JCAP {\bf
  01}, 070}~(2024).
\newblock  \href{http://arxiv.org/abs/2307.05428}{arXiv:2307.05428}.

\bibitem{Fiore:2007vg}
Gaetano Fiore and Julius Wess.
\newblock ``{Full twisted Poincaré symmetry and QFT on Moyal-Weyl spaces}''.
\newblock \href{https://dx.doi.org/10.1103/PhysRevD.75.105022}{Phys. Rev. D
  {\bf 75}, 105022}~(2007).
\newblock  \href{http://arxiv.org/abs/hep-th/0701078}{arXiv:hep-th/0701078}.

\bibitem{Szabo:2001kg}
Richard~J. Szabo.
\newblock ``{Quantum field theory on noncommutative spaces}''.
\newblock \href{https://dx.doi.org/10.1016/S0370-1573(03)00059-0}{Phys. Rept.
  {\bf 378}, 207--299}~(2003).
\newblock  \href{http://arxiv.org/abs/hep-th/0109162}{arXiv:hep-th/0109162}.

\bibitem{Lizzi:2021rlb}
Fedele Lizzi and Flavio Mercati.
\newblock ``{$\kappa$-Poincar\'e-comodules, Braided Tensor Products and
  Noncommutative Quantum Field Theory}''.
\newblock \href{https://dx.doi.org/10.1103/PhysRevD.103.126009}{Phys. Rev. D
  {\bf 103}, 126009}~(2021).
\newblock  \href{http://arxiv.org/abs/2101.09683}{arXiv:2101.09683}.

\bibitem{DiLuca:2022idu}
Maria~Grazia Di~Luca and Flavio Mercati.
\newblock ``{New class of plane waves for \ensuremath{\kappa}-noncommutative
  quantum field theory}''.
\newblock \href{https://dx.doi.org/10.1103/PhysRevD.107.105018}{Phys. Rev. D
  {\bf 107}, 105018}~(2023).
\newblock  \href{http://arxiv.org/abs/2211.11627}{arXiv:2211.11627}.

\bibitem{Fabiano:2023xke}
Giuseppe Fabiano and Flavio Mercati.
\newblock ``{Multiparticle states in braided lightlike
  \ensuremath{\kappa}-Minkowski noncommutative QFT}''.
\newblock \href{https://dx.doi.org/10.1103/PhysRevD.109.046011}{Phys. Rev. D
  {\bf 109}, 046011}~(2024).
\newblock  \href{http://arxiv.org/abs/2310.15063}{arXiv:2310.15063}.

\bibitem{Woronowicz:1987vs}
S.~L. Woronowicz.
\newblock ``{Compact matrix pseudogroups}''.
\newblock \href{https://dx.doi.org/10.1007/BF01219077}{Commun. Math. Phys. {\bf
  111}, 613--665}~(1987).

\bibitem{Mikusch:2021kro}
Marion Mikusch, Luis~C. Barbado, and \v{c}aslav Brukner.
\newblock ``{Transformation of spin in quantum reference frames}''.
\newblock \href{https://dx.doi.org/10.1103/PhysRevResearch.3.043138}{Phys. Rev.
  Res. {\bf 3}, 043138}~(2021).
\newblock  \href{http://arxiv.org/abs/2103.05022}{arXiv:2103.05022}.

\bibitem{Major:1995yz}
Seth Major and Lee Smolin.
\newblock ``{Quantum deformation of quantum gravity}''.
\newblock \href{https://dx.doi.org/10.1016/0550-3213(96)00259-3}{Nucl. Phys. B
  {\bf 473}, 267--290}~(1996).
\newblock  \href{http://arxiv.org/abs/gr-qc/9512020}{arXiv:gr-qc/9512020}.

\bibitem{Freidel:1998pt}
Laurent Freidel and Kirill Krasnov.
\newblock ``{Spin foam models and the classical action principle}''.
\newblock \href{https://dx.doi.org/10.4310/ATMP.1998.v2.n6.a1}{Adv. Theor.
  Math. Phys. {\bf 2}, 1183--1247}~(1999).
\newblock  \href{http://arxiv.org/abs/hep-th/9807092}{arXiv:hep-th/9807092}.

\bibitem{Smolin:2002sz}
Lee Smolin.
\newblock ``{Quantum gravity with a positive cosmological constant}''~(2002).
\newblock  \href{http://arxiv.org/abs/hep-th/0209079}{arXiv:hep-th/0209079}.

\bibitem{Girelli:2022foc}
Florian Girelli and Matteo Laudonio.
\newblock ``{Group field theory on quantum groups}''~(2022).
\newblock  \href{http://arxiv.org/abs/2205.13312}{arXiv:2205.13312}.

\bibitem{Bianchi:2011uq}
Eugenio Bianchi and Carlo Rovelli.
\newblock ``{Note on the geometrical interpretation of quantum groups and
  non-commutative spaces in gravity}''.
\newblock \href{https://dx.doi.org/10.1103/PhysRevD.84.027502}{Phys. Rev. D
  {\bf 84}, 027502}~(2011).
\newblock  \href{http://arxiv.org/abs/1105.1898}{arXiv:1105.1898}.

\bibitem{Calmet:2021sws}
Xavier Calmet and Stephen D.~H. Hsu.
\newblock ``{Fundamental limit on angular measurements and rotations from
  quantum mechanics and general relativity}''.
\newblock \href{https://dx.doi.org/10.1016/j.physletb.2021.136763}{Phys. Lett.
  B {\bf 823}, 136763}~(2021).
\newblock  \href{http://arxiv.org/abs/2108.11990}{arXiv:2108.11990}.

\bibitem{Xing-ChangSong_1992}
Xing-Chang Song.
\newblock ``Spinor analysis for quantum group suq(2)''.
\newblock \href{https://dx.doi.org/10.1088/0305-4470/25/10/021}{Journal of
  Physics A: Mathematical and General {\bf 25}, 2929}~(1992).

\bibitem{Schmidt:2007yp}
Alexander Schmidt and Hartmut Wachter.
\newblock ``{Spinor calculus for q-deformed quantum spaces. I.}''~(2007).
\newblock  \href{http://arxiv.org/abs/0705.1640}{arXiv:0705.1640}.

\bibitem{franz2013idempotent}
Uwe Franz, Adam Skalski, and Reiji Tomatsu.
\newblock ``Idempotent states on compact quantum groups and their
  classification on $\mathrm{U}_q(2)$, $\mathrm{SU}_q(2)$, and
  $\mathrm{SO}_q(3)$''.
\newblock \href{https://dx.doi.org/10.4171/jncg/115}{Journal of Noncommutative
  Geometry {\bf 7}, 221–254}~(2013).
\newblock  \href{http://arxiv.org/abs/0903.2363}{arXiv:0903.2363}.

\bibitem{Chari:1994pz}
V.~Chari and A.~Pressley.
\newblock ``A guide to quantum groups''.
\newblock Cambridge University Press. ~(1994).

\bibitem{Woronowicz:1987wr}
S.~L. Woronowicz.
\newblock ``{Twisted SU(2) group: An Example of a noncommutative differential
  calculus}''.
\newblock \href{https://dx.doi.org/10.2977/prims/1195176848}{Publ. Res. Inst.
  Math. Sci. Kyoto {\bf 23}, 117--181}~(1987).

\bibitem{Murphy:1990}
Gerard~J. Murphy.
\newblock ``C*-algebras and operator theory''.
\newblock \href{https://dx.doi.org/10.1016/C2009-0-22289-6}{Academic Press}.
  ~(1990).

\bibitem{Lizzi:2018qaf}
Fedele Lizzi, Mattia Manfredonia, Flavio Mercati, and Timoth\'e Poulain.
\newblock ``{Localization and Reference Frames in $\kappa$-Minkowski
  Spacetime}''.
\newblock \href{https://dx.doi.org/10.1103/PhysRevD.99.085003}{Phys. Rev. D
  {\bf 99}, 085003}~(2019).
\newblock  \href{http://arxiv.org/abs/1811.08409}{arXiv:1811.08409}.

\bibitem{Lizzi:2019wto}
Fedele Lizzi, Mattia Manfredonia, and Flavio Mercati.
\newblock ``{Localizability in $\kappa$-Minkowski spacetime}''.
\newblock \href{https://dx.doi.org/10.1142/S0219887820400101}{Int. J. Geom.
  Meth. Mod. Phys. {\bf 17}, 2040010}~(2020).
\newblock  \href{http://arxiv.org/abs/1912.07098}{arXiv:1912.07098}.

\bibitem{Lizzi:2022hcq}
Fedele Lizzi, Luca Scala, and Patrizia Vitale.
\newblock ``{Localization and observers in \ensuremath{\varrho}-Minkowski
  spacetime}''.
\newblock \href{https://dx.doi.org/10.1103/PhysRevD.106.025023}{Phys. Rev. D
  {\bf 106}, 025023}~(2022).
\newblock  \href{http://arxiv.org/abs/2205.10862}{arXiv:2205.10862}.

\bibitem{Giacomini:2017zju}
Flaminia Giacomini, Esteban Castro-Ruiz, and \v{C}aslav Brukner.
\newblock ``{Quantum mechanics and the covariance of physical laws in quantum
  reference frames}''.
\newblock \href{https://dx.doi.org/10.1038/s41467-018-08155-0}{Nature Commun.
  {\bf 10}, 494}~(2019).
\newblock  \href{http://arxiv.org/abs/1712.07207}{arXiv:1712.07207}.

\bibitem{Vanrietvelde:2018pgb}
Augustin Vanrietvelde, Philipp~A. Höhn, Flaminia Giacomini, and Esteban
  Castro-Ruiz.
\newblock ``{A change of perspective: switching quantum reference frames via a
  perspective-neutral framework}''.
\newblock \href{https://dx.doi.org/10.22331/q-2020-01-27-225}{Quantum {\bf 4},
  225}~(2020).
\newblock  \href{http://arxiv.org/abs/1809.00556}{arXiv:1809.00556}.

\bibitem{Demichev:1997sg}
A.~Demichev.
\newblock ``{Physical meaning of quantum space-time symmetries}''~(1997).
\newblock  \href{http://arxiv.org/abs/hep-th/9701079}{arXiv:hep-th/9701079}.

\bibitem{Pati_2015}
Arun~Kumar Pati, Uttam Singh, and Urbasi Sinha.
\newblock ``Measuring non-hermitian operators via weak values''.
\newblock \href{https://dx.doi.org/10.1103/physreva.92.052120}{Physical Review
  A{\bf 92}}~(2015).
\newblock  \href{http://arxiv.org/abs/1406.3007}{arXiv:1406.3007}.

\bibitem{Kempe:2001}
J.~Kempe, D.~Bacon, D.~A. Lidar, and K.~B. Whaley.
\newblock ``Theory of decoherence-free fault-tolerant universal quantum
  computation''.
\newblock \href{https://dx.doi.org/10.1103/PhysRevA.63.042307}{Phys. Rev. A
  {\bf 63}, 042307}~(2001).
\newblock
  \href{http://arxiv.org/abs/quant-ph/0004064}{arXiv:quant-ph/0004064}.

\bibitem{Bartlett_2003}
Stephen~D. Bartlett, Terry Rudolph, and Robert~W. Spekkens.
\newblock ``Classical and quantum communication without a shared reference
  frame''.
\newblock \href{https://dx.doi.org/10.1103/physrevlett.91.027901}{Physical
  Review Letters{\bf 91}}~(2003).
\newblock
  \href{http://arxiv.org/abs/quant-ph/0302111}{arXiv:quant-ph/0302111}.

\bibitem{Bartlett:2006tzx}
Stephen~D. Bartlett, Terry Rudolph, and Robert~W. Spekkens.
\newblock ``{Reference frames, superselection rules, and quantum
  information}''.
\newblock \href{https://dx.doi.org/10.1103/RevModPhys.79.555}{Rev. Mod. Phys.
  {\bf 79}, 555--609}~(2007).
\newblock
  \href{http://arxiv.org/abs/quant-ph/0610030}{arXiv:quant-ph/0610030}.

\bibitem{Poulin:2006ryq}
David Poulin and Jon Yard.
\newblock ``{Dynamics of a quantum reference frame}''.
\newblock \href{https://dx.doi.org/10.1088/1367-2630/9/5/156}{New J. Phys. {\bf
  9}, 156--156}~(2007).
\newblock
  \href{http://arxiv.org/abs/quant-ph/0612126}{arXiv:quant-ph/0612126}.

\bibitem{Girelli:2007xn}
Florian Girelli and David Poulin.
\newblock ``{Quantum reference frames and deformed symmetries}''.
\newblock \href{https://dx.doi.org/10.1103/PhysRevD.77.104012}{Phys. Rev. D
  {\bf 77}, 104012}~(2008).
\newblock  \href{http://arxiv.org/abs/0710.4393}{arXiv:0710.4393}.

\bibitem{Aharonov1984}
Y.~Aharonov and T.~Kaufherr.
\newblock ``Quantum frames of reference''.
\newblock \href{https://dx.doi.org/10.1103/PhysRevD.30.368}{Phys. Rev. D {\bf
  30}, 368--385}~(1984).

\bibitem{Toller:1996ki}
M.~Toller.
\newblock ``{Quantum reference frames and quantum transformations}''.
\newblock Nuovo Cim. B {\bf 112}, 1013--1026~(1997).
\newblock  \href{http://arxiv.org/abs/gr-qc/9605052}{arXiv:gr-qc/9605052}.

\bibitem{Ballesteros:2020lgl}
Angel Ballesteros, Flaminia Giacomini, and Giulia Gubitosi.
\newblock ``{The group structure of dynamical transformations between quantum
  reference frames}''.
\newblock \href{https://dx.doi.org/10.22331/q-2021-06-08-470}{Quantum {\bf 5},
  470}~(2021).
\newblock  \href{http://arxiv.org/abs/2012.15769}{arXiv:2012.15769}.

\bibitem{Ballesteros:2025ypr}
Angel Ballesteros, Diego Fernandez-Silvestre, Flaminia Giacomini, and Giulia
  Gubitosi.
\newblock ``{Quantum Galilei group as quantum reference frame
  transformations}''~(2025).
\newblock  \href{http://arxiv.org/abs/2504.00569}{arXiv:2504.00569}.

\bibitem{delaHamette:2020dyi}
Anne-Catherine de~la Hamette and Thomas~D. Galley.
\newblock ``{Quantum reference frames for general symmetry groups}''.
\newblock \href{https://dx.doi.org/10.22331/q-2020-11-30-367}{Quantum {\bf 4},
  367}~(2020).
\newblock  \href{http://arxiv.org/abs/2004.14292}{arXiv:2004.14292}.

\bibitem{Muller_2021}
Markus~P. Müller.
\newblock ``{Probabilistic theories and reconstructions of quantum theory}''.
\newblock \href{https://dx.doi.org/10.21468/SciPostPhysLectNotes.28}{SciPost
  Phys. Lect. Notes {\bf 28}, 1}~(2021).
\newblock  \href{http://arxiv.org/abs/2011.01286}{arXiv:2011.01286}.

\bibitem{CHSH}
John~F. Clauser, Michael~A. Horne, Abner Shimony, and Richard~A. Holt.
\newblock ``Proposed experiment to test local hidden-variable theories''.
\newblock \href{https://dx.doi.org/10.1103/PhysRevLett.23.880}{Phys. Rev. Lett.
  {\bf 23}, 880--884}~(1969).

\bibitem{Cirelson1980QuantumGO}
B.~S. Cirel'son.
\newblock ``Quantum generalizations of bell's inequality''.
\newblock \href{https://dx.doi.org/10.1007/BF00417500}{Letters in Mathematical
  Physics {\bf 4}, 93--100}~(1980).

\bibitem{Pawlowski_2009}
Marcin Pawłowski, Tomasz Paterek, Dagomir Kaszlikowski, Valerio Scarani,
  Andreas Winter, and Marek Żukowski.
\newblock ``Information causality as a physical principle''.
\newblock \href{https://dx.doi.org/10.1038/nature08400}{Nature {\bf 461},
  1101–1104}~(2009).
\newblock  \href{http://arxiv.org/abs/0905.2292}{arXiv:0905.2292}.

\end{thebibliography}
\end{document}